\documentclass[11pt,preprintnumbers,superscriptaddress,nofootinbib]{revtex4}

\usepackage{url}
\usepackage{graphicx}
\usepackage{dcolumn}
\usepackage{bm}
\usepackage{amssymb}
\usepackage{amsmath}
\usepackage{dsfont}   
\usepackage{color}
\usepackage{slashed}
\usepackage{hyperref}
\usepackage{comment}
\usepackage[utf8]{inputenc}
\usepackage{booktabs}
\usepackage{array,multirow,makecell}
\setcellgapes{1pt}
\makegapedcells
\newcolumntype{R}[1]{>{\raggedleft\arraybackslash }b{#1}}
\newcolumntype{L}[1]{>{\raggedright\arraybackslash }b{#1}}
\newcolumntype{C}[1]{>{\centering\arraybackslash }b{#1}}
\allowdisplaybreaks

\begin{document} 


\title{Theory and Phenomenology of Two Higgs Doublet Type-II Seesaw \\ model at the LHC Run-2
}

\author{B.~Ait Ouazghour}
\email{brahim.aitouazghour@edu.uca.ac.ma}
\affiliation{LPHEA, Faculty of Science Semlalia, Cadi Ayyad University, P.O.B. 2390 Marrakech, Morocco}

\author{A.~Arhrib}
\email{aarhrib@gmail.com}
\affiliation{Faculty of Sciences and techniques, Abdelmalek Essaadi University, B.P. 416. Tanger, Morocco}

\author{R.~Benbrik}
\email{r.benbrik@uca.ac.ma}
\affiliation{LPHEA, Faculty of Science Semlalia, Cadi Ayyad University, P.O.B. 2390 Marrakech, Morocco}
\affiliation{MSISM Team, Facult\'e Polydisciplinaire de Safi, Sidi Bouzid, B.P. 4162, Safi, Morocco}

\author{M.~Chabab}
\email{mchabab@uca.ac.ma}
\affiliation{LPHEA, Faculty of Science Semlalia, Cadi Ayyad University, P.O.B. 2390 Marrakech, Morocco}

\author{L.~Rahili}
\email{rahililarbi@gmail.com}
\affiliation{LPHEA, Faculty of Science Semlalia, Cadi Ayyad University, P.O.B. 2390 Marrakech, Morocco}
\affiliation{EPTHE, Faculty of Sciences, Ibn Zohr University, B.P 8106, Agadir, Morocco}


\begin{abstract}
\vspace*{5mm}
\begin{center}
{\bf Abstract}
\end{center}
We study the most popular scalar extension of the Standard Model, namely the Two Higgs doublet model, extended by a complex triplet scalar (2HDMcT). Such considering model with a very small vacuum expectation value, provides a solution to the massive neutrinos through the so-called type II seesaw mechanism. We show that the 2HDMcT enlarged parameter space allow for a rich and interesting phenomenology compatible with current experimental constraints. In this paper the 2HDMcT is subject to a detailed scrutiny. Indeed, a complete set of tree level unitarity constraints on the coupling parameters of the potential is determined, and the exact tree-level boundedness from below constraints on these couplings are generated for all directions. We then perform an extensive parameter scan in the 2HDMcT parameter space, delimited by the above derived theoretical constraints as well as by experimental limits. We find that an important triplet admixtures are still compatible with the Higgs data and investigate which observables will allow to restrict the triplet nature most effectively in the next runs of the LHC. Finally, we emphasize new production and decay channels and their phenomenological relevance and treatment at the LHC.
\end{abstract}

\maketitle


\newpage
\section{Introduction}
After the discovery of a Standard-Model-like Higgs boson at the Large Hadron Collider (LHC) in
2012~\cite{Aad:2012tfa,Chatrchyan:2012ufa}, the Standard Model (SM) of particle physics has been established as the most
successful theory describing the elementary particles and their
interactions. Despite its success, the SM has several drawbacks that have suggested theoretical investigations as well as
experimental searches of physics beyond it. 
As an example,  the observed neutrino oscillation cannot be explained within the SM 
~\cite{Patrignani:2016xqp}. Indeed, although the SM Higgs field
is responsible for the generation of the masses of all known fundamental
particles, it is unable to accommodate the tiny observed neutrino masses. 
Within a renormalisable theory where new heavy fields are introduced, the neutrino
masses are generated via the dimension-five Weinberg
operator~\cite{Weinberg:1979sa}, this is the so called seesaw
mechanism. Different realisations of such a mechanism can be classified into
three types: type~I~\cite{Minkowski:1977sc,GellMann:1980vs,Yanagida:1979as,Mohapatra:1979ia,Schechter:1980gr} in which only right-handed neutrinos
coupling to the Higgs field, type~II~\cite{Mohapatra:1980yp,Lazarides:1980nt,
Wetterich:1981bx,Schechter:1981cv} where a new scalar field 
in the adjoint representation of $SU(2)_L$ and type~III~\cite{Foot:1988aq}
which involves two extra fermionic fields. In the above seesaw mechanisms heavy
fields are supplemented to the SM spectrum in such a way the desired neutrino
properties are reproduced once  the electroweak symmetry is broken.

In the  type II seesaw model, also dubbed Higgs  triplet models (HTM), \cite{Arhrib:2011uy,Arhrib:2011vc,Chabab:2014ara,Chabab:2015nel,Arhrib:2014nya,Camargo:2018uzw}, the  SM  Lagrangian is augmented by a $SU(2)_L$  scalar triplet field $\Delta$ with hypercharge $Y_\Delta = 2$. In HTM, neutrino masses are proportional to the vacuum expectation
value (vev) of the triplet field. Hence, the
small values of neutrino masses is guaranteed by the smallness of the triplet vev
 assumed to be less than 1 GeV and the non-conservation of lepton
number which is explicitly broken by a trilinear coupling $\mu$ term  in the HTM scalar potential. The latter, protected by symmetry, is naturally small, thus ensuring  small neutrino
masses. The model spectrum contains several scalar particles, including a pair of singly 
charged Higgs boson $(H^{\pm})$ and  doubly charged Higgs boson
$(H^{\pm\pm})$. In addition, it also predicts a CP-odd neutral scalar $(A^0)$
as well as  two CP even neutral scalars, $h$ and $H$.   The lightest scalar $h$ has
essentially the same couplings to the fermions and vector bosons as the Higgs boson of the SM within a large region of 
 the HTM parameter space.

By the new discovery of 125 scalar boson at the LHC, the phenomenology of
two-Higgs-doublet mod (2HDM) has been investigated broadly in the
literature. In the present work, due to the similarity in mass generation mechanism between
type-II seesaw and the Higgs mechanism, we extend HTM  and focus on the two Higgs
double model extension to the type-II seesaw model, displaying
phenomenological  characteristics  notably  different  from  the scalar sector
emerging from the HTM. In this context, we study several Higgs processes
giving rise to the production times branching ratios of heavy Higgs bosons
and focus on. Unlike most of the earlier studies,  we consider here a framework where both
type~I and type~II seesaw mechanisms are implemented and contribute to
neutrino mass generation. We consider a high integrated luminosity of LHC
collisions at a centre-of-mass energy of 13~TeV. 

The content of the paper is laid out as follows. In Sec. \ref{sec:thdmt}, we derive some
crucial features of the 2HDMcT, with a focus on the particle content and the
scalar potential of the model, followed by discussions on the minimization
conditions of the scalar potential and the scalar mass spectra.  Besides, in this work we investigate the impact of two new terms introduced in the model scalar potential. Sec. \ref{sec:theo-constraints}, is devoted to
the study of the theoretical constraints on the scalar potential parameters from
tree-level vacuum stability and  perturbative  unitarity  of the  scalar
sector. In Sec. \ref{sec:exp-constraints}, we further impose the LHC constraints associated with the 125 GeV Higgs boson and its signal strength to delimit the parameter space. Finally, we present some phenomenological aspects in Sec.\ref{sec:higgs-phynomenology}, benchmark points in Sec.\ref{sec:benchmark} and summarise our findings in Sec. \ref{sec:conlusion}.

\section{General considerations of 2HDM with triplet}
\label{sec:thdmt}
\subsection{The Higgs sector}
\label{sec:higgs-sector}
In a model with two Higgs doublets $H_{1,2}$, the Two Higgs Doublet Type-II Seesaw Model (2HDMcT) contains an additional $SU(2)_{L}$ triplet Higgs field $\Delta$ with hypercharge $Y = 2$ and lepton number $L = -2$,
\begin{eqnarray}
H_1&=&\left(
          \begin{array}{c}
          \phi_1^+ \\
           \phi^0_1 \\
          \end{array}
           \right){,}~~~H_2=\left(
                    \begin{array}{c}
                      \phi_2^+ \\
                      \phi^0_2 \\
                    \end{array}
                  \right){,}~~~
\Delta =\left(
\begin{array}{cc}
\delta^+/\sqrt{2} & \delta^{++} \\
\delta^0 & -\delta^+/\sqrt{2}\\
\end{array}
\right)
\end{eqnarray}
The most general renormalizable and gauge invariant Lagrangian of the 2HDMcT scalar sector is given by,
\begin{eqnarray}
\mathcal{L} &=&
(D_\mu{H})^\dagger(D^\mu{H})+Tr(D_\mu{\Delta})^\dagger(D^\mu{\Delta}) -V(H, \Delta) + \mathcal{L}_{\rm Yukawa}
\label{eq:2THMT}
\end{eqnarray}
where the scalar potential $V(H, \Delta)$, symmetric under a group $SU(2)_L \times U(1)_Y$, reads as \cite{Chen-Nomura-2014}
\begin{eqnarray}
V(H_i, \Delta) = V(H_1, H_2) + V(\Delta) + V_{int}(H_1, H_2, \Delta)
\label{eq:Vpot}
\end{eqnarray}
where :
\begin{eqnarray}
V(H_1, H_2) &=& m^2_{1}\, H_1^\dagger H_1 + m^2_{2}\, H_2^\dagger H_2 - m_{3}^2\, \left(H_1^\dagger H_2 + H_2^\dagger H_1\right) +\frac{\lambda_1}{2} (H_1^\dagger H_1)^2 + \frac{\lambda_2}{2}  (H_2^\dagger H_2)^2 \nonumber\\
&& + \lambda_3\, H_1^\dagger H_1\, H_2^\dagger H_2 + \lambda_4\, H_1^\dagger
H_2\, H_2^\dagger H_1 +\frac{\lambda_5}{2} \left[(H_1^\dagger H_2)^2+
  (H_2^\dagger H_1)^2 \right] \label{eq:VH1H2}\\
V(\Delta) &=& m^2_{\Delta}\, Tr(\Delta^{\dagger}{\Delta}) +\bar{\lambda}_8(Tr\Delta^{\dagger}{\Delta})^2 + \bar{\lambda}_9Tr(\Delta^{\dagger}{\Delta})^2\\
 V_{int}(H_1, H_2, \Delta)&=& [\mu_1 H_1^T{i}\sigma^2\Delta^{\dagger}H_1 + \mu_2H_2^T{i}\sigma^2\Delta^{\dagger}H_2 + \mu_3 H_1^T{i}\sigma^2\Delta^{\dagger}H_2  + {\rm h.c.}]\nonumber\\
&& +\lambda_6\,H_1^\dagger H_1 Tr\Delta^{\dagger}{\Delta} +\lambda_7\,H_2^\dagger H_2 Tr\Delta^{\dagger}{\Delta} + \lambda_8\,H_1^\dagger{\Delta}\Delta^{\dagger} H_1 + \lambda_9\,H_2^\dagger{\Delta}\Delta^{\dagger} H_2 \label{eq:VH1H2Delta}
\label{eq:VDelta}
\end{eqnarray}
\noindent
In the above, $m^2_i$, i=1,2,3 and $m^2_\Delta$ are mass squared parameters, $\lambda_i$, i=1,...,5
are dimensionless couplings not related to the triplet, 
${\bar\lambda_i}$, i=8,9  are dimensionless couplings related to the
triplet field, while $\mu_i$, i=1,2,3 with $\lambda_i$, i=6,...,9, are dimensionless
couplings that mixe all three Higgs fields. In  Eq. \ref{eq:VDelta}, $Tr$ denotes the trace over $2\times2$ matrices, where for convenience we have used the $2 \times 2$ traceless matrix representation for the triplet. Also, the potential defined in Eq. \ref{eq:VDelta} exhausts all possible gauge invariant renormalizable operators. For instance, a terms of the form $\lambda_{10}H_{1}^{\dagger}\Delta^{\dagger}{\Delta} H_1$ and $\lambda_{11}H_{2}^{\dagger}\Delta^{\dagger}{\Delta} H_2$ \cite{Chen-Nomura-2014}, which would be legitimate to add if $\Delta$ contained a singlet component, can actually be projected on the $\lambda_{6,7}$ and $\lambda_{8,9}$ operators appearing in Eq. \ref{eq:VDelta} thanks to the identity $H_{i}^{\dagger}\Delta^{\dagger}\Delta\,H_{i} + H_{i}^{\dagger}\Delta\Delta^{\dagger}H_{i}$ $=H_{i}^{\dagger} H_{i} Tr(\Delta^{\dagger}{\Delta})$ which is valid because $\Delta$ is a traceless $2\times 2$ matrix.

Subsequently, we will assume that all these parameters are real 
valued. Indeed, apart from the $\mu_i$ terms, all the other operators in 
$V(H_i, \Delta)$ are self-conjugate so that, by hermicity of the potential,
only the real parts of the $\lambda$'s and the $m_{1}^2, m_{2}^2, m_\Delta^2$
mass parameters are relevant. As for $\mu_i$, the only parameters that
can pick up a would be CP-phases, these phases are unphysical and can always
be absorbed in a redefinition of the fields $H_i$ and $\Delta$. One thus concludes that the 2HDMcT Lagrangian is CP conserving.
The electroweak symmetry is spontaneously  broken when the neutral components
of the Higgs fields acquire vacuum expectation values $v_1$, $v_2$ and
$v_t$. Thus we can  shift the Higgs fields in the following way,
\begin{eqnarray}
\phi^0_1 = \frac{v_1+\rho_1+i\eta_1}{\sqrt{2}}, \quad\quad \phi^0_2 =\frac{v_2+\rho_2+i\eta_2}{\sqrt{2}},\quad\quad \delta^0 = \frac{v_t+\rho_0+i\eta_0}{\sqrt{2}}
\end{eqnarray}
finding minimization conditions, or tree-level tadpole equations, given by
\begin{eqnarray}
V_{linear} = T_1 \rho_1 + T_2 \rho_2 + T_3 \rho_0 =0,
\end{eqnarray}
where it is safe to take $T_i = 0$ ( i=1,2,3) which leads to, 
\begin{eqnarray}
m_{1}^2 &=& \frac{2\,m_{3}^2\,v_2 + \sqrt{2}(2\,\mu_1{v_1} + \mu_3\,v_2)\,v_t - v_1\,(\lambda_1\,v_1^2 + \lambda_{345}\,v_2^2 + \lambda_{68}^{+}\,v_t^2)}{2\,v_1}  \label{eq:ewsb1}\\
m_{2}^2 &=& \frac{2\,m_{3}^2\,v_1 + \sqrt{2}(2\,\mu_2{v_2} + \mu_3\,v_1)\,v_t - v_2\,(\lambda_2\,v_2^2 + \lambda_{345}\,v_1^2 + \lambda_{79}^{+}\,v_t^2)}{2\,v_2}   \label{eq:ewsb2}\\
m^2_\Delta &=& \frac{\sqrt{2}(\mu_1{v_1^2} + v_2\,(\mu_3{v_1}+\mu_2{v_2})) -v_t(\lambda_{68}^{+} v_1^2 + 2 \bar{\lambda}_{89}^{+} v_t^2 + \lambda_{79}^{+} v_2^2)}{2\,v_t} \label{eq:ewsb3}
\end{eqnarray}
with $\lambda_{ij}^{+} = \lambda_i+\lambda_j$, $\bar{\lambda}_{ij}^{+} = \bar{\lambda}_i+\bar{\lambda}_j$ and $\lambda_{345} = \lambda_3+\lambda_4+\lambda_5$.
\noindent  
 If the terms associated to $v_t^3$ are omitted in Eq. \ref{eq:ewsb3}, then we can derive a new expression for $v_t$ as a function of the triplet scalar mass, 
\begin{eqnarray}
v_t \approx \frac{\mu_1 v_1^2 + \mu_3 v_1 v_2 + 
  \mu_2 v_2^2 }{\sqrt{2} (m^2_\Delta + \lambda_{68}^+ v_1^2/2 + \lambda_{79}^+ v_2^2/2)}
\label{MDelta_vs_vt}
\end{eqnarray}
Furthermore, for $m_\Delta$ sufficiently large compared to $v_{1,2}$, we see that the above formula reduces to, 
$v_t \sim \mu_1 v^2_1 / \sqrt{2} m^2_\Delta$, which is referred as type II seesaw
mechanism. \footnote{In the absence of $\mu_1$, $\mu_2$ and $\mu_3$ the
  $m^2_{\Delta}$ becomes negative leading to a spontaneous violation of
lepton number. The resulting Higgs spectrum contains a massless triplet
scalar, called Majoron $J$. This model was excluded by LEP.}

The 2HDMcT model has altogether $24$ degrees of freedom: 21 parameters originating from the scalar potential given by
Eq. \ref{eq:Vpot} and tree vacuum expectation values of the Higgs doublets and triplet fields. However, thanks to the three minimisation conditions, the $W$ gauge boson mass and the correct electroweak scales, the parameters $m_1^2$, $m_2^2$, $m_\Delta^2$ and $v$ can be eliminated.

\subsection{Higgs masses and mixing angles}
\label{sec:masses-mixing}
In what follows, we will use Eqs. \ref{eq:ewsb1},
\ref{eq:ewsb2} and \ref{eq:ewsb3} to trade the
mass parameters $m_{11}^2$, $m_{22}^2$ and $m_{\Delta}^2$ for the rest of
parameters given in the potential. Thus, the $14 \times 14$ squared mass matrix is given by,
\begin{equation}
 {\mathcal M}^2=\frac{1}{2} \frac{\partial^2 V}{\partial \eta_i^2} |_{H_{i} = \langle H_{i} \rangle , \Delta = \langle \Delta \rangle}
\end{equation}
by denoting the corresponding VEV's
\begin{eqnarray}
\langle H_1 \rangle =\left(
                    \begin{array}{c}
                      0 \\
                       v_1/\sqrt{2}\\
                    \end{array}
                  \right)~~~~{\rm ,}~~~~\langle H_2 \rangle =\left(
                    \begin{array}{c}
                      0 \\
                       v_2/\sqrt{2}\\
                    \end{array}
                  \right)~~~{\rm and}~~~~\langle \Delta \rangle &=\left(
\begin{array}{cc}
0 & 0 \\
v_t/\sqrt{2} & 0\\
\end{array}
\right)~~~,~~~~~
\label{eq:VEVs}
\end{eqnarray}
\noindent
Eq. \ref{eq:Vpot} can be recast in a block diagonal form of one doubly degenerate eigenvalue $m_{H^{\pm\pm}}^2$
and three $3 \times 3$ matrices denoted in the following by ${\mathcal{M}}_{\pm}^2$, ${\mathcal{M}}_{{\mathcal{CP}}_{odd}}^2$ and ${\mathcal{M}}_{{\mathcal{CP}}_{even}}^2$. The bilinear part of the Higgs potential is then given by:
\begin{eqnarray}
V^{(2)}_{H_1,H_2,\Delta}&=&\frac{1}{2} \begin{pmatrix} \rho_1,\rho_2, \delta^0 \end{pmatrix} 
\mathcal{M}_{{\mathcal{CP}}_{even}}^2
\begin{pmatrix} \rho_1 \\ \rho_2 \\ \rho_0 \end{pmatrix} +
\frac{1}{2} \begin{pmatrix} \eta_1, \eta_2, \eta_0 \end{pmatrix}
\mathcal{M}_{{\mathcal{CP}}_{odd}}^2
\begin{pmatrix} \eta_1 \\ \eta_2 \\ \eta_0 \end{pmatrix} \nonumber\\
&&+\begin{pmatrix} \phi^-_1,\phi^-_2,\delta^-  \end{pmatrix}
\mathcal{M}_{\pm}^2
\begin{pmatrix} \phi^+_1 \\ \phi^+_2\\ \delta^+  \end{pmatrix} +
\delta^{++}{\mathcal{M}_{\pm\pm}^2}\delta^{--} \cdots, 
\end{eqnarray}
at tree-level. The elements of these mass matrices are explicitly presented below.
\subsubsection*{Mass of the doubly charged field}
%
The double eigenvalue $m_{H^{\pm\pm}}^2$, corresponding to the doubly charged eigenstate $\delta^{\pm\pm}$, can simply be determined by collecting all the coefficients of $\delta^{++}\delta^{--}$ in the scalar potential. It is given by reads,
\begin{eqnarray}
m_{H^{\pm\pm}}^2=\frac{\sqrt{2}\mu_1 v_1^2 + \sqrt{2}\mu_3 v_1 v_2 + \sqrt{2}\mu_2 v_2^2 - \lambda_8 v_1^2 v_t
- \lambda_9 v_2^2 v_t - 2 \bar{\lambda}_9 v_t^3}{2v_t}  \label{eq:mHpmpm}
\end{eqnarray}
\subsubsection*{Mass of the simply charged field}
%
The mass-squared matrix for the simply charged field in the ($\phi_1^{-},\phi_2^{-},\delta^{-}$) basis reads as:
\begin{eqnarray}
{\mathcal{M}}_{\pm}^2= \left(
\begin{array}{ccc}
{\mathcal{M}}^{\pm}_{11} & \frac{1}{2} \left(\lambda_{45}^+ v_1 v_2-2 m_{3}^2\right) & \frac{1}{4} \left(v_1 A - 2 \mu_3 v_2\right) \\
\frac{1}{2} \left(\lambda_{45}^+ v_1 v_2-2 m_{3}^2\right) & {\mathcal{M}}^{\pm}_{22} & \frac{1}{4} \left(v_2 B-2 \mu_3 v_1\right) \\
\frac{1}{4} \left(v_1 A - 2 \mu_3 v_2\right) & \frac{1}{4} \left(v_2 B-2 \mu_3 v_1\right) & {\mathcal{M}}^{\pm}_{33}\\
\end{array}
\right)
\label{matrix_charged}
\end{eqnarray}
where $A=\sqrt{2} \lambda_8 v_t-4\mu_1$, $B=\sqrt{2} \lambda_9 v_t-4\mu_2$ and the diagonal terms are given by,
\begin{eqnarray}
{\mathcal{M}}^{\pm}_{11} &=& \frac{2 m_{3}^2 v_2+v_1 v_t \left(2 \sqrt{2} \mu_1-\lambda_8 v_t\right)+\sqrt{2} \mu_3 v_2 v_t- \lambda_{45}^+ v_1 v_2^2}{2 v_1} \nonumber\\
{\mathcal{M}}^{\pm}_{22} &=& \frac{2 m_{3}^2 v_1+v_2 v_t \left(2\sqrt{2} \mu_2-\lambda_9 v_t \right)+\sqrt{2} \mu_3 v_1 v_t - \lambda_{45}^+ v_1^2 v_2}{2 v_2} \nonumber\\
{\mathcal{M}}^{\pm}_{33} &=& \frac{v_1^2 \left(2\sqrt{2} \mu_1-\lambda_8 v_t\right)+v_2^2 \left(2\sqrt{2} \mu_2-\lambda_9 v_t\right)+2 \sqrt{2} \mu_3 v_2 v_1-2\bar{\lambda_9}v_t^3}{4 v_t}  
\label{diago-charged}
\end{eqnarray}
Among the three eigenvalues of this matrix, one is zero and corresponds to the charged Goldstone
bosons $G^\pm$, while the two others correspond to the singly
charged Higgs bosons denoted by $m_{H_{1}^\pm}^2$ and $m_{H_{2}^\pm}^2$ given by,
\begin{eqnarray}
m^2_{H^\pm_{1,2}}=\frac{1}{4 v_0^2 v_t}\Big[-v_0 \left(v_0 \left(2 {\mathcal{M}}_{12}^\pm \text{cs}_{\beta } \text{se}_{\beta } v_t+\kappa \right)+2 \sqrt{2} v_t^2 \left({\mathcal{M}}_{23}^\pm \text{cs}_{\beta }+{\mathcal{M}}_{13}^\pm \text{se}_{\beta }\right)\right)\mp\text{cs}_{\beta } \text{se}_{\beta }\sqrt{ \mathcal{Y}}\Big]
\end{eqnarray}
where $\text{c}_x$, $\text{s}_x$, $\text{cs}_x$, $\text{se}_x$ stand for the $\cos(x)$, $\sin(x)$, $\csc(x)$, $\sec(x)$ respectively, while $v_0=\sqrt{v_1^2+v_2^2}$, $v=\sqrt{v_1^2+v_2^2+2 v_t^2}$, $\kappa=\sqrt{2} v_0 \left({\mathcal{M}}_{13}^\pm c_{\beta }+{\mathcal{M}}_{23}^\pm s_{\beta }\right)$ and
\begin{eqnarray}
\mathcal{Y}&=&v_0^2 \Big(\big(v_0 \left(\kappa  c_{\beta } s_{\beta }+2 {\mathcal{M}}^{\pm}_{12} v_t \right)+2 \sqrt{2} v_t^2 \left({\mathcal{M}}^{\pm}_{23} c_{\beta }+{\mathcal{M}}^{\pm}_{13} s_{\beta }\right)\big)^2\nonumber\\
&-&4 v^2 s_{2 \beta } v_t \big(\kappa  {\mathcal{M}}^{\pm}_{12}+2 {\mathcal{M}}^{\pm}_{13} {\mathcal{M}}^{\pm}_{23} v_t\big)\Big)
\end{eqnarray}
\noindent 
The above symmetric squared matrix ${\mathcal{M}}_{\pm}^2$ is diagonalized via ${\mathcal{C}}$ as:
\begin{eqnarray}
{\mathcal{C}}{\mathcal{M}}_{\pm}^2{\mathcal{C}}^T&=&diag(m^2_{G^\pm},m^2_{H_1^\pm},m^2_{H_2^\pm})
\label{rota-matrix-charged}
\end{eqnarray}
where the ${\mathcal{C}}$ rotation matrix is described by three mixing angles $\theta^\pm_1$, $\theta^\pm_2$ and $\theta^\pm_3$, and the corresponding expressions
for the ${\mathcal{C}}$ elements are give in Appendix \ref{sec-diagonal} as a function of the parameters inputs of our model.

\subsubsection*{Mass of the neutral pseudo-scalar field}
As to the mass-squared matrix for the neutral ${{\mathcal{CP}}_{odd}}$ pseudoscalar field in the basis ($\eta_{1},\eta_{2},\eta_{0}$), it is expressed as:
\begin{eqnarray}
{\mathcal{M}}_{odd}^2=\left(
  \begin{array}{ccc}
   {\mathcal{M}}_{11}^{odd}  & \frac{-2 m_{3}^2+\sqrt{2} \mu_3 v_t+2 \lambda_5 v_1 v_2}{2}&-\frac{2 \mu_1 v_1+\mu_3 v_2}{\sqrt{2}}\\
   \frac{-2 m_{3}^2+\sqrt{2} \mu_3 v_t+2 \lambda_5 v_1 v_2}{2}  &  {\mathcal{M}}_{22}^{odd}  &  -\frac{2 \mu_2 v_2+\mu_3 v_1}{\sqrt{2}}\\
   -\frac{2 \mu_1 v_1+\mu_3 v_2}{\sqrt{2}}&  -\frac{2 \mu_2 v_2+\mu_3 v_1}{\sqrt{2}}&  {\mathcal{M}}_{33}^{odd}      
  \end{array}
\right)
\label{matrix-cp-odd}
\end{eqnarray}
the diagonal terms are given by,
\begin{eqnarray}
{\mathcal{M}}_{11}^{odd} &=& \frac{2 m_{3}^2 v_2+ 4 \sqrt{2} \mu_1v_1 v_t +\sqrt{2} \mu_3 v_2 v_t-2 \lambda_5 v_1 v_2^2}{2 v_1} \nonumber \\
{\mathcal{M}}_{22}^{odd} &=& \frac{2 m_{3}^2 v_1+4 \sqrt{2} \mu_2 v_2 v_t+\sqrt{2} \mu_3 v_1 v_t-2 \lambda_5 v_1^2 v_2}{2 v_2}  \nonumber \\
{\mathcal{M}}_{33}^{odd} &=& \frac{\eta_1 v_1^2+\mu_3 v_2 v_1+\mu_2 v_2^2}{\sqrt{2} v_t} 
\label{diago-cp-odd}
\end{eqnarray}
here, ${\mathcal{M}}_{{\mathcal{CP}}_{odd}}^2$ has three eigenvalues, one is zero corresponding to the neutral Goldstone boson $G^0$ while the two others are the physical states ${{\mathcal{CP}}_{odd}}$ $A_1$ and $A_2$, \textcolor{black}{by setting,
\begin{eqnarray}
\varrho &=& {\mathcal{M}}_{23}^{odd} c_{\beta}+{\mathcal{M}}_{13}^{odd} s_{\beta}\nonumber\\
\digamma &=& {\mathcal{M}}_{13}^{odd} c_{\beta}+{\mathcal{M}}_{23}^{odd} s_{\beta}\\
\mathcal{X} &=& \Big[v_0^2 \left(\left(v_0^2 \digamma  c_{\beta } s_{\beta }+2 {\mathcal{M}}_{12}^{odd} v_0 v_t+4 \varrho  v_t^2\right){}^2-4 s_{2 \beta } v_t \left(4 v_t^2+v_0^2\right) \left(2 {\mathcal{M}}_{13}^{odd} {\mathcal{M}}_{23}^{od} v_t+{\mathcal{M}}_{12}^{odd} v_0 \digamma \right)\right)\Big]\nonumber
\end{eqnarray}
theirs masses read as,
\begin{eqnarray}
m^2_{A_{1,2}}=\frac{1}{4 v_0^2 v_t}\Big[-v_0 \left(2 \text{cs}_{\beta} \text{se}_{\beta} v_t \left({\mathcal{M}}_{12}^{odd} v_0+2 \varrho  v_t\right)+v_0^2 \digamma \right)\mp\text{cs}_{\beta } \text{se}_{\beta}\sqrt{ \mathcal{X}}\Big]
\end{eqnarray}
while the diagonalization of such matrix is done in this case by the introduction of an unitary matrix ${\mathcal{O}}$ described by three mixing angles $\beta_1$, $\beta_2$ and $\beta_3$ whose expressions can be found in the appendix \ref{sec-diagonal} as,
\begin{eqnarray}
{\mathcal{O}}{\mathcal{M}}_{{{\mathcal{CP}}_{odd}}}^2{\mathcal{O}}^T&=&diag(m^2_{G^0},m^2_{A_1},m^2_{A_2})
\label{rota-matrix-cp-odd}
\end{eqnarray}}
%
\subsubsection*{Mass of the neutral scalar field}
In the basis ($\rho_{1},\rho_{2},\rho_{0}$) the neutral scalar mass matrix reads :
\begin{eqnarray}
{\mathcal{M}}_{{\mathcal{CP}}_{even}}^2=\left(
  \begin{array}{ccc}
   m_{\rho_{1}\rho_{1}}^2  &  m_{\rho_{2}\rho_{1}}^2  &  m_{\rho_0\rho_{1}}^2   \\
   m_{\rho_{1}\rho_{2}}^2  &  m_{\rho_{2}\rho_{2}}^2  &  m_{\rho_0\rho_{2}}^2   \\
   m_{\rho_{1}\rho_0}^2  &  m_{\rho_{2}\rho_0}^2  &  m_{\rho_0\rho_0}^2      
  \end{array}
\right)
\label{matrix-cp-even}   
\end{eqnarray}
Its diagonal terms are,
\begin{eqnarray}
m_{\rho_{1}\rho_{1}}^2 &=& \lambda_1 v_1^2+\frac{v_2 \left(\sqrt{2} m_{3}^2 + \mu_3 v_t\right)}{\sqrt{2} v_1} \nonumber\\
m_{\rho_{2}\rho_{2}}^2 &=& \lambda_2 v_2^2+\frac{v_1 \left(\sqrt{2} m_{3}^2+ \mu_3 v_t\right)}{\sqrt{2} v_2} \nonumber\\
m_{\rho_0\rho_0}^2 &=& \frac{4 \left(\bar{\lambda}_{8}+\bar{\lambda}_{9}\right) v_t^3+\sqrt{2} \left(\mu_1 v_1^2+\mu_3 v_2 v_1+\mu_2 v_2^2\right)}{2 v_t}
\label{diago-cp-even}
\end{eqnarray}
while the off-diagonal terms are given by,
\begin{eqnarray}
m_{\rho_{2}\rho_{1}}^2 &=& m_{\rho_{1}\rho_{2}}^2 = \frac{1}{\sqrt{2}} \left(\sqrt{2} v_1 v_2 \lambda_{345} - \sqrt{2} m_{3}^2-\mu_3 v_t\right) \nonumber\\
m_{\rho_0\rho_{1}}^2 &=& m_{\rho_{1}\rho_0}^2 = \frac{1}{\sqrt{2}} \left(\sqrt{2} v_1 v_t (\lambda_6+\lambda_8) - (2 \mu_1 v_1 + \mu_3 v_2)\right) \nonumber\\
m_{\rho_0\rho_{2}}^2 &=& m_{\rho_{2}\rho_0}^2 = \frac{1}{\sqrt{2}} \left(\sqrt{2} v_2 v_t (\lambda_7+\lambda_9) - (2 \mu_2 v_2 + \mu_3 v_1)\right) 
\label{off-diago-cp-even} 
\end{eqnarray}
The mass matrix can be diagonalised by an orthogonal matrix ${\mathcal{E}}$ which we parametrise as
\begin{eqnarray}
{\mathcal{E}} =\left( \begin{array}{ccc}
c_{\alpha_1} c_{\alpha_2} & s_{\alpha_1} c_{\alpha_2} & s_{\alpha_2}\\
-(c_{\alpha_1} s_{\alpha_2} s_{\alpha_3} + s_{\alpha_1} c_{\alpha_3})
& c_{\alpha_1} c_{\alpha_3} - s_{\alpha_1} s_{\alpha_2} s_{\alpha_3}
& c_{\alpha_2} s_{\alpha_3} \\
- c_{\alpha_1} s_{\alpha_2} c_{\alpha_3} + s_{\alpha_1} s_{\alpha_3} &
-(c_{\alpha_1} s_{\alpha_3} + s_{\alpha_1} s_{\alpha_2} c_{\alpha_3})
& c_{\alpha_2}  c_{\alpha_3}
\end{array} \right)
\label{eq:mixingmatrix}
\end{eqnarray}
where the mixing angles $\alpha_1$, $\alpha_2$ and $\alpha_3$ can be chosen in the range
\begin{eqnarray}
- \frac{\pi}{2} \le \alpha_{1,2,3} \le \frac{\pi}{2} \;.
\end{eqnarray}
the rotation between the two basis ($\rho_{1},\rho_{2},\rho_{0}$) and ($h_1,h_2,h_3$) diagonalises the mass matrix ${\mathcal{M}}_{{\mathcal{CP}}_{even}}^2$ as,
\begin{eqnarray}
{\mathcal{E}}{\mathcal{M}}_{{{\mathcal{CP}}_{even}}}^2{\mathcal{E}}^T&=&diag(m^2_{h_1},m^2_{h_2},m^2_{h_3})
\label{rota-matrix-cp-even}
\end{eqnarray}
and leads to three mass eigenstates, ordered by ascending mass as:
\begin{eqnarray}
m^2_{h_1} < m^2_{h_2} < m^2_{h_3} \;.
\end{eqnarray}
One choice of input parameters implemented in 2HDMcT consistes to use the
following hybrid parameterisation,
\begin{eqnarray}
\mathcal{P}_I = \left\{\alpha_1,\alpha_2,\alpha_3,m_{h_1},m_{h_2},m_{h_3},m_{H^{\pm\pm}}, \lambda_{1},\lambda_{3},\lambda_{4},\lambda_{6},\lambda_{8},\bar{\lambda}_{8},\bar{\lambda}_{9},\mu_1,v_t,\tan\beta \right\}
\label{eq:set-para1}
\end{eqnarray}
in which $\tan\beta=v_2/v_1=\tan\theta^\pm_1=\tan\beta_1$. 

In Appendix \ref{appendice:A}, we discuss the second choice of
input parameters in the physical basis 2HDMcT. Using the Eqs. \ref{rota-matrix-cp-even} and \ref{eq:mHpmpm}, one can easily express the reset of Lagrangian parameters in terms of those given by. \ref{eq:set-para1}. These are given by
\begin{eqnarray}
&& \lambda_2 = \frac{-\mathcal{B} c_{\beta }^2+\lambda_1 v_0^2 c_{\beta }^4+\left({\mathcal{E}}_{12}^2 m_{h_1}^2+{\mathcal{E}}_{22}^2 m_{h_2}^2+{\mathcal{E}}_{32}^2 m_{h_3}^2\right)s_{\beta }^2}{v_0^2 s_{\beta }^4}\nonumber\\
&& \lambda_5 =\frac{\left(\mathcal{B} - \lambda^+_{34} v_0^2 s_{\beta }^2\right)c_{\beta }-\lambda_1v_0^2c_{\beta }^3+\left({\mathcal{E}}_{11} {\mathcal{E}}_{12} m_{h_1}^2+{\mathcal{E}}_{21} {\mathcal{E}}_{22} m_{h_2}^2+{\mathcal{E}}_{31} {\mathcal{E}}_{32} m_{h_3}^2\right)s_{\beta }}{v_0^2 c_{\beta } s_{\beta }^2}\nonumber\\
&& \lambda_9 = \frac{-\lambda_8 v_0^2 c_{\beta }^2+2 \mathcal{F} - 2 m_{H^{\pm\pm}}^2-2 \left(2 \bar{\lambda}_8+3 \bar{\lambda}_9\right) v_t^2}{v_0^2 s_{\beta }^2}\nonumber\\
&& \lambda_7 =\frac{v_0 \left(\mathcal{A} c_{\beta }+\mathcal{M} s_{\beta }\right)+v_t \left(-\lambda_6 v_0^2 c_{\beta }^2-2 \mathcal{F}+2 m_{H^{\pm\pm}}^2+2 \left(2 \bar{\lambda}_8+3 \bar{\lambda}_9 \right) v_t^2\right)}{v_0^2 s_{\beta }^2 v_t}\nonumber\\
&& \mu_2 = \frac{c_{\beta } \left(2 \mathcal{A}+v_0 c_{\beta } \left(\sqrt{2} \mu _1-2 \lambda_{68}^+ v_t\right)\right)}{\sqrt{2} v_0 s_{\beta }^2}\nonumber\\
&& \mu_3 =\frac{\sqrt{2} \left(-\mathcal{A} v_0 c_{\beta }+v_0^2 c_{\beta }^2 \left(\lambda_{68}^+ v_t-\sqrt{2} \mu _1\right)+v_t \left({\mathcal{E}}_{13}^2 m_{h_1}^2+{\mathcal{E}}_{23}^2 m_{h_2}^2+{\mathcal{E}}_{33}^2 m_{h_3}^2-2 \bar{\lambda}_{89}^+ v_t^2\right)\right)}{v_0^2 c_{\beta } s_{\beta }}\nonumber\\
&& m_{3}^2 = \frac{\mathcal{A} v_t+v_0 c_{\beta } \left(\mathcal{B}-v_t \left(\lambda_{68}^+ v_t-\sqrt{2} \mu _1\right)\right)-\lambda_1 v_0^3 c_{\beta }^3}{v_0 s_{\beta }}
\end{eqnarray}
where $v_0=\sqrt{v_1^2+v_2^2}$ and
\begin{eqnarray}
&& \mathcal{A}={\mathcal{E}}_{11} {\mathcal{E}}_{13} m_{h_1}^2+{\mathcal{E}}_{21} {\mathcal{E}}_{23} m_{h_2}^2+{\mathcal{E}}_{31} {\mathcal{E}}_{33} m_{h_3}^2\nonumber\\
&& \mathcal{B} = {\mathcal{E}}_{11}^2 m_{h_1}^2+{\mathcal{E}}_{21}^2 m_{h_2}^2+{\mathcal{E}}_{31}^2 m_{h_3}^2\nonumber\\
&& \mathcal{F} = {\mathcal{E}}_{13}^2 m_{h_1}^2+{\mathcal{E}}_{23}^2 m_{h_2}^2+{\mathcal{E}}_{33}^2 m_{h_3}^2\nonumber\\
&& \mathcal{M} = {\mathcal{E}}_{12} {\mathcal{E}}_{13} m_{h_1}^2+{\mathcal{E}}_{22} {\mathcal{E}}_{23} m_{h_2}^2+{\mathcal{E}}_{32} {\mathcal{E}}_{33} m_{h_3}^2
\end{eqnarray}
the remaining 10 parameters consist of the 6 charged and ${{\mathcal{CP}}_{odd}}$ sectors mixing angles given respectively by $\theta^\pm_i$ (i=1,2,3),
\begin{eqnarray}
&& \theta^\pm_2 = \sin^{-1}\left( {\mathcal{C}}_{13} \right)\nonumber\\
&& \theta^\pm_1 = \cos^{-1}\left( {\mathcal{C}}_{11}/\cos\theta^\pm_2 \right)\nonumber\\
&& \theta^\pm_3 = \cos^{-1}\left( {\mathcal{C}}_{33}/\cos\theta^\pm_2 \right)
\end{eqnarray}
and $\beta_j$ (j=1,2,3)
\begin{eqnarray}
&& \beta_2 = \sin^{-1}\left( {\mathcal{O}}_{13} \right)\nonumber\\
&& \beta_1 = \cos^{-1}\left( {\mathcal{O}}_{11}/\cos\beta_2 \right)\nonumber\\
&& \beta_3 = \cos^{-1}\left( {\mathcal{O}}_{33}/\cos\beta_2 \right)
\end{eqnarray}

\noindent
and 4 Higgs bosons masses, two of them correspond to the charged states $H_{1,2}^\pm$ masses, while the two others are matched to ${{\mathcal{CP}}_{odd}}$ states $A_{1,2}$ as discussed previously.

\subsection{Yukawa and gauge bosons textures}
\label{sec:yukawa-gbosons}
$\mathcal{L}_{\rm Yukawa}$ contains all the Yukawa sector
of the SM plus one extra Yukawa term that leads after spontaneous symmetry breaking to (Majorana) mass terms for
the neutrinos, without requiring right-handed neutrino states,
\begin{equation}
-\mathcal{L}_{\rm Yukawa} \supset  - Y_{\nu} L^T C \otimes i \sigma^2 \Delta L  + {\rm h.c.} \label{eq:yukawa}
\end{equation}

\noindent
\textcolor{black}{
where $L$ denotes $SU(2)_L$ doublets of left-handed leptons, $Y_{\nu}$ denotes neutrino Yukawa couplings,  
$C$ the charge conjugation operator. The $\mathbb{Z}_2$ symmetry is imposed in order to avoid tree-level FCNCs. Furthermore, and in terms of the various $\alpha_i$ which appear in the expressions of $\mathcal{E}_{ij}$ matrix elements, we liste in table-\ref{Ycoupl}, all the ${\mathcal{CP}}_{even}$ $h_i$ (i=1,2,3) Yukawa couplings for both type I and type II in the model.}

\begin{table}[!h]
\begin{center}
\setcellgapes{4pt}
\begin{tabular}{|C{1.5cm}|C{1cm}|C{1cm}|C{1cm}|C{1cm}|C{1cm}|C{1cm}|}
\hline  & $C^{h_1}_U$    & $C^{h_1}_D$   &   $C^{h_2}_U$   &   $C^{h_2}_D$   &   $C^{h_3}_U$  &   $C^{h_3}_D$ \\
\hline  type-I & $\displaystyle\frac{\mathcal{E}_{12}}{s_\beta} $ & $\displaystyle\frac{\mathcal{E}_{12}}{s_\beta} $& $\displaystyle\frac{\mathcal{E}_{22}}{s_\beta} $ & $\displaystyle\frac{\mathcal{E}_{22}}{s_\beta} $ & $\displaystyle\frac{\mathcal{E}_{32}}{s_\beta} $ & $\displaystyle\frac{\mathcal{E}_{32}}{s_\beta} $ \\
\hline  type-II &$\displaystyle\frac{\mathcal{E}_{12}}{s_\beta} $& $\displaystyle\frac{\mathcal{E}_{11}}{c_\beta} $ & $\displaystyle\frac{\mathcal{E}_{22}}{s_\beta} $ & $\displaystyle\frac{\mathcal{E}_{21}}{c_\beta} $ & $\displaystyle\frac{\mathcal{E}_{32}}{s_\beta} $ &$\displaystyle\frac{\mathcal{E}_{31}}{c_\beta} $ \\
\hline 
\end{tabular}
\end{center} 
\caption{Normalized Yukawa couplings coefficients of the neutral Higgs bosons $h_i$ 
to the up-quarks, down-quarks ($u,d$) in 2HDMcT.}
\label{Ycoupl}   
\end{table}

\begin{table}[!h]
\begin{center}
\setcellgapes{4pt}
\begin{tabular}{|C{1cm}|C{4cm}|C{4cm}|}
\hline  
& $C^{h_i}_W$    & $C^{h_i}_Z$  \\
\hline  $h_1$ & $\displaystyle{\frac{v_1}{v} \mathcal{E}_{11} + \frac{v_2}{v} \mathcal{E}_{21} + 2\,\frac{v_t}{v} \mathcal{E}_{31}}$ & 
$\displaystyle{\frac{v_1}{v} \mathcal{E}_{11} + \frac{v_2}{v} \mathcal{E}_{21} + 4\,\frac{v_t}{v} \mathcal{E}_{31}}$  \\
\hline  $h_2$ & $\displaystyle{\frac{v_1}{v} \mathcal{E}_{12} + \frac{v_2}{v} \mathcal{E}_{22} + 2\,\frac{v_t}{v} \mathcal{E}_{32}}$ & 
$\displaystyle{\frac{v_1}{v} \mathcal{E}_{12} + \frac{v_2}{v} \mathcal{E}_{22} + 4\,\frac{v_t}{v} \mathcal{E}_{32}}$  \\
\hline  $h_3$ & $\displaystyle{\frac{v_1}{v} \mathcal{E}_{13} + \frac{v_2}{v} \mathcal{E}_{23} + 2\,\frac{v_t}{v} \mathcal{E}_{33}}$ & 
$\displaystyle{\frac{v_1}{v} \mathcal{E}_{13} + \frac{v_2}{v} \mathcal{E}_{23} + 4\,\frac{v_t}{v} \mathcal{E}_{33}}$  \\
\hline 
\end{tabular}
\caption{The normalized couplings of the neutral ${\mathcal{CP}}_{even}$ $H_i$ Higgs bosons
to the massive gauge bosons $V=W,Z$ in 2HDMcT.}
\label{table3}
\end{center}
\end{table}

On the other hand, expanding the covariant derivative ${\rm D}_\mu$, and performing the usual transformations on the
gauge and scalar fields to obtain the physical fields, one can identify the Higgs couplings $H_i$ to the massive gauge bosons $V=W,Z$ as given in table-\ref{table3}. Note that in our model, The triplet field $\Delta$ does directly couple to the SM particles, so a new contribution will be appears, and the two couplings $C^{h_i}_V$ ($V=W^\pm, Z$) differs from one to another by a factor 2 associated to $v_t$.
%
%

\section{Theoretical Constraints}
\label{sec:theo-constraints}

\subsection{Unitarity}
\label{sec:unita-const}
Physics beyond the Standard Model (BSM) refers to the theoretical developments needed to explain the deficiencies of the SM, and any SM extension might tend to reproduce the entirety of current phenomena. Here the question usually addressed is 
which theory BSM is the right one, can only be settled
via experiments. In our paper, we will globally scrutinise the 2HDMcT model,
looking for the space of parameters allowed by all the theoretical constraints
as well as experimental ones. In this subsection, we apply perturbative unitarity to a complex triplet field $\Delta$ coupled to the two Higgs doublet field of 2HDM, and as usual we consider the elastic $2 \to 2$ scattering processes for this purpose. The explicit formulas for all the eigenvalues are given by,
\begin{eqnarray}
&& a_{1}^\pm = \frac{1}{2}(\lambda_1+\lambda_2 \pm \sqrt{(\lambda_1-\lambda_2)^2 + 4\lambda_5^2},
\hspace{0.4cm}
a_{2}^\pm= \frac{1}{2} \left(\lambda_1+\lambda_2\pm\sqrt{\lambda_1^2-2 \lambda_2 \lambda_1+\lambda_2^2+4\lambda_5^2}\right) \nonumber\\
&& a_{3}^\pm = \lambda_3 \pm \lambda_4, \hspace{0.4cm} 
a_{4}^\pm = \lambda_3 \pm \lambda_5,\hspace{0.4cm}  
a_{5}^\pm = \lambda_3 +2\lambda_4 \pm 3\lambda_5,\hspace{0.4cm} 
a_{6} = \lambda_6,\hspace{0.4cm} 
a_{7} = \lambda_7,\hspace{0.4cm} 
a_{8} = 2\bar{\lambda}_8  \nonumber\\
&& a_{9} = \lambda_6 + \lambda_8,\hspace{0.4cm}  
a_{10} = \lambda_7 + \lambda_9, \hspace{0.4cm}  
a_{11} = \lambda_6 + \frac{3}{2}\lambda_8, \hspace{0.4cm}
a_{12} = \lambda_6 - \frac{1}{2}\lambda_8, \hspace{0.4cm}
a_{13} = \lambda_7 + \frac{3}{2}\lambda_9  \nonumber\\
&& a_{14} = \lambda_7 - \frac{1}{2}\lambda_9, \hspace{0.4cm} 
a_{15} = 2(\bar{\lambda}_8 + \bar{\lambda}_9), \hspace{0.4cm} 
a_{16} = 2\bar{\lambda}_8+\bar{\lambda}_9 
\end{eqnarray}
in addition with three other $a_{17,18,19}$ eigenvalues originating from the cubic polynomial equation,
\begin{eqnarray}
&& x^3-\big(\lambda_1+\lambda_2+2\bar{\lambda}_8+4\bar{\lambda}_9)\,x^2+(\lambda_1 \lambda_2-\lambda_4^2 - \lambda_8^2+2\lambda_1 \bar{\lambda}_8+2\lambda_2 \bar{\lambda}_8-\lambda_9^2+4\lambda_1 \bar{\lambda}_9+4\lambda_2 \bar{\lambda}_9)\,x \nonumber\\
&&+(\lambda_2\lambda_8^2 - 2 \lambda_1 \lambda_2 \bar{\lambda}_8 + 2 \lambda_4^2 \bar{\lambda}_8 - 2 \lambda_4 \lambda_8 \lambda_9 + \lambda_1 \lambda_9^2 - 4 \lambda_1 \lambda_2\bar{\lambda}_9 + 4 \lambda_4^2 \bar{\lambda}_9 ) = 0
\label{eq:cubic-polynom}
\end{eqnarray}
for which the solutions have been extensively detailed in Appendix \ref{unita-appendix}, Eq. \ref{eq:a19}. 
To ensure the unitarity constraints, the absolute value of the above eigenvalues must be bounded from above as
\cite{Arhrib:2000is, Akeroyd:2000wc},
\begin{eqnarray}
|a_{i}|  \le 0.5 \,\quad\quad\quad i = 1,...,19 
\end{eqnarray}
In Appendix \ref{unita-appendix}, we describe in more detail the
scattering matrix of all the two-body processes in the scalar sector of the
2HDMcT model.

\subsection{Boundedness From Below (BFB)}
\label{sec:bfb-const}
In order to derive the BFB constraints, we require that the vacuum is stable at tree level. Generically, this means that the scalar potential has to be bounded from below at large scalar fields values in any directions of the field space. Thus, the potential is asymptotically dominated by the quartic terms,
\begin{eqnarray}
V^{(4)}(H_1,H_2,\Delta) &=& \frac{\lambda_1}{2} (H_1^\dagger H_1)^2 + \frac{\lambda_2}{2} (H_2^\dagger H_2)^2 + \lambda_3\, H_1^\dagger H_1\, H_2^\dagger H_2 + \lambda_4\, H_1^\dagger H_2\, H_2^\dagger H_1\nonumber\\
&+&\frac{\lambda_5}{2} \left[(H_1^\dagger H_2)^2+ (H_2^\dagger H_1)^2 \right] + \lambda_6\,H_1^\dagger H_1Tr\Delta^{\dagger}{\Delta} + \lambda_7\,H_2^\dagger H_2Tr\Delta^{\dagger}{\Delta}\nonumber\\
&+& \lambda_8\,H_1^\dagger{\Delta}\Delta^{\dagger} H_1 + \lambda_9\,H_2^\dagger{\Delta}\Delta^{\dagger} H_2 + \bar{\lambda}_8 (Tr\Delta^{\dagger}{\Delta})^2+\bar{\lambda}_9 Tr(\Delta^{\dagger}{\Delta})^2
\label{eq:Vquartic4}
\end{eqnarray}
Therefore, to obtain  in this model the full set of BFB conditions valid for all directions, it is suitable to only consider $V^{(4)}(H_1,H_2,\Delta) $ than to study the full scalar potential. As an illustration, consider for instance the case where there is no coupling
between doublets $H_{i}$ and triplet $\Delta$ Higgs bosons, i.e. $\lambda_3 = \lambda_4 = \lambda_5 = \lambda_6 = \lambda_7  = \lambda_8 = \lambda_9 =0$. Obviously, one can see that,  
\begin{equation}
\lambda_1 > 0 \;\;{\rm \&}\;\; \lambda_2 > 0 \;\;{\rm \&}\;\; \bar{\lambda}_8 > 0  \;\;{\rm \&}\;\; \bar{\lambda}_9 > 0 
\end{equation}
Given this,  all the other necessary and sufficient conditions for stability are listed in Appendix \ref{bfb-appendix} and can be read as,
\begin{eqnarray}
\Omega_{\text{2HDM}} \cup \Omega_1 \cup \Omega_2 \cup \Omega_3 \cup \Omega_4 \cup \Omega_5
\end{eqnarray}
with
\begin{eqnarray}
\Omega_{\text{2HDM}} &=& \Bigg\{ \lambda_1\,,\,\lambda_2\, > 0\,,\,\lambda_3 + \sqrt{\lambda_1\lambda_2} > 0 \,,\,\lambda_3 + \lambda_4 - |\lambda_5| + \sqrt{\lambda_1\lambda_2} > 0 \Bigg\} \label{eq:omega1}
\end{eqnarray}
the corresponding \text{2HDM} BFB constraints well known in the literature, whereas the $\Omega_i\,(i=1,..,5)$ stand for the new contraints added as follows, 
\begin{eqnarray}
\Omega_1 &=& \Bigg\{ \lambda_6\,,\,\lambda_7 > 0\,,\,\lambda_6 + \lambda_8 > 0\,,\,\lambda_7 + \lambda_9 > 0\,,\,\bar{\lambda}_8 + \bar{\lambda}_9 > 0\,,\,\bar{\lambda}_8 + \frac{\bar{\lambda}_9}{2} > 0 \Bigg\} \label{eq:omega1}\nonumber\\
\Omega_2 &=& \Bigg\{ \bigg(\bar{\lambda}_9\sqrt{2\lambda_1}\ge |\lambda_8|\sqrt{\bar{\lambda}_8 + \bar{\lambda}_9} \bigg) \,\text{or}\, \bigg(2\lambda_6+\lambda_8+\sqrt{\big(4\lambda_1\bar{\lambda}_9-\lambda_8^2\big)\big(1+2\frac{\bar{\lambda}_8}{\bar{\lambda}_9}\big)} \bigg) \Bigg\}\label{eq:omega2}\nonumber\\
\Omega_3 &=& \Bigg\{ \bigg(\bar{\lambda}_9\sqrt{2\lambda_2}\ge |\lambda_9|\sqrt{\bar{\lambda}_8 + \bar{\lambda}_9} \bigg) \,\text{or}\, \bigg(2\lambda_7+\lambda_9+\sqrt{\big(4\lambda_2\bar{\lambda}_9-\lambda_9^2\big)\big(1+2\frac{\bar{\lambda}_8}{\bar{\lambda}_9}\big)} \bigg) \Bigg\}\label{eq:omega3}\nonumber\\
\Omega_4 &=& \Bigg\{ \lambda_6+\sqrt{2\lambda_1(\bar{\lambda}_8 + \bar{\lambda}_9)}>0\,,\, \lambda_6+\lambda_8+\sqrt{2\lambda_1(\bar{\lambda}_8 + \bar{\lambda}_9)}>0 \Bigg\}\label{eq:omega4}\nonumber\\
\Omega_5 &=& \Bigg\{ \lambda_7+\sqrt{2\lambda_2(\bar{\lambda}_8 + \bar{\lambda}_9)}>0\,,\, \lambda_7+\lambda_9+\sqrt{2\lambda_2(\bar{\lambda}_8 + \bar{\lambda}_9)}>0 \Bigg\}\label{eq:omega5}
\end{eqnarray}
besides other are mentioned in Appendix \ref{bfb-appendix}.
\subsection{Bounds from theoretical constraints}
\label{sec:result-the}
\begin{figure}[!h]
\hspace{-0.2cm}
\begin{minipage}{5.62cm}
\begin{center}
\includegraphics[height =5.6cm,width=5.6cm]{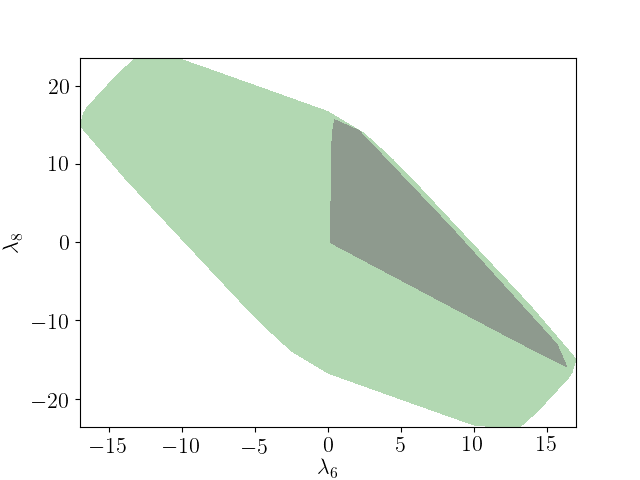}
\end{center}
\end{minipage}
\hspace{-0.4cm}
\begin{minipage}{5.62cm}
\begin{center}
\includegraphics[height =5.6cm,width=5.6cm]{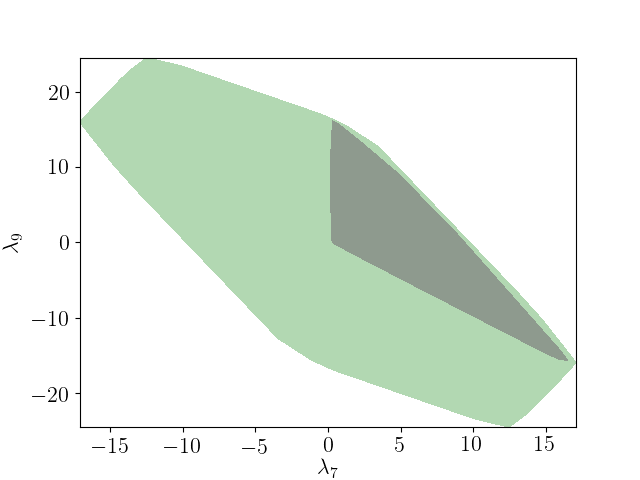}
\end{center}
\end{minipage}
\hspace{-0.4cm}
\begin{minipage}{5.62cm}
\begin{center}
\includegraphics[height =5.6cm,width=5.6cm]{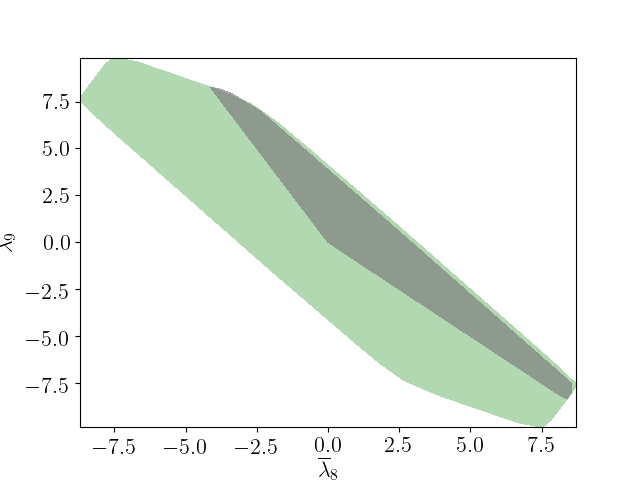}
\end{center}
\end{minipage}
\caption{The allowed ranges of $(\lambda_6,\lambda_8)$ (left), 
$(\lambda_7,\lambda_9)$ (middle) and $(\bar{\lambda}_8,\bar{\lambda}_9)$
(right) by imposing unitarity (green), combined unitarity and BFB (grey) constraints.}
\label{fig:0}
\end{figure}

In order to validate our rough analytical understanding and to further explore
the impact of the unitarity and BFB constraints we use the numerical
machinery. There are many possibilities what to use as input
parameters. Naively using the initial Lagrangian parameters will hardly
produce points which are in agreement with the Higgs measurements. Therefore,
we trade hybrid parameterization. With that choice, the full set of parameters
is given by Eq. \ref{eq:set-para1}. As first step, we show in Fig. \ref{fig:0},
all generated points in the planes $\lambda_6$ vs $\lambda_8$,
$\lambda_7$ vs $\lambda_9$ and $\bar{\lambda}_8$ vs
$\bar{\lambda}_9$. 

In the 2HDMcT, the presence of the triplet field implies a new scalar couplings $\lambda_{6,7,8,9}$ et $\bar{\lambda}_{8,9}$ and the vacuum stability condition requires that not only $\lambda_{1,2}\ge 0$ but also $\lambda_{7,8}\ge 0$ with the conditions in Eq. (\ref{eq:omega5}). By varying $\lambda_{1,...5}$ in $[-8\pi:8\pi]$, we show in Fig. \ref{fig:0}, the allowed domains on $\lambda_{6,7,8,9}$ et $\bar{\lambda}_{8,9}$ plans without conflicting with the theoretical constraints. We assume that the seesaw mechanism at the TeV scale that we consider here, is a "low-energy" effective phenomenological manifestation at high energy scale. We therefore assume that the couplings remain perturbative up to GUT scale.

\section{Limits from experimental constraints}
\label{sec:exp-constraints}
\subsection{Oblic parameters}
\label{sec:bfb-const}
Strong indirect probe of physics beyond SM is provided by the oblique parameters S, T and U. More than that, the calculations of several observables check their dependences on those oblique parameters, for example, and not as a limitation, the $\rho$ parameter \cite{veltman75}, i.e. $\rho=m_W^2/m_Z^2c_W^2$. In the SM, this parameter is equal to 1 at tree level. In the THDMcT, the new triplet contributions to $W$ and $Z$ masses readily form Eq. \ref{eq:VEVs} and the kinetic terms in Eq. \ref{eq:2THMT} leads to write,
\begin{eqnarray}
&& m_Z^2=\frac{g^2\,(v_1^2+v_2^2+4v_t^2)}{4\,\cos^2\theta_W} \label{eq:mZ2}\\
&& m_W^2=\frac{g^2\,(v_1^2+v_2^2+2v_t^2)}{4}=\frac{g^2\,v^2}{4}\label{eq:mW2}
\end{eqnarray}
\textcolor{black}{the modified form of the $\rho$ parameter reads}
\begin{eqnarray}
\rho = \frac{v_0^2+2v_t^2}{v_0^2+4v_t^2} \approx 1-2 \frac{v_t^2}{v_0^2} =1+\delta\rho \ne 0 \label{eq:rho-thdmt}
\end{eqnarray}
The impact of a 2HDMcT in the so-called electroweak precision requires
that $\rho$ to be close to its SM value: $\rho=1.0004^{+0.0003}_{-0.0004}$
\cite{pdg2016}. Then, one gets an upper bound for $v_t
< 5$ GeV. Furthermore, the major contribution to the T-parameter comes from
the loops involving the scalar triplet when $v_t$ equal to zero or
less. Also, since deviations from the Standard Model expectations in $U$ are negligible
\cite{negU}, then we will assume the latter to be zero and consider
only $S$ and $T$. We compute their $\chi^2_{ST}$ contribution through, 
\begin{eqnarray}
   \chi^2_{ST}=\left(\begin{array}{cc}
   S-S^{best} & T-T^{best}\end{array} \right)\left(\begin{array}{cc}
  0.0085 & 0.0063\\
  0.0063 & 0.0057\end{array}\right)^{-1}\left(\begin{array}{c}
  S-S^{best}\\
  T-T^{best}\end{array}\right)
\end{eqnarray}
For $U=0$,the electroweak fit gives the values.
\begin{eqnarray}
S^{best}=0.06\;\;\;\;\;\;,\;\;\;\;\;T^{best}=0.097
\end{eqnarray}
The contribution of the scalar triplet to $S$ and $T$ reads as \cite{sharma2012}

\begin{eqnarray}
&& \text{T} = \text{T}_{2HDMcT}=\frac{1}{4\pi s^2_w m^2_W}\bigg[\sum_{+} F(m_{++}^2,m_+^2)+\sum_{+,\,0}F(m_{+}^2,m_0^2)\bigg] \\
&& \text{S} = \text{S}_{2HDMcT}=-\frac{1}{3\pi}\bigg[\sum_{0}\log\left(\frac{m_{++}^2}{m_{0}^2}\right)+6\sum_{0}\xi\left(\frac{m_{0}^2}{m_Z^2},\frac{m_{0}^2}{m_Z^2}\right)+6(1-2s^2_w)^2\xi\left(\frac{m_{++}^2}{m_Z^2},\frac{m_{++}^2}{m_Z^2}\right)\nonumber\\
&&\hspace{0.6cm}+6 s^4_w \sum_{+}\xi\left(\frac{m_{+}^2}{m_Z^2},\frac{m_{+}^2}{m_Z^2}\right)\bigg]
\end{eqnarray}

where, $m_{++}=m_{H^{\pm\pm}}$, $m_+=m_{H_1^\pm},m_{H_2^\pm}$ and $m_0=m_{h_2},m_{h_3}$, while $s_w$ stands for the sinus of the Weinberg angle $\theta_w$. The function $\xi(x,y)$ and $F(x,y)$ are defined by,
\begin{eqnarray}
\hspace*{-9cm}F(x,y)=\frac{x+y}{2}-\frac{xy}{x-y}\ln\left(\frac{x}{y}\right)
\end{eqnarray}

\begin{eqnarray}
\hspace*{-3cm}\xi(x,y)&=&\frac{4}{9}-\frac{5}{12}\left(x+y\right)+\frac{1}{6}\left(x-y\right)^2+\frac{1}{4}\Big[x^2-y^2-\frac{1}{3}\left(x-y\right)^3-\frac{x^2+y^2}{x-y}\Big]\ln\frac{x}{y}\nonumber\\
&-&\frac{1}{12}d\left(x,y\right)f\left(x,y\right)
\end{eqnarray}
\begin{eqnarray}
\hspace*{-2cm}f\left(x,y\right)=\left\lbrace\begin{matrix}
-2\sqrt{d\left(x,y\right)}\Big[\arctan\frac{x-y+1}{\sqrt{d\left(x,y\right)}}-\arctan\frac{x-y-1}{\sqrt{d\left(x,y\right)}}&for&d\left(x,y\right)>0\\
0&for&d\left(x,y\right)=0\\
\sqrt{-d\left(x,y\right)}\ln\Big[\frac{x+y-1+\sqrt{-d\left(x,y\right)}}{x+y-1-\sqrt{-d\left(x,y\right)}}\Big]&for&d\left(x,y\right)<0
\end{matrix}\right.
\end{eqnarray}
\begin{eqnarray}
\hspace*{-8cm}d\left(x,y\right)=-1+2\left(x+y\right)-\left(x-y\right)^2
\end{eqnarray}
\subsection{Direct LHC and LEP constraints}
The HiggsBounds code \cite{HiggsBounds} is used to test a model against experimental data from LEP, Tevatron and  the LHC. In our analysis we use part function of HiggsBounds version $5.2.0beta$ with the latest constraints for heavy Higgs bosons. The required input for HiggsBounds program are masses for all the scalar, the effective Higgs couplings, the total
decay widths for all scalar and the branching ratios. The exclusion test at $2\sigma$ is then performed on the five physical scalars of our model. HiggsBounds returns a binary result indicating if the specific model point has been excluded at 95$\%$ C.L. or not. 

Both experiments ATLAS and CMS at LHC reported the discovery of a scalar particule with mass around 125.09 GeV \cite{masse_higgs}. Meanwhile, this has been strengthen  further by ATLAS and CMS with the first 13 TeV results. In our model, we identify the Higgs field $H_1$ with the observed SM-like Higgs boson with a mass:
\begin{eqnarray}
m_{h_1} = 125.09 \pm 0.4 \,\text{GeV}
\end{eqnarray}
We include both the measured signal rates from the ATLAS and CMS Run I and Run II and their combinaisons in our study via the public code HiggsSignals-2.2.1beta \cite{HiggsSignals}. 

\noindent
The global $\chi$-square is defined by:
\begin{eqnarray}
\chi^2=\chi^{2\,(8\,TeV)}_{HS}+\chi^{2\,(\,13TeV)}_{HS}+\chi^2_{ST}
\end{eqnarray}
we then determine the minimal $\chi^2$ value over the scanned parameter space, $\chi^2_{min}$, and keep the allowed parameter space that features a $\chi^2$ value within $\Delta\chi^2=\chi^2-\chi^2_{min}\leq2.3,\,5.99,\,11.8$ (which corresponds respectively to 68$\%$ CL, 95.5$\%$ CL and 99.7$\%$ CL.)

\section{Light and heavy Higgs Phenomenology}
\label{sec:higgs-phenomenology}
In this section we study the influence of the  constraints presented in the previous section
(indirect, LEP, Tevatron and LHC constraints) on the free parameters. For this purpose
we generate a set of $10^9$ points randomly for each of the two different types of model defined in Tables \ref{Ycoupl} and \ref{table3} with random values for each of the free parameters. The available ranges we use in the
simulation are given:
\begin{eqnarray}
\begin{matrix}
m_{h_1}\leq m_{h_2}\leq m_{h_3}\leq 1\,\text{TeV},\,\,80\,\text{GeV}\leq m_{H^{\pm\pm}}\leq 1\,\text{TeV},\,\,\\ \frac{-\pi}{2}\leq \alpha_{1,2,3}\leq \frac{\pi}{2}\\
1\leq \tan\beta\leq 40,\,\,-10^2\leq \mu_1\leq 10^2,\,\,v_t=1 \,\text{GeV},\,\,\,\lambda_1\approx0.15,\,\,\lambda_3\approx1.6,\,\,\lambda_4\approx 1.6
\end{matrix}
\end{eqnarray} 

\begin{figure}[!h]
\hspace{-0.2cm}
\begin{minipage}{5.62cm}
\begin{center}
\includegraphics[height =5.6cm,width=5.6cm]{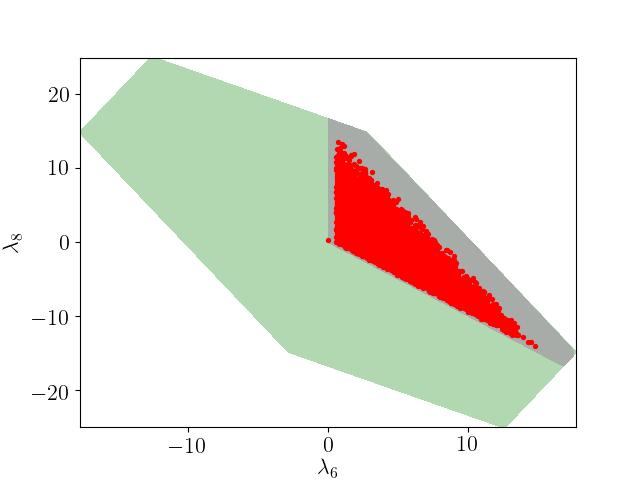}
\end{center}
\end{minipage}
\hspace{-0.4cm}
\begin{minipage}{5.62cm}
\begin{center}
\includegraphics[height =5.6cm,width=5.6cm]{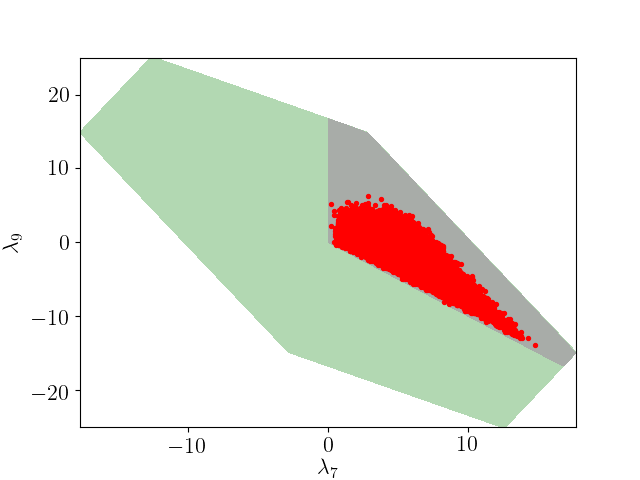}
\end{center}
\end{minipage}
\hspace{-0.4cm}
\begin{minipage}{5.62cm}
\begin{center}
\includegraphics[height =5.6cm,width=5.6cm]{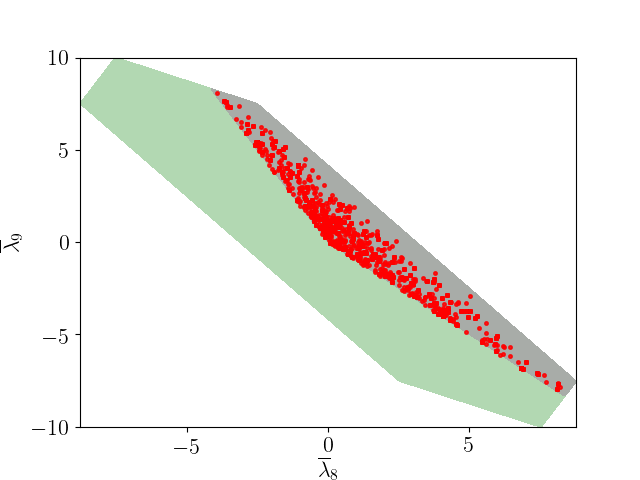}
\end{center}
\end{minipage}
\caption{The largest $(\lambda_6,\lambda_8)$ (left), 
$(\lambda_7,\lambda_8)$ (center) and $(\bar{\lambda}_8,\bar{\lambda}_9)$ (right) domain allowed  by $C_1$ (gray) and  $C_2$ (red)}
\label{fig:1}
\end{figure}

$\lambda_1$, $\lambda_3$ and $\lambda_4$ are set respectively to the values $0.15$, $1.6$ and $1.6$ for the sake of simplification. The ranges for $\lambda_6$, $\lambda_7$, $\lambda_8$, $\lambda_9$, $\bar{\lambda}_8$ and $\bar{\lambda}_9$ resulted from
the unitarity and boundedness constraints as can be seen from Fig. \ref{fig:1}. For simplicity, let us classify the dimensionless parameters in the scalar potential into two different sets according to the following two types of constraints:
\begin{itemize}
	\item First set of constraints includes the unitarity, vacuum stability and BFB
	constraints as well as non-tachyonic masses. We
	refer to this set as $C_1$.
	\item The second set of constraints contains $C_1$ and constraints from Higgs data. We refer to this set as $C_2$.
\end{itemize}
%
In Fig. \ref{fig:1}, all the generated points are plotted in the planes $\lambda_6$ vs $\lambda_8$, $\lambda_7$ vs $\lambda_9$ and $\bar{\lambda}_8$ vs $\bar{\lambda}_9$,
as we can see $C_2$ set of constraints reduce the above domain of $\lambda_6$, $\lambda_7$, $\lambda_8$, $\lambda_9$, $\bar{\lambda}_8$ and $\bar{\lambda}_9$. 

Looking now at the plane $\bar{\lambda}_8$ vs
$\bar{\lambda}_9$  are not very much restricted by $C_2$ constraint due to the fact that
$\bar{\lambda}_8$ and $\bar{\lambda}_9$  are always dependent of the vev of scalar triplet.

\begin{figure}[!h]
\centering	
\includegraphics[height =7.6cm,width=7.6cm]{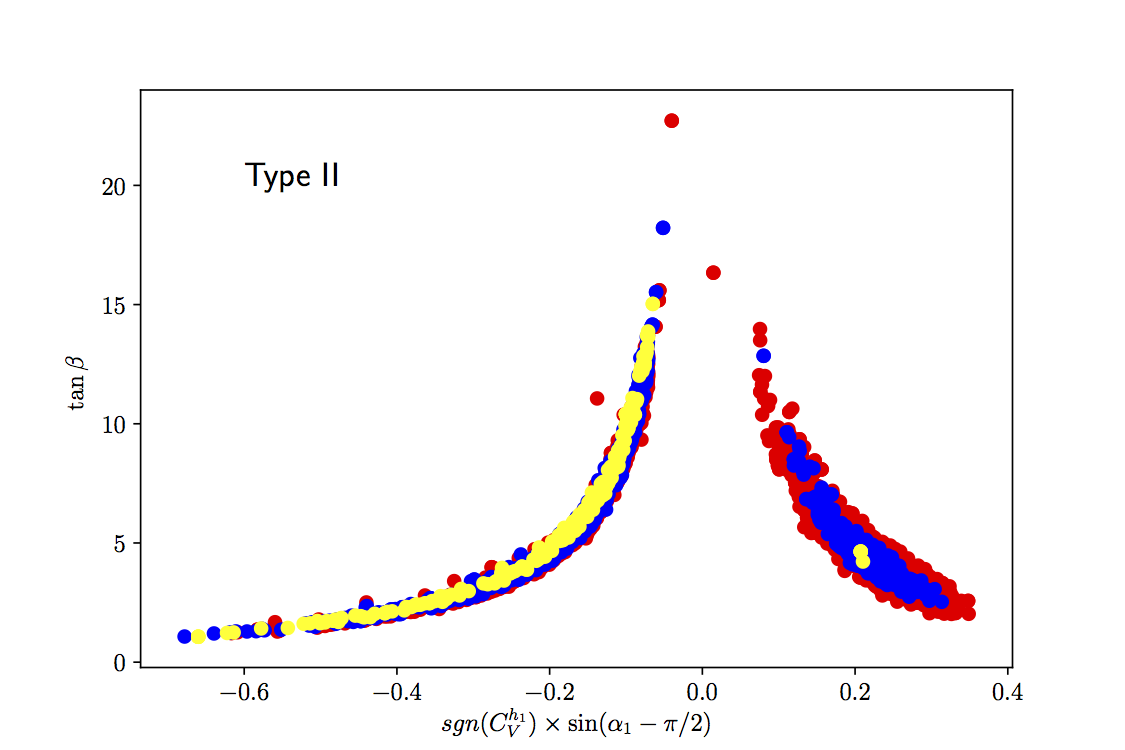}
\includegraphics[height =7.6cm,width=7.6cm]{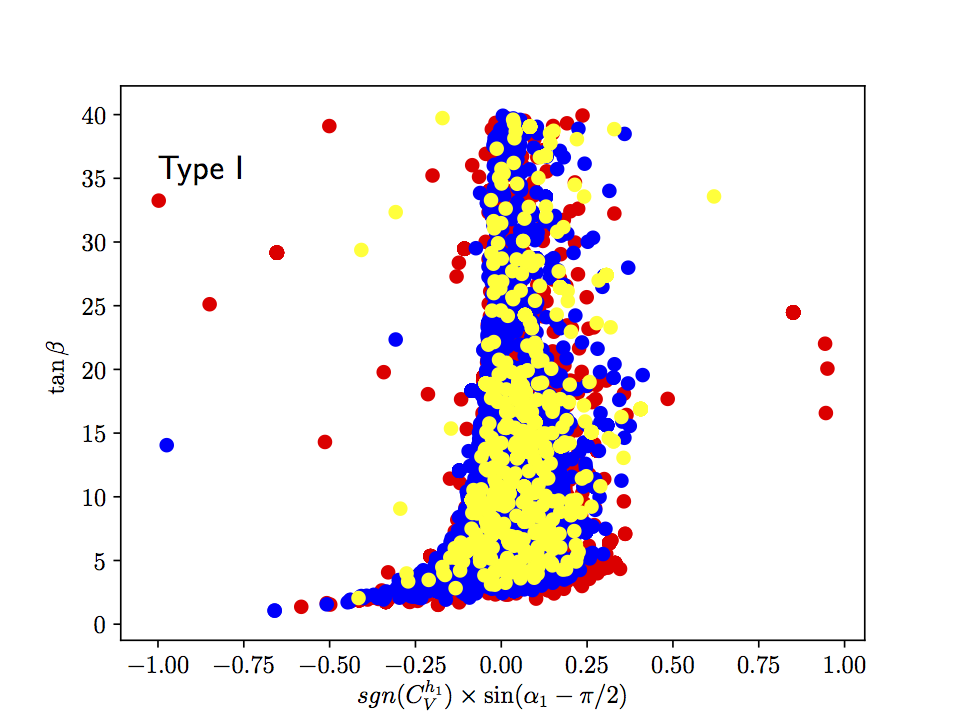}
\caption[titre court]{The allowed regions in ($sgn(C_V^{h_1})\sin(\alpha_1-\pi/2)$, $\tan\beta$) after imposing $C_1$ and $C_2$ constraints, where the left and right panels represent the allowed values
in type-II(left), type-I(right), respectively. The errors for
$\chi$-square fit are 99.7\% CL (red), 95.5\% CL (blue) and 68\% CL (yellow).}
\label{fig:2}
\end{figure}

In Fig. \ref{fig:2}, we present the allowed points in the  $(\rm{sign}(C_V^{h_1})\sin(\alpha_1-\pi/2)\,,\,\tan\beta)$ plane, that passes all constraints in type II (left) and type I (right) at $1\sigma$, $2\sigma$ and $3\sigma$. 
In type-II, one can appreciate that the mixing angle $\alpha_1$ seems more constrained than in type-I. Results are shown by imposing the conditions $C_1$ and $C_2$. The later has a strong impact on how the mixing angles are constrained. Fig.\ref{fig:2}(left) displays wrong-sing Yukawa couplings scenario at 2$\sigma$. 

\begin{figure}
\begin{minipage}{6.5cm}
\begin{center}
\includegraphics[height =6cm,width=6.5cm]{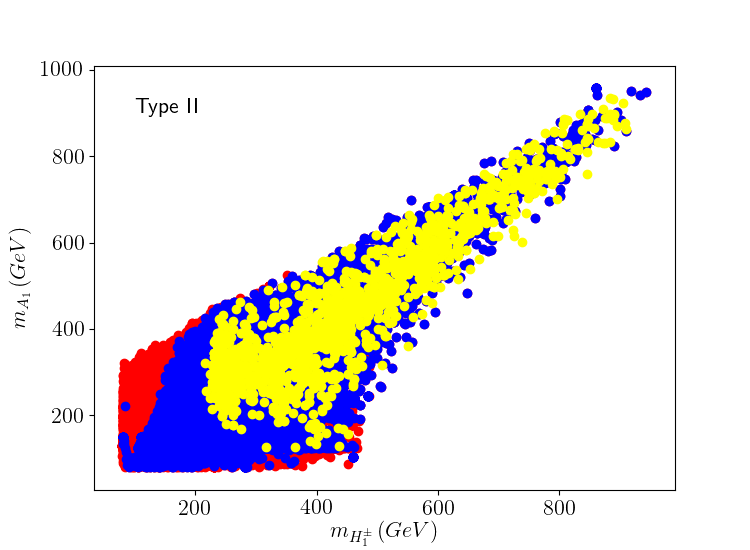}
\end{center}
\end{minipage}
\begin{minipage}{6.5cm}
\begin{center}
\includegraphics[height =6cm,width=6.5cm]{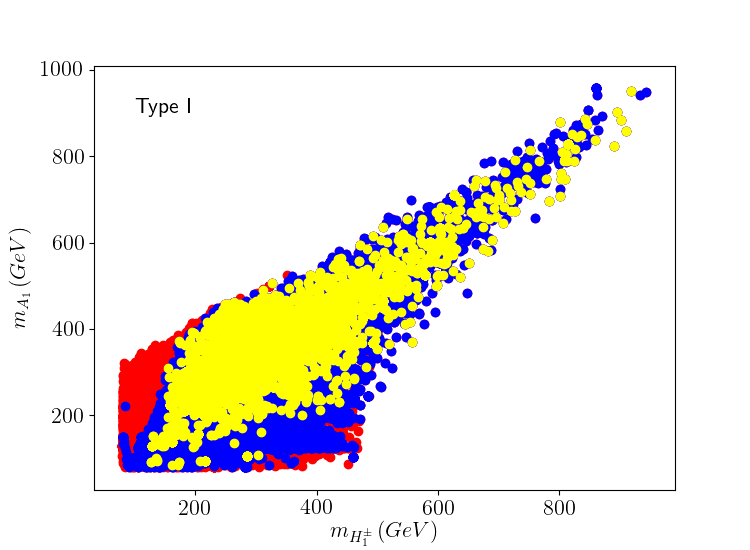}
\end{center}
\end{minipage}\\
\begin{minipage}{6.5cm}
\begin{center}
\includegraphics[height =6cm,width=6.5cm]{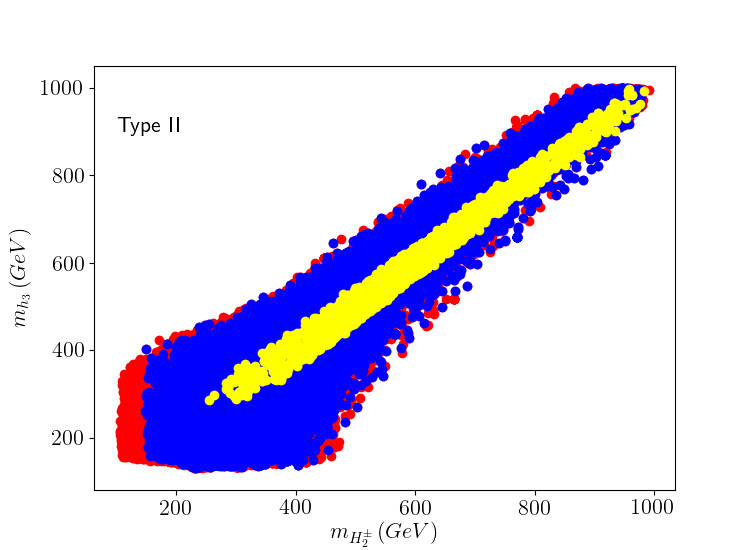}
\end{center}
\end{minipage}
\begin{minipage}{6.5cm}
\begin{center}
\includegraphics[height =6cm,width=6.5cm]{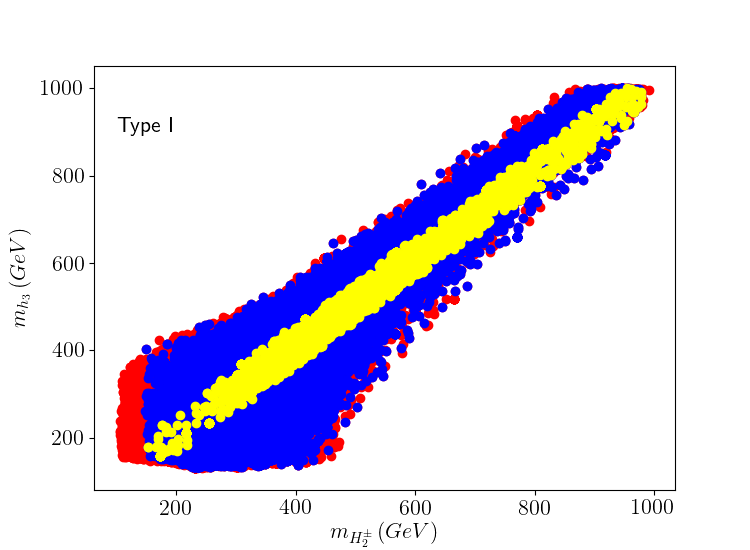}
\end{center}
\end{minipage}\\
\begin{minipage}{6.5cm}
\begin{center}
\includegraphics[height =6cm,width=6.5cm]{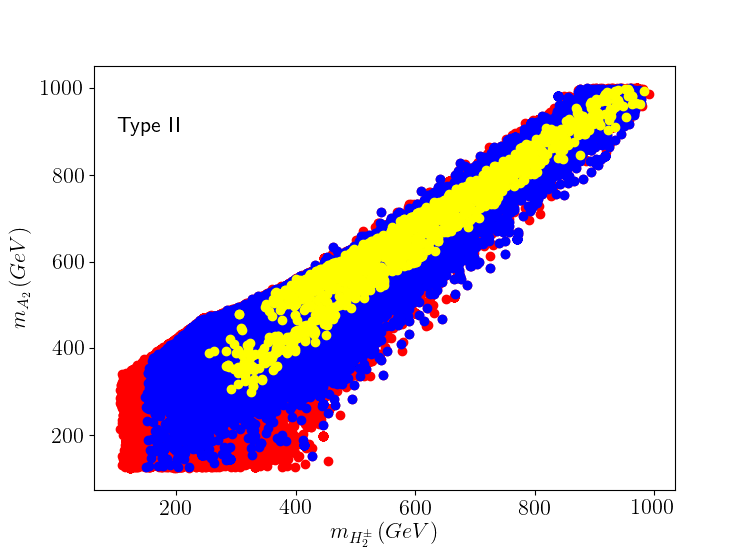}
\end{center}
\end{minipage}
\begin{minipage}{6.5cm}
\begin{center}
\includegraphics[height =6cm,width=6.5cm]{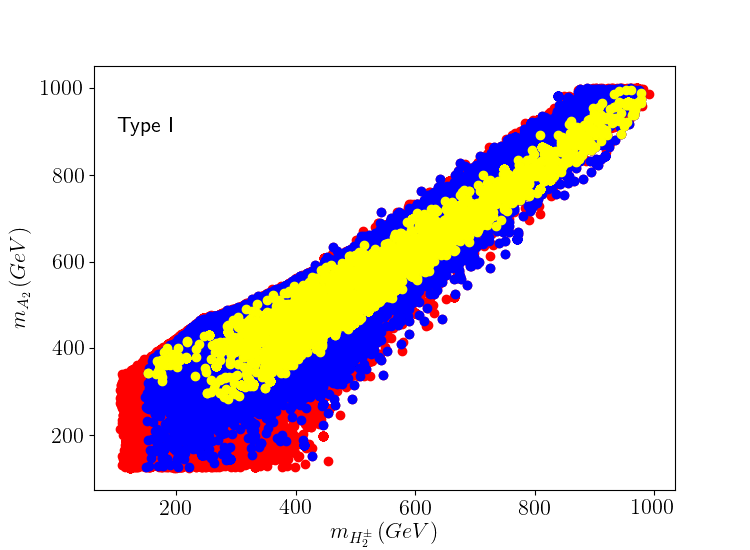}
\end{center}
\end{minipage}\\
\begin{minipage}{6.5cm}
\begin{center}
\includegraphics[height =6cm,width=6.5cm]{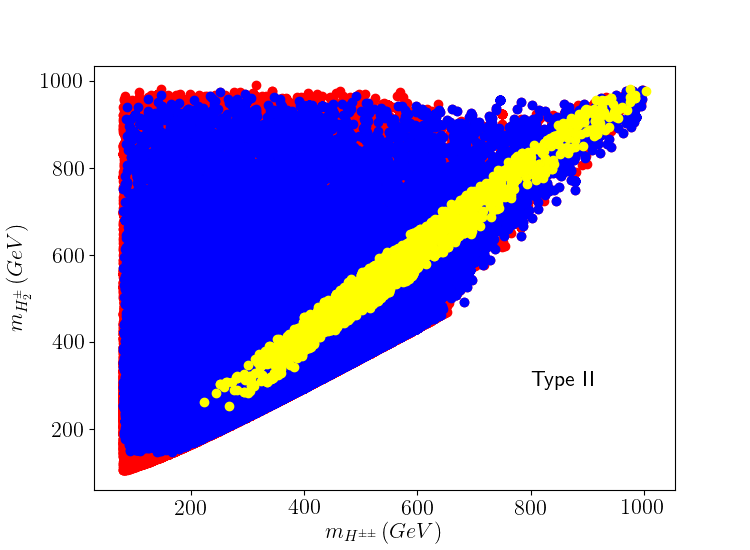}
\end{center}
\end{minipage}
\begin{minipage}{6.5cm}
\begin{center}
\includegraphics[height =6cm,width=6.5cm]{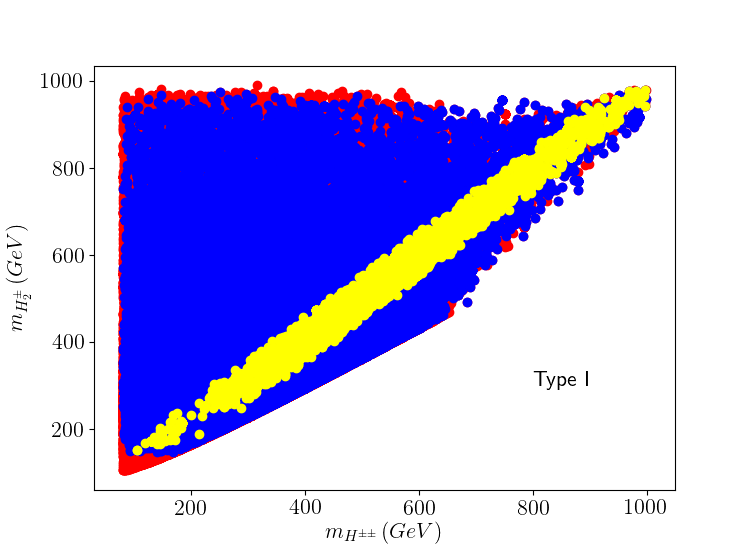}
\end{center}
\end{minipage}
\caption{Allowed mass ranges in 2HDMcT type-II (left column) and type-I (right column) taking into account all theoretical and experimental constraints.}
\label{fig:4}
\end{figure}

In Fig. \ref{fig:4}, the allowed ranges are plotted in the planes $m_{A_1}$, $m_{A_2}$, $m_{h_3}$ and $m_{H^{\pm\pm}}$ vs $m_{H^\pm_2}$. The left-columns panels corresponds to type-II, the right-column to the type-I, all points passed the constraints mentioned above at 1$\sigma$ (yellow), 2$\sigma$ (blue) and 3$\sigma$ (red). As can be seeing most of the masses can be light in type-I less than 300 GeV. 
Looking only at the yellow points, those which pass the $C_2$ constraint, we can see that the $m_{A_2}$, $m_{H^\pm_2}$, $m_{h_3}$ and $m_{H^{\pm\pm}}$ masses are bounded in type II where we find that most of the yellow points lie in the ranges $m_{H_1^\pm, H_2^\pm, A_1, A_2, h_2, h_3,H^{\pm\pm} }\in[200, 1000]$ GeV.

\begin{figure}[!h]
\centering
\includegraphics[height =6.5cm,width=7cm]{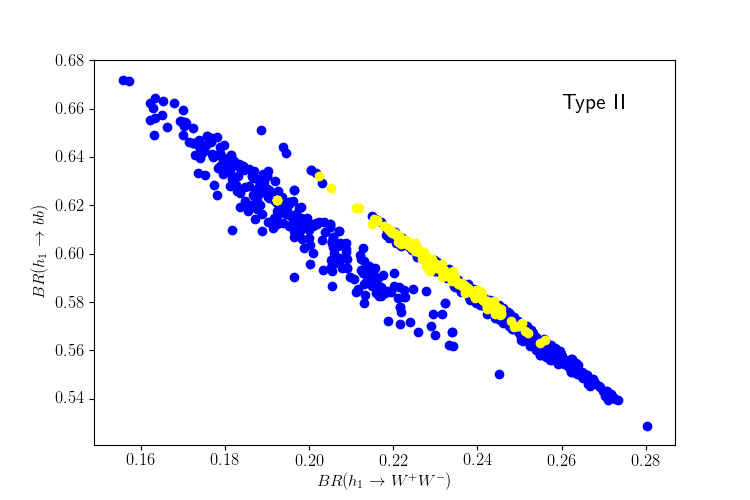}
\includegraphics[height =6.5cm,width=7cm]{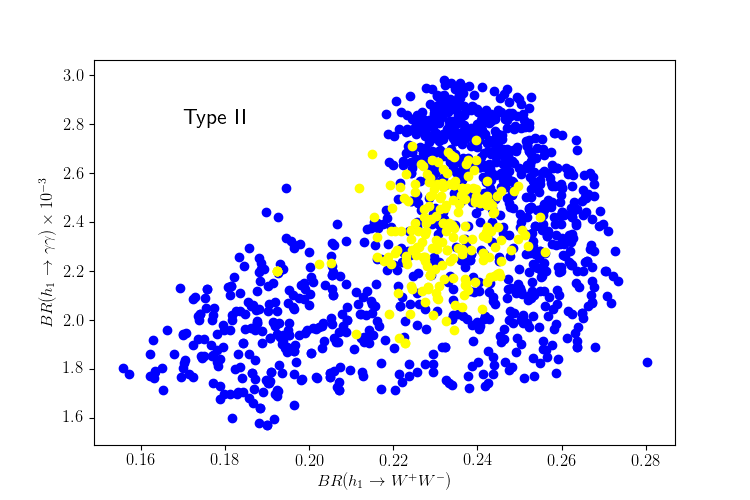}\\
\includegraphics[height =6.5cm,width=7cm]{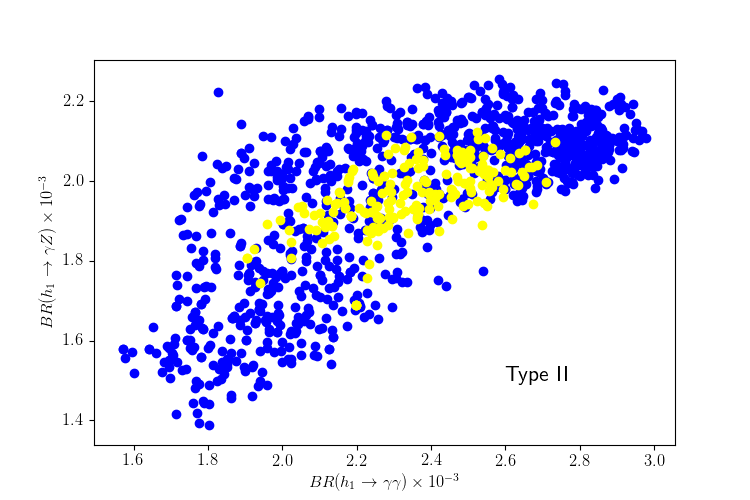}
\includegraphics[height =6.5cm,width=7cm]{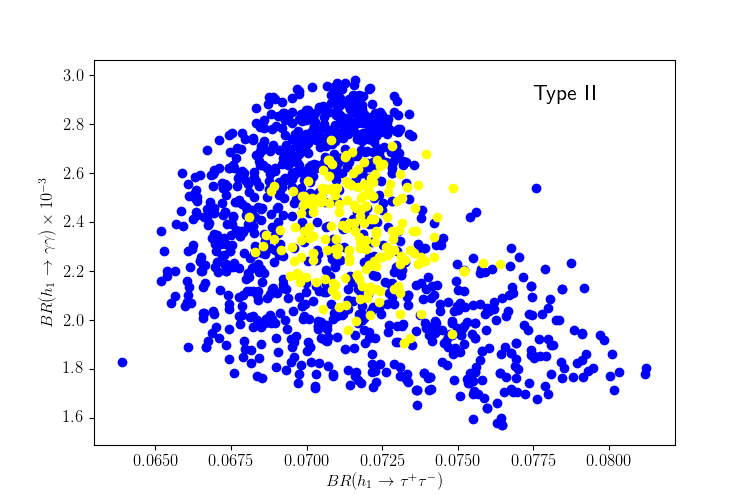}\\
\caption{Correlation plots between the BRs of (a) $h_1 \to WW$ vs $h_1 \to \gamma\gamma$, (b) $h_1 \to WW$ vs $h_1 \to b\bar b$, (c) $h_1 \to \tau\tau$ vs $h_1 \to \gamma\gamma$ and (d) $h_1 \to \gamma\gamma$ vs $h_1 \to h_1 \to Z \gamma$ in 2HDMcT type-II by considering the BRs of the lightest $h_1$ boson consistent within 95.5\% CL (blue) and 68\% CL (yellow)}
\label{fig:5}
\end{figure}

The sensitivity of the Higgs couplings of $h_1$ at the LHC is not appreciably better than 20$\%$,  leaving thus a significant window of opportunity for new physics. Here we investigate the correlations among relevant couplings within the $C1$ and $C_2$ constraints. In Fig \ref{fig:5}, we show these correlations which are consistent within the BRs of $m_{h_1} = 125$ GeV within 1$\sigma$ (yellow) and 2$\sigma$(blue). This is related to the fact that the central values of some Higgs couplings deviate from the SM, which strongly restrict the range of deviations from the SM. While the decay $h_1 \to Z\gamma$ is not yet observed at the LHC, the correlation between the $h_1 \to WW, \gamma\gamma$ and $h\to b\bar b$ couplings can now be measured by the experiments. The ratio of the BR of $WW$ to that $\gamma\gamma$ can be measured with an accuracy better than 5$\%$ and an integrated luminosity that is expected to be accumulated by the High Luminosity LHC.
In the following, we will scrutinize the impact of the searches for heavy Higgs particles in 2HDMcT ordered by their decay products. First we will address the bosonic decays to $VV$ ($V=\gamma, Z, W^\pm$) and fermionic mode $\tau^-\tau^+$ branching ratios of the neutral Higgs ($h_2$, $h_3$, $A_1$ and $A_2$) using  both CMS and ATLAS Higgs data for 8 TeV \cite{ATLAS1, ATLAS2, ATLAS3} and 13 TeV \cite{ATLAS4,ATLAS5,ATLAS6,ATLAS7,ATLAS88}. After that, we will turn towards the pair production of $h_i h_j$. The narrow width approximation will be applied throughout this section, we will comment on its validity at the end of the text section. We define the cross section as, 
\begin{eqnarray}
\sigma^{2HDMcT}(pp\rightarrow\phi)=k^2_{g}\sigma^{SM}(pp\rightarrow\phi), \,\,\ \phi = h_{2,3} \,\, or \,\,A_{1,2}
\end{eqnarray}
where $\sigma^{SM}(gg\rightarrow\phi)$ is the cross section for Higgs production in gluon fusion in the SM, 
and $k^2_{g}=\Gamma_{2HDMcT}(\phi\rightarrow gg)/\Gamma_{SM}(\phi\rightarrow gg)$, with $\Gamma_{2HDMcT}(\phi\rightarrow gg)$ and $\Gamma_{SM}(\phi\rightarrow gg)$ are the partial decay rates  in 2HDMcT and SM respectively. 

\subsubsection*{\bf{$h_{2,3}\to ZZ$ and $WW$}}
\begin{figure}[h]
\centering
\includegraphics[height =6.5cm,width=7cm]{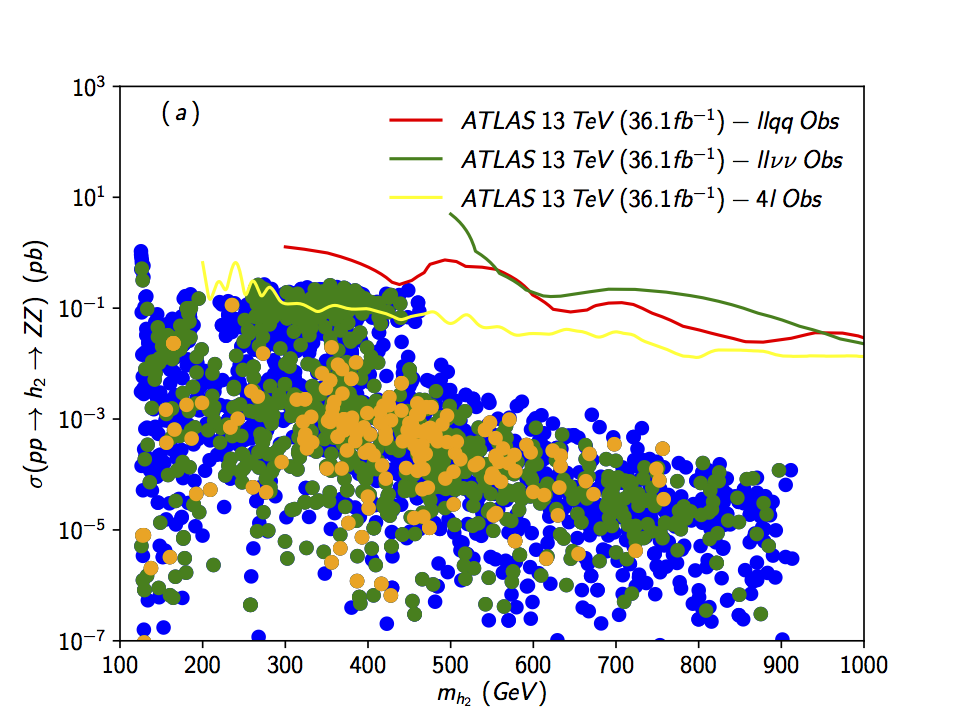}
\includegraphics[height =6.5cm,width=7cm]{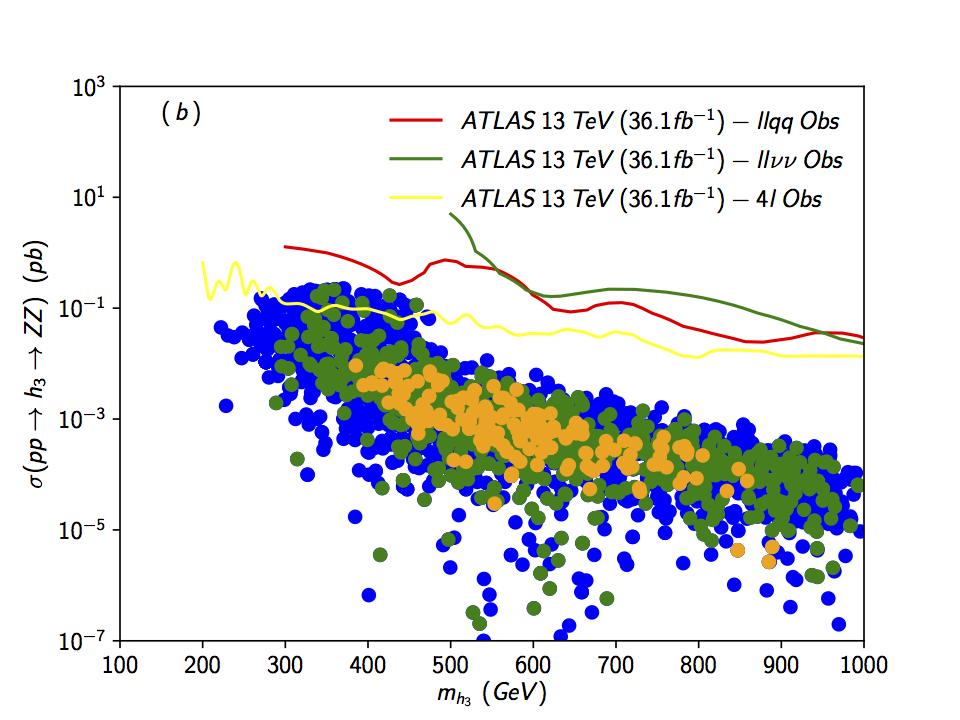} \\
\includegraphics[height =6.5cm,width=7cm]{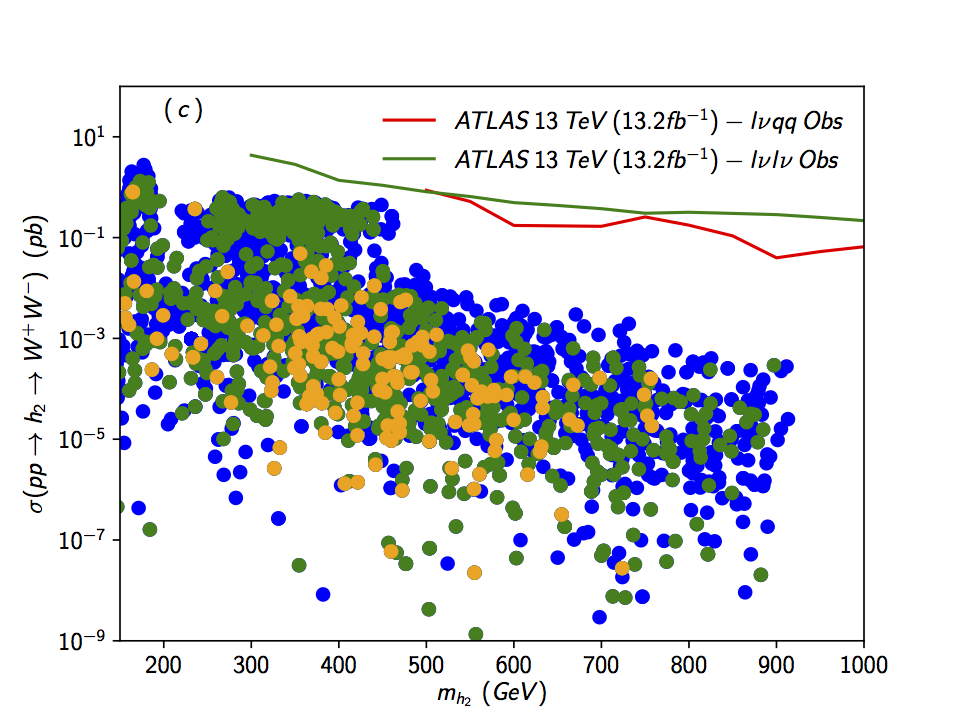}
\includegraphics[height =6.5cm,width=7cm]{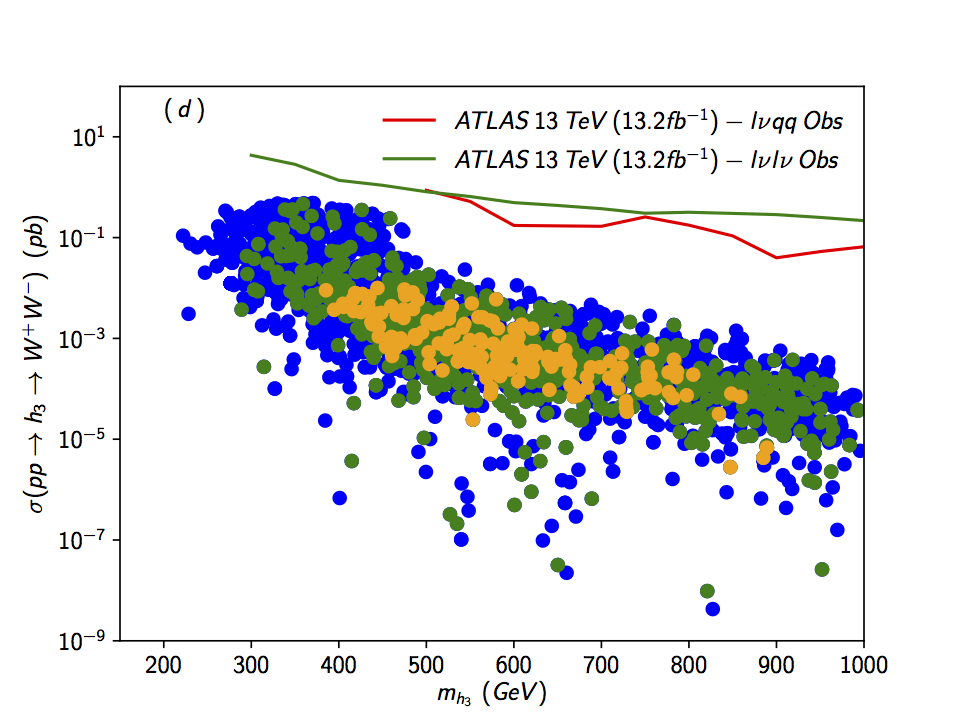}
\caption{Scatter plot for $\sigma(pp\to h_{2,3}) \times Br(h_{2,3} \to ZZ, WW)$ as a function $m_{h_{2,3}}$ after imposing $C_1$ and $C_2$ constrains assuming  99.7\% CL (red), 95.5\% CL (blue) and 68\% CL (yellow) in 2HDMcT type-II}
\label{fig.sc_h2}
\end{figure}

We use Sushi v1.6.0 public code \cite{Sushi1,Sushi2} at NNLO QCD to perform the calculation of the cross sections for Higgs production in gluon fusion $(ggF)$  and bottom-quark annihilation in the SM at 13 TeV. The relative coupling of $h_{2,3}$ to the two vector bosons $ZZ$ and $WW$ is universal and type independent. However, the production of the $h_{2,3}$ differs between the types. 
In Fig. (\ref{fig.sc_h2}) we plot the production cross sections in proton proton fusion times $ZZ$ and $WW$ branching ratios of $h_{2,3}$ Higgs after imposing $C_1$ and $C_2$ constrains within 99.7\% CL (red), 95.5\% CL (blue) and 68\% CL(yellow) of the Higgs data. 
It is well seeing from Fig. (\ref{fig.sc_h2}) that $h_{2,3} \to WW/ZZ$ rates are compared to what would be expected at Run-II LHC. The colored lines in Fig. (\ref{fig.sc_h2}) denote the observed limits in $ZZ/WW$ final states from 13 TeV ATLAS data \cite{ATLAS10, ATLAS11, ATLAS12, ATLAS13}. The direct LHC searches for this channels yield a strong suppression of $\sigma\times BR$. For $ZZ$ searches the $m_{h_2} \le 250$ GeV region is constrained by Run-I data whereas Run-II data determine the dominant limits for the rest of the mass region. For the $WW$ searches, Run-I data dictate the limit until 350 GeV and the high mass range is dominated by Run-II data.

\subsubsection*{\bf{$h_{2,3}, A_{1,2}\to \gamma\gamma$} }

\begin{figure}[!ht]
\centering
\includegraphics[height =6.5cm,width=7cm]{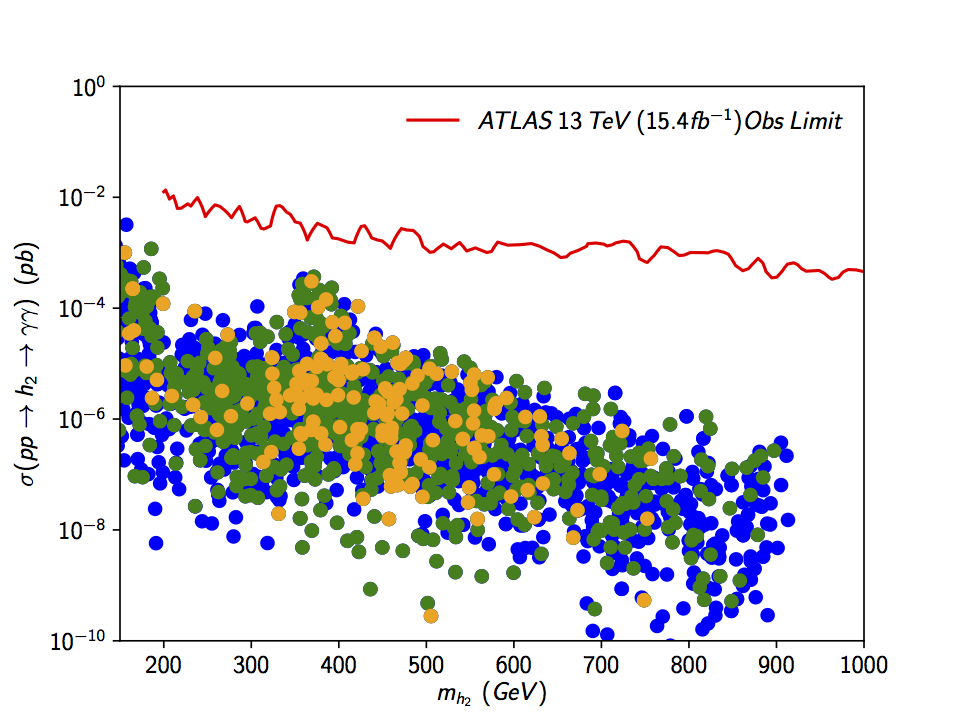}
\includegraphics[height =6.5cm,width=7cm]{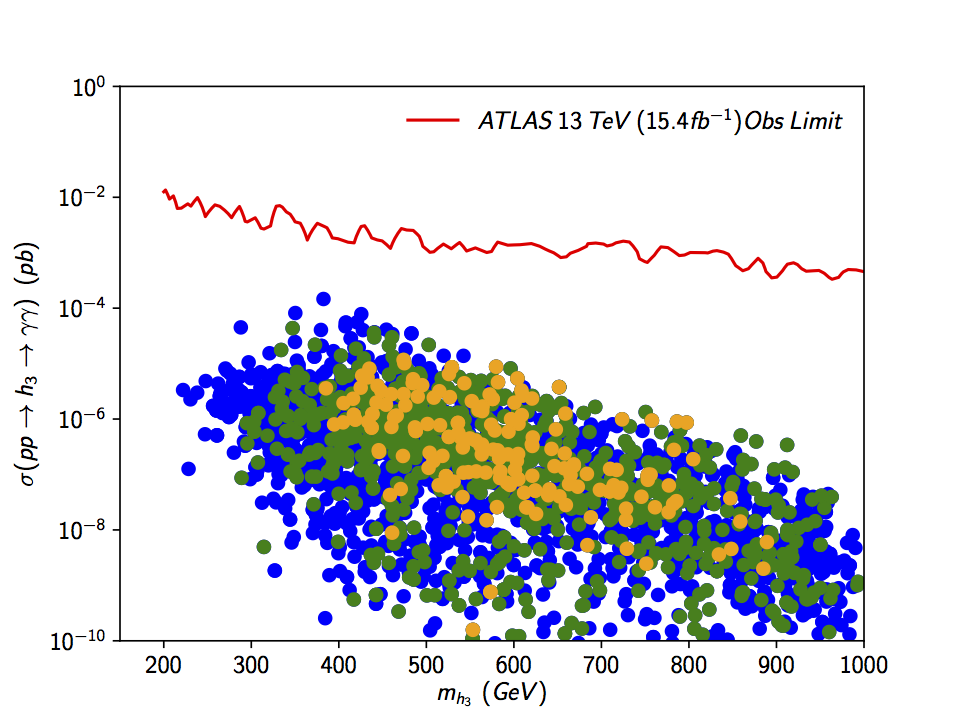}\\
\includegraphics[height =6.5cm,width=7cm]{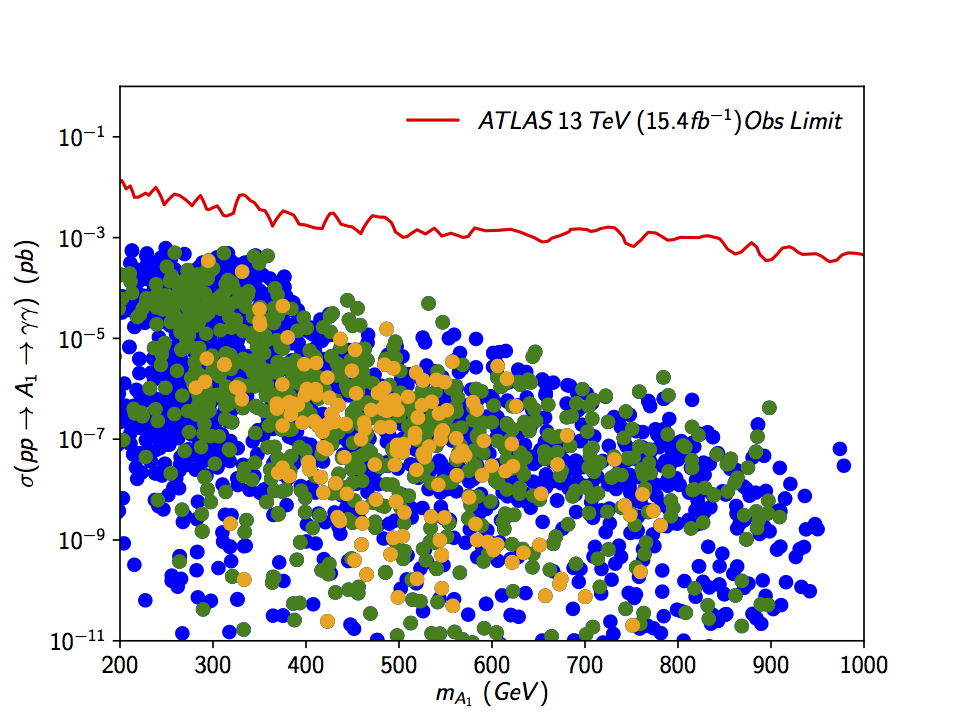}
\includegraphics[height =6.5cm,width=7cm]{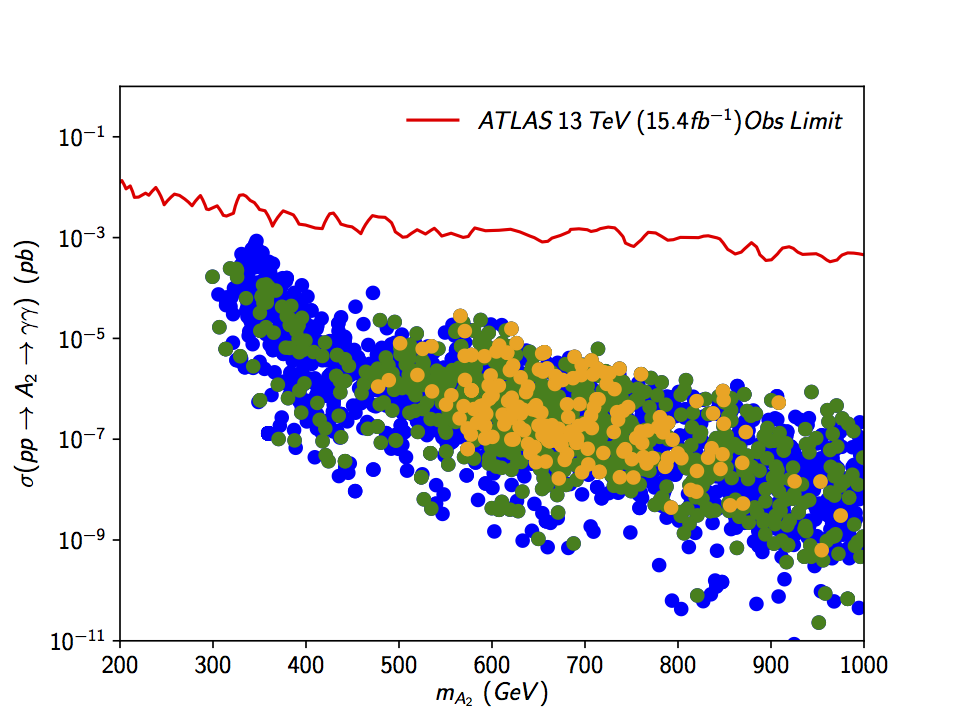}
\caption{Scatter plot in the $m_{\phi}$-$[\sigma \times Br(\phi\rightarrow \gamma\gamma)]$ ($\phi = h_{2,3}$ or $A_{2,3}$) planes after imposing $C_3$ constrains at $\sqrt{s} = 13$ TeV. The errors for $\chi$-square fit are 99.7\% CL (red), 95.5\% CL (blue) and 68\% CL (yellow) in 2HDMcT type-II.}
\label{fig.sc_h3}
\end{figure}

Direct searches for a heavy Higgs decaying to two photons constrain $\sigma\times BR$ by roughly one magnitude compared the ATLAS limit in CP-even decay modes (see Fig. \ref{fig.sc_h3}). The searches in the di-photon decay channel of a pseudoscalar Higgs yield a suppression of   $\sigma\times BR$ one to three orders of magnitude compared to the ATLAS limit for $m_{A_{1,2}} \le 400$ GeV certain intermediate $\sigma\times BR$ regions for low  $m_{A_{1,2}}$ are disfavoured by the prior.

\begin{figure}
\centering
\includegraphics[height =6.5cm,width=7cm]{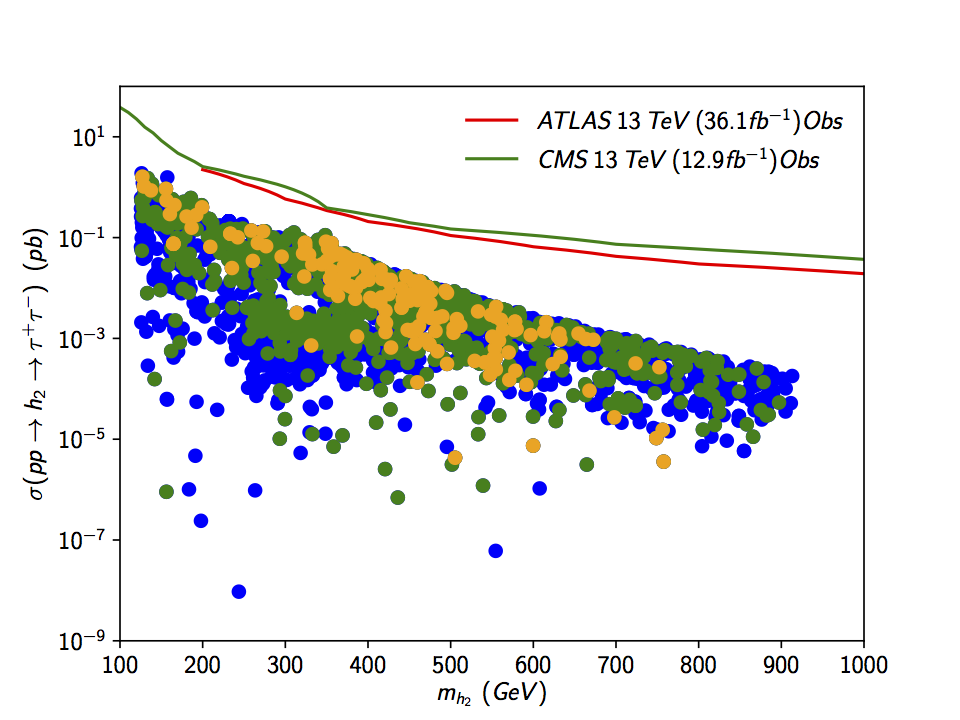}
\includegraphics[height =6.5cm,width=7cm]{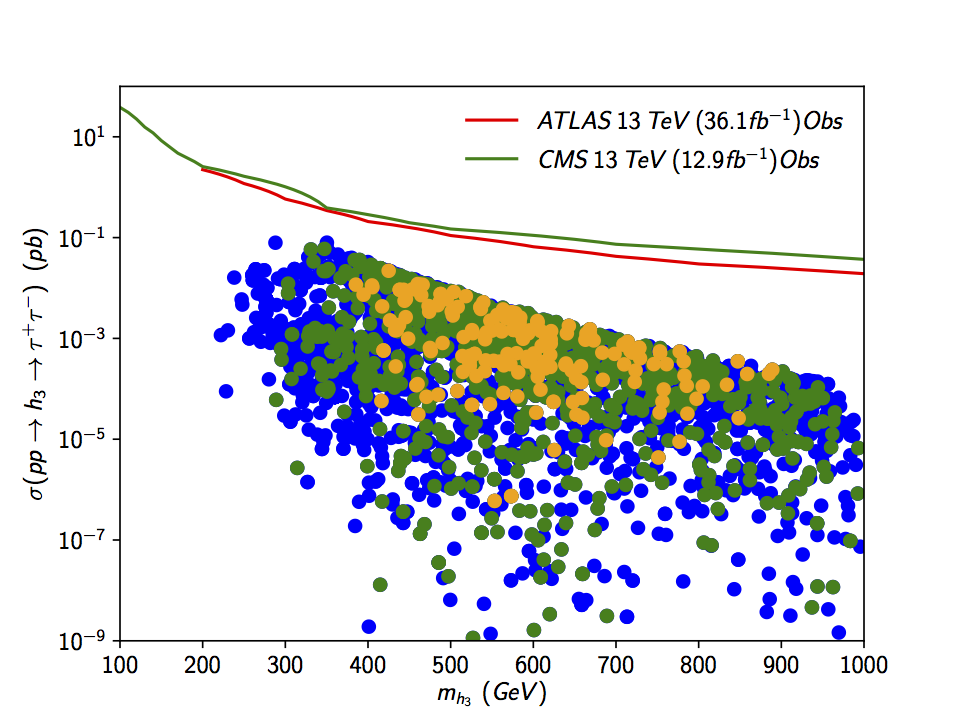}\\
\includegraphics[height =6.5cm,width=7cm]{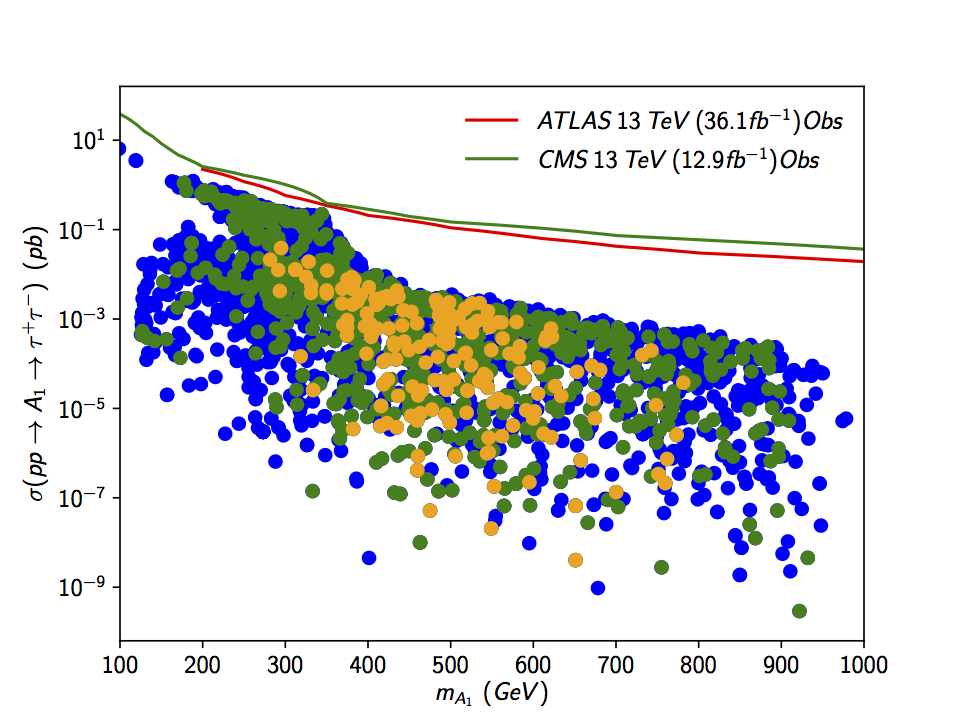}
\includegraphics[height =6.5cm,width=7cm]{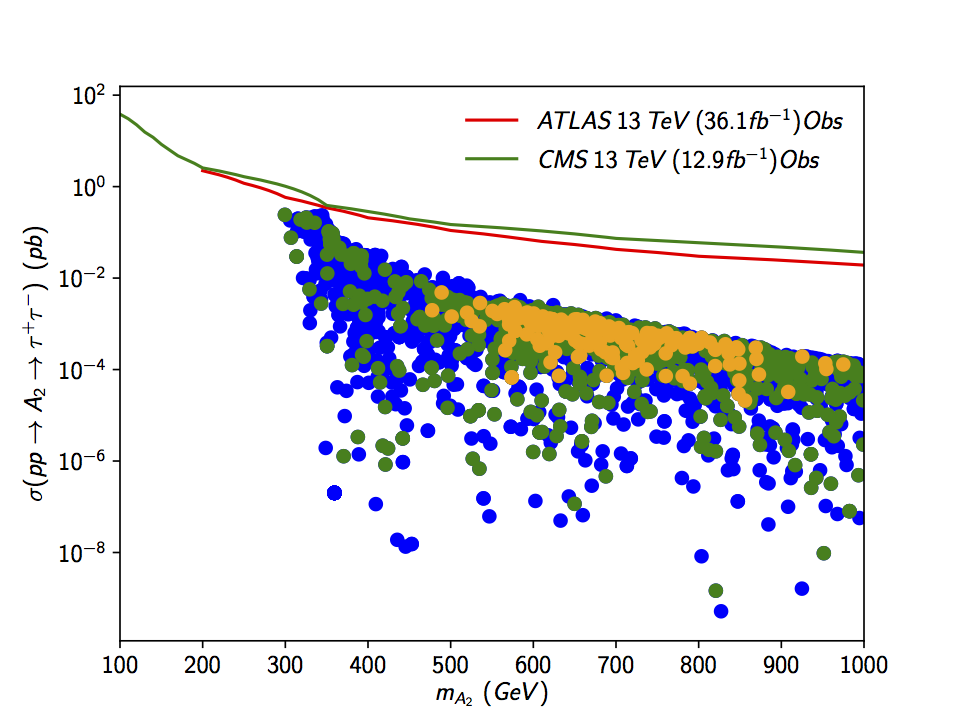}
\caption{The same as Fig. (\ref{fig.sc_h3}) but for $\tau\tau$ pair production at $\sqrt{s} = 13$ TeV in 2HDMcT type-II}\label{fig.sc_h4}
\end{figure}

\begin{figure}
\centering
\includegraphics[height =6.5cm,width=7cm]{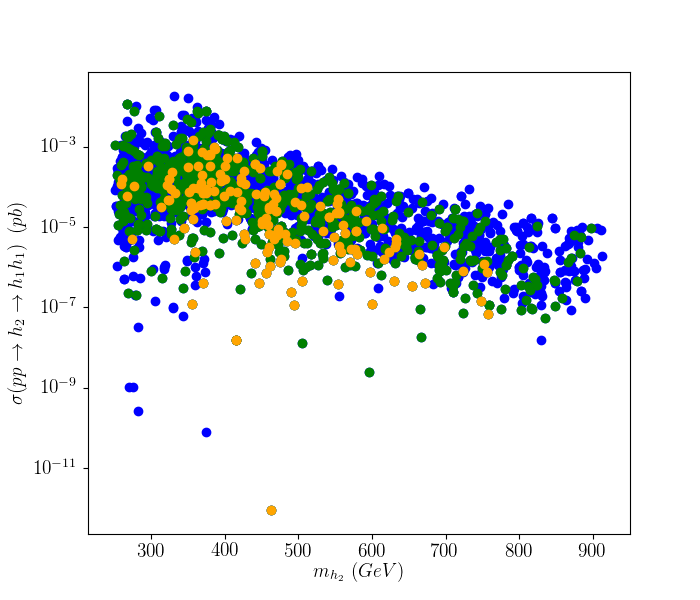}
\includegraphics[height =6.5cm,width=7cm]{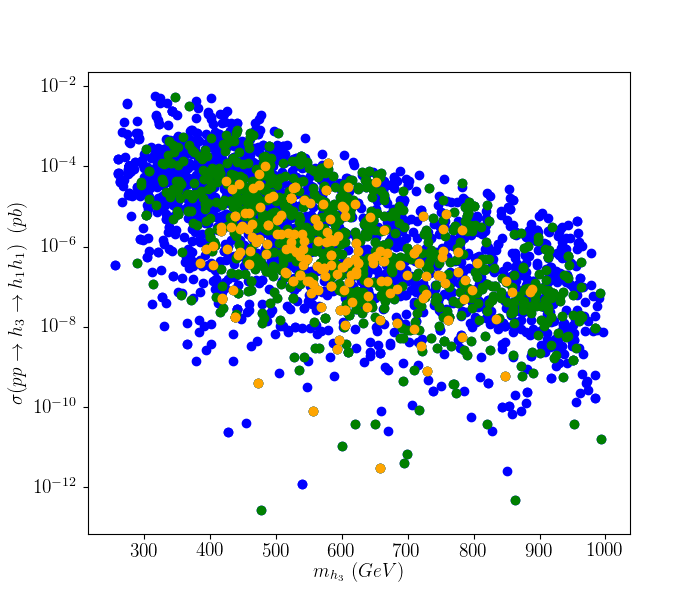}\\
\includegraphics[height =6.5cm,width=7cm]{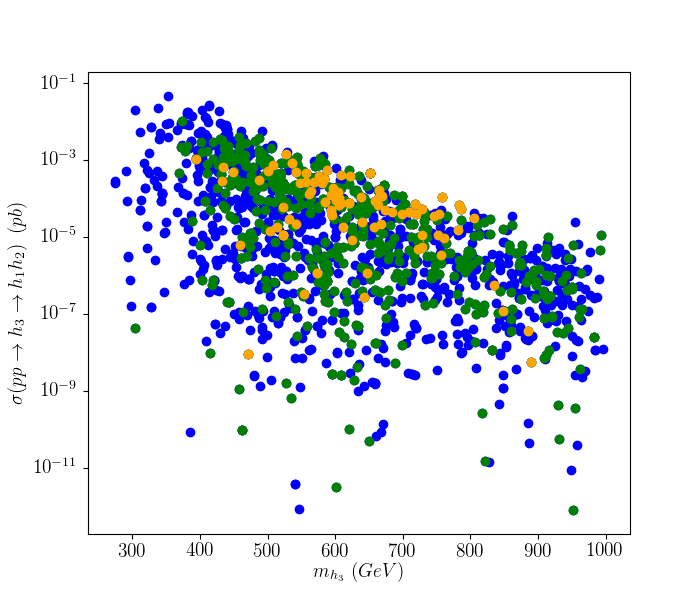}
\includegraphics[height =6.5cm,width=7cm]{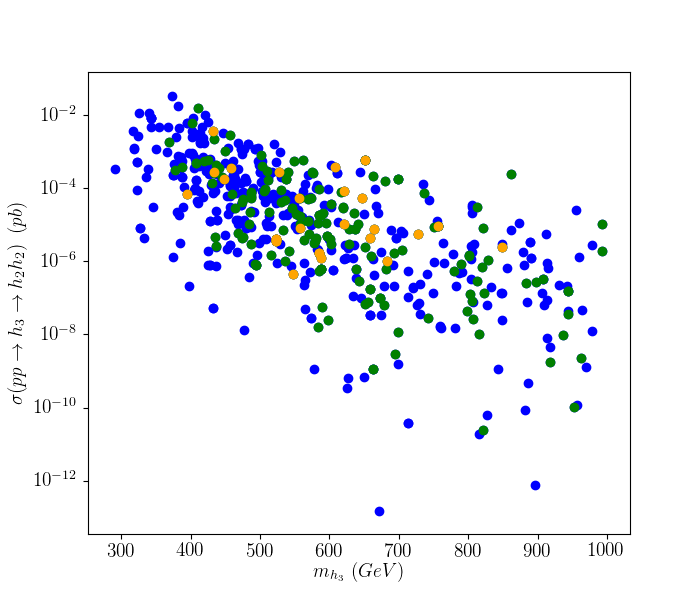}
\caption{The same as Fig. (\ref{fig.sc_h3}) but for $h_i h_j$ pair production at $\sqrt{s} = 13$ TeV in 2HDMcT type-II}
\label{fig.sc_h33}
\end{figure}

In Fig. \ref{fig.sc_h4} we plot the production cross section in proton proton fusion times  $\tau\tau$ branching ration after imposing $C_3$ constrains. The errors for
$\chi$-square fit are 99.7\% CL (red), 95.5\% CL (blue) and 68\% CL (yellow). The red(green) solid line in Fig. \ref{fig.sc_h4} is the upper limit on the cross-section times branching ration from the $ATLAS$ 13 TeV results \cite{ATLAS14} ($CMS$ 13 TeV results \cite{ATLAS15}). The colored lines in Fig. \ref{fig.sc_h3} denote the observed limits from 13 TeV ATLAS data \cite{ATLAS16}.

\subsubsection*{\bf{$h_{2,3} \to h_i h_j$ with $(i,j) = (1,2)$} }
Finally, in Fig. (\ref{fig.sc_h33}) we present a sample of points generated for Higgs pair production in the scenario where the observed Higgs boson is the lightest scalar $h_1$ (top panels) and the next-to-lighest scalar (bottom panels). We present cross sections as a function of the masses of the two new scalars that can be involved in the chain decay contributions. We overlay three layers of point for which the total cross section is within $3\sigma$ (red), $2\sigma(blue)$ and $1\sigma$ respectively at the LHC Run-2. In the upper plots, where $h_1$ chain decays become possible above 250 GeV. In Fig.(\ref{fig.sc_h33}) at nearly $m_{h_1}\sim m_{h_2}$, which means that some opening decay channels contribute to chain decay $h_3 \to h_1 + h_1, h_1 + h_2$ and $h_2 + h_2$. This means that imposing a constraint on the parameter space with including the chain decay contribution would be too strong at the LHC Run-2. Overall, the shape of the three different $\sigma$ regions in the various panels is the results of an interplay between the kinematics and the applied constraints. To certain extent the structure can be directly related to the exclusion curves from the collider searches imposed by HiggsBound. The strong increase at $m_{h_3} = 250$ GeV in the lower left panel is due the opening of the decay $h_3 \to h_2 + h_2$ and the decrease in the number of points for large values of $m_{h_3}$ is due the S and T parameters.
\section{Benchmark points for the LHC Run-2 in 2HDMcT}
\label{sec:benchmark}Benchmark points
In this section we provide a set of benchmark points in 2HDMcT. There are chosen to cover various physical situations.  From a phenomenological perspective we are interested in maximising various visibility of the new scalars in the LHC Run-2 and in covering, simultaneously, many kinematically different possibilities. In particular, we are interested in scenarios where $h_{SM} = h_1$, $h_{SM} = h_2$ and $h_{SM} = h_3$ while preserving consistency with LHC Run-2 measurements. Thus, in many of the points presented below we have tried to maximise the cross section for the cross sections. At the same time we required the rates of the SM-like Higgs within $2\sigma$ from the global signal strength provided by the ATLAS and CMS data from LHC Run-2. Table (\ref{tab:benchm}), contains the parameters that define the chosen benchmark points and the production rate of the lightest as well as next-to-lightest Higgs bosons $h_2$ and $h_3$ in the various final states.

\begin{table}[t!]
\begin{center}
\scriptsize
\begin{tabular}{|c|c|c|c|c|c|}
\hline
 	 & BP1 	 & BP2 	 & BP3 	 & BP4 	 & BP5\\ \hline \hline
 $m_{h1}$ (GeV)          	 & $125.1$ 	 & $125.1$ 	 & 12.75 	 & 77.27 	 & 21.8\\
 $m_{h_2}$ (GeV) 	 & 415.46 	 & 354.18 	 & $125.1$ 	 & $125.1$  & 101.7\\
 $m_{h_3}$ (GeV) 	 &426.17  	 & 937.02	 & 155.96 &  448.59	 & $125.1$\\
 $m_{A_1}$ (GeV) 	 & 258.76  	 &311.72 	 & 152.46 	 &  174.34	 &  104 \\
$m_{A_2}$ (GeV) 	 &417.81  	 & 936.06 	 &  270.28	 &  457	 &    327    \\
$m_{H^\pm_{1}}$ (GeV) 	 &270.22  	 & 267.60 	 &  147.13	 &232.8  	 &  143      \\
$m_{H^\pm_{2}}$ (GeV) 	 &386.35  	 & 906.42 	 &  164.24	 &434.7  	 & 174.87       \\
$m_{H^{\pm\pm}}$ (GeV) 	 &352.06  	 & 875.54 	 &  161.01	 &  397.2	 &  174.89      \\ 
 $\alpha_1$ 	 &-1.26  	 &  -1.25     &  -0.46	 & -0.58      & -0.08\\
 $\alpha_2$  	 &-0.011  	 &  -0.086	 &  -0.16     &0.45  	 &-0.06 \\
 $\alpha_3$ 	 &-1.08  	 & -0.12 	 & -0.13	 &-0.08 	 &-1.25 \\
$v_t$ (GeV) 	         &1  	 & 1	 & 1	 &  1	 &1 \\
$\lambda_6$ 	         &5.56 	 & 0.89 	 & 6.24      & 4.17 	 & 5.92\\
$\lambda_7$       	 &1.13 	 & 0.69 	 & 5.52 	 &4.57  	 &7.08 \\
$\lambda_8$ 	                 &2.75   	 & 1.54 	 &-2.92 	 &  1.5     &-4.64\\
$\lambda_9$ 	 &1.50  	 & 3.36 	 & 0.90	 & -1.04 	 & -0.6\\
$\tan\beta$ 	 &2.57 	 &  2.53	 &  1.7	 &  1.07	 & 8.83\\
$\mu_{1}$ 	 &26.22  	 & 16.42 	 &  -21.49	 &-36  	 &56.69 \\
$\mu_{2}$ 	 &  	11.9 &  263.7	 &  	6.19 &  -451	 &1.19 \\
$\mu_{3}$ 	 &  	 -28.95&  	-615.2 &  3.41	 &  525.42	 & -14.69\\
\hline \hline
\hline \hline
$\mu_{h_1}$ 	 &0.88  	 & 0.86 	 &  0.043	 &0.069  	 &0.78 \\
 \hline
$\sigma_1\equiv \sigma(g g\rightarrow h_1)$ 	 & 132.9 [pb] 	 & 131.7 [pb] 	 & 10003.81 [pb] 	 & 175.45 [pb] 	 &  423.9[pb]\\
$\sigma_1\times {\rm BR}(h_1\rightarrow WW)$ 	 & 11.69 [pb] 	 & 10.70 [pb] 	 & 0 [fb] 	 &  0[fb] 	 &  0[pb]\\
$\sigma_1\times {\rm BR}(h_1\rightarrow ZZ)$ 	 & 1.73$\times 10^{3}$ [fb] 	 & 1.59$\times 10^{3}$[fb] 	 & 0 [fb] 	 & 0 [fb] 	 & 0 [pb]\\
$\sigma_1\times {\rm BR}(h_1\rightarrow bb)$ 	 &  35.1 [pb] 	 & 35.96 [pb] 	 & 4811 [pb] 	 & 71.62 [pb] 	 & 177.45 [pb]\\
$\sigma_1\times {\rm BR}(h_1\rightarrow \tau \tau)$ 	 & 4.24 [pb] 	 &  4.35[pb] 	 & 452$\times 10^{3}$[fb] 	 & 7.8$\times 10^{3}$ [fb] 	 &17.7$\times 10^{3}$[fb]\\
$\sigma_1\times {\rm BR}(h_1\rightarrow \gamma \gamma)$ 	 & 0.12$\times 10^{3}$ [fb] 	 &  0.12$\times 10^{3}$[fb] 	 & 7 [fb] 	 & 8.4 [fb] 	 & 0.34 [fb]\\
\hline \hline
$\mu_{h_2}$ 	 &  	0.03 &  0.095	 &  	0.78 &  0.6	 &0.01 \\ \hline
$\sigma_2\equiv \sigma(g g\rightarrow h_2)$ 	 & 0.53 [pb] 	 & 4.74 [pb] 	 & 106.18 [pb] 	 & 94.83 [pb] 	 & 18.31 [pb]\\
$\sigma_2\times {\rm BR}(h_2\rightarrow WW)$ 	 & 0.11 [pb] 	 & 0.56 [pb] 	 &  9.55$\times 10^{3}$[fb] 	 & 11.4$\times 10^{3}$ [fb] 	 &  4.98$\times 10^{-3}$[pb]\\
$\sigma_2\times {\rm BR}(h_2\rightarrow ZZ)$ 	 & 52 [fb] 	 & 260[fb] 	 &  1.41$\times 10^{3}$[fb] 	 & 1.69$\times 10^{3}$ [fb] 	 & 9.66$\times 10^{-5}$ [pb]\\
$\sigma_2\times {\rm BR}(h_2\rightarrow bb)$ 	 & 6.2$\times 10^{-4}$ [pb] 	 &  4.1$\times 10^{-3}$[pb] 	 &  19.06[pb] 	 & 23.12 [pb] 	 &  0.35[pb]\\
$\sigma_2\times {\rm BR}(h_2\rightarrow \tau \tau)$ 	 & 9.38$\times 10^{-5}$ [pb] 	 &  6.1$\times 10^{-4}$[pb] 	 &2.3 $\times 10^{3}$[fb] 	 & 2.79$\times 10^{3}$ [fb] 	 &0.041$\times 10^{3}$[fb]\\
$\sigma_2\times {\rm BR}(h_2\rightarrow \gamma \gamma)$ 	 & 7.401$\times 10^{-4}$ [fb] 	 &  7.86$\times 10^{-3}$[fb] 	 & 0.1$\times 10^{3}$ [fb] 	 & 0.08$\times 10^{3}$ [fb] 	 & 1.86 [fb]\\
${\rm BR}(h_2\rightarrow h_1h_1)$ \% 	 &  	5.1$\times 10^{-2}$ & 1.6 $\times 10^{-2}$	 &  25.69	 &  0	 & 94.68\\ 
\hline
\hline \hline
$\mu_{h_3}$ 	 &  	 0.11&  	0.0003 &  4.69$\times 10^{-5}$	 &  0.51	 & 1.07\\ \hline
$\sigma_3\equiv \sigma(g g\rightarrow h_3)$ 	 &  1.43[pb] 	 & 4.6$\times 10^{-4}$ [pb] 	 & 0.15 [pb] 	 & 2.79 [pb] 	 & 93.87 [pb]\\
$\sigma_3\times {\rm BR}(h_3\rightarrow WW)$ 	 &  0.35[pb] 	 & 1.52$\times 10^{-6}$[pb] 	 & 7 [fb] 	 & 2.4 [fb] 	 & 8.86 [pb]\\
$\sigma_3\times {\rm BR}(h_3\rightarrow ZZ)$ 	 &  1.6$\times 10^{2}$[fb] 	 & 1.52$\times 10^{-3}$[fb] 	 & 0.8 [fb] 	 & 2.2 [fb] 	 &  1.32[pb]\\
$\sigma_3\times {\rm BR}(h_3\rightarrow bb)$ 	 &  1.8$\times 10^{-3}$[pb] 	 &  2.7$\times 10^{-8}$[pb] 	 & 0.004 [pb] 	 &  4.2$\times 10^{-4}$[pb] 	 &  19.14[pb]\\
$\sigma_3\times {\rm BR}(h_3\rightarrow \tau \tau)$ 	 &  2.7$\times 10^{-4}$[pb] 	 &  4.6$\times 10^{-9}$[pb] 	 & 0.56[fb] 	 &  6.46$\times 10^{-2}$[fb] 	 &2.31[fb]\\
$\sigma_3\times {\rm BR}(h_3\rightarrow \gamma \gamma)$ 	 &  2.027$\times 10^{-3}$ [fb] 	 & 2.25$\times 10^{-8}$ [fb] 	 & 1.54$\times 10^{-3}$ [fb] 	 & 5.05$\times 10^{-3}$ [fb] 	 & 0.14 [fb]\\
${\rm BR}(h_3\rightarrow \sum_{i=1,2} h_i h_j)$ \% 	 &  	6.9$\times 10^{-2}$ &  	6.1$\times 10^{-2}$ & 81.3	 &  0.79	 &18.42 \\ 
\hline
\end{tabular}
\end{center}
\caption{Benchmark points for LHC Run-2 in 2HDMcT. The cross sections are for $\sqrt{s}\equiv 13$~TeV. }
\label{tab:benchm}
\end{table}

We also give the signal rates $\mu_{h_i}$ $(i=1,2,3)$ we have chosen two points where the SM-like Higgs is the lightest Higgs boson (BP1 and BP2) two points where it is the next-to-lightest Higgs boson (BP3 and BP4) and one poin where it is the heaviest (BP5). 
 
Most points were chosen such that the cross sections for the indirect decay channels of the new scalars can compete with the direct decays. In particular we have tried to maximise $h_3 \to h_1 + h_2$ where all decays scalars could be observed at once. We have furthermore chosen points with large cross sections for the new scalars, so that they can be detected directly in their decays. In BP1, the lightest Higgs boson is $h_1$ and decay channel $h_3 \to h_2 + h_2$ is eventually closed. The presented point has a maximum cross section so that all new scalars can be expected to be observed.  In BP2, the SM Higgs-like still $h_1$ but we allow $h_3 \to h_2 + h_2$ now is open with a large possible branching ratio. For the BP3 where $h_2$ is the SM-like Higgs boson and $m_{h_1} < m_{h_2}/2$, all kinematic situations for the scalar decays ara available while the spectrum remains light. So that Higgs-Higgs decays present an interesting discovery option for the heavy Higgs states. Furthermore, large production cross sections have been required for the new light scalar $h_1$ so it will be visible in tes direct decays in addition to chain production from heavier scalars. With BP4, the situation is different from the previous one, so that the channel $h_2 \to h_1 + h_1$ is kinematically closed. At the same time the direct $h_1$ production rates are increased. Lastly, in BP5 benchmark the spectrum is very light. So that the production  of $h_1$ is possible through $h_3 \to h_1 + h_1$ or $h_3 \to h_1 + h_2$. This benchmark point has been required to have large branching ratios for $h_1 h_1$ and $h_1 h_2$, however the direct production rates of the heavier $h_2$ are expected to be small, but still accessible at the LHC Run-2.

\section{Summary}
\label{sec:conlusion}
In this paper we presented a comprehensive analysis of the scalar sector in 2HDMcT, when the bosons mix with arbitrary angle $\alpha_i$ (i = 1, 2, 3). Of the bare states in the model, one is usual neutral component of the SM Higgs doublet. We started by presenting the model and the theoretical
constraints that were imposed. Indeed we have discussed the questions of unitarity and vacuum stability in this model and showed the allowed parameter ranges. We have shown that, in the case of vanishing new extra parameters, the spectrum of the scalar Higgs is reduced to HTM. Later on
we studied the phenomenological aspects of the scalar sector by by imposing both theoretical and experimental constraints from combined LHC results by identifying the lighter of them with the 125
Higgs boson of standard model. In the numerical analysis, we first investigated how important the contributions from Higgs to gauge bosons decays can be compared to its direct production and decay. Depending on the mixing angles and the scenarios, the production rates can attain a maximum values ranging between 0.1 $fb^{-1}$ and 100 $fb^{-1}$ while remaining compatible with the Higgs signal measurements within 2$\sigma$. We have tested the main chain decays of heavy Higgs bosons scenario where the power to discriminate them from other extended models like 2HDM depends critically on differentiating their couplings and decays.

\section*{Acknowledgments}
R. Benbrik thanks Uppsala University for the warm hospitality. Part of this work was done within H2020-MSCA-RISE-2014 no. 645722 (NonMinimalHiggs) project. This work is also supported also by the Moroccan Ministry of Higher Education and Scientific Research MESRSFC and  CNRST: Projet PPR/2015/6. 

\clearpage
\appendix
\begin{center}
{\huge{Appendices}}
\end{center}
\section{Diagonalization Matrices}
\label{sec-diagonal}
The rotation between the physical and non-physical states are given by: 
\begin{eqnarray}
\begin{matrix}
\left(\begin{matrix}
G^\pm\\
H_1^\pm\\
H_2^\pm
\end{matrix}\right)={\mathcal{C}}\left(\begin{matrix}
\phi_1^\pm\\
\phi_2^\pm\\
\delta^\pm
\end{matrix}\right)\quad\,{\rm and}\quad\,
\left(\begin{matrix}
G^0\\
A_1\\
A_2
\end{matrix}\right)
={\mathcal{O}}
\left(\begin{matrix}
\eta_1\\
\eta_2\\
\eta_0
\end{matrix}\right)
\end{matrix}
\end{eqnarray}
The matrices elements for each sector are given below.
\subsection{charged sector}
In terms of the mixing angles $\theta_i^\pm$
\begin{eqnarray}
{\mathcal{C}} =\left( \begin{array}{ccc}
{\mathcal{C}}_{11} & {\mathcal{C}}_{12} & {\mathcal{C}}_{13}\\
{\mathcal{C}}_{21} & {\mathcal{C}}_{22} & {\mathcal{C}}_{23} \\
{\mathcal{C}}_{31} & {\mathcal{C}}_{32} & {\mathcal{C}}_{33}
\end{array} \right)=\left( \begin{array}{ccc}
c_{\theta^\pm_1} c_{\theta^\pm_2} & s_{\theta^\pm_1} c_{\theta^\pm_2} & s_{\theta^\pm_2}\\
-(c_{\theta^\pm_1} s_{\theta^\pm_2} s_{\theta^\pm_3} + s_{\theta^\pm_1} c_{\theta^\pm_3})
& c_{\theta^\pm_1} c_{\theta^\pm_3} - s_{\theta^\pm_1} s_{\theta^\pm_2} s_{\theta^\pm_3}
& c_{\theta^\pm_2} s_{\theta^\pm_3} \\
- c_{\theta^\pm_1} s_{\theta^\pm_2} c_{\theta^\pm_3} + s_{\theta^\pm_1} s_{\theta^\pm_3} &
-(c_{\theta^\pm_1} s_{\theta^\pm_3} + s_{\theta^\pm_1} s_{\theta^\pm_2} c_{\theta^\pm_3})
& c_{\theta^\pm_2}  c_{\theta^\pm_3}
\end{array} \right)
\label{eq:mixingmatrix}
\end{eqnarray}
where in the hybrid parameterization $\mathcal{P}_I$, these elements are given by,
\begin{eqnarray}
&& {\mathcal{C}}_{11}=\frac{v_1 }{v},\hspace{3.6cm}{\mathcal{C}}_{12}=\frac{v_2}{v},\hspace{3.6cm}
{\mathcal{C}}_{13}=\sqrt{2}\frac{v_t}{v} \\
&& {\mathcal{C}}_{21}=\frac{x_1}{\sqrt{\mathcal{N}}},\hspace{3.25cm}{\mathcal{C}}_{22}=\frac{x_2}{\sqrt{\mathcal{N}}},\hspace{3.2cm}{\mathcal{C}}_{22}=\frac{1}{\sqrt{\mathcal{N}}}\\
&& {\mathcal{C}}_{31}={\mathcal{C}}_{21}[m^2_{H_1^\pm}\to m^2_{H_2^\pm}],
\hspace{1cm}
{\mathcal{C}}_{32}={\mathcal{C}}_{32}[m^2_{H_1^\pm}\to m^2_{H_2^\pm}],
\hspace{1cm}
{\mathcal{C}}_{33}={\mathcal{C}}_{23}[m^2_{H_1^\pm}\to m^2_{H_2^\pm}]
\end{eqnarray}
where $v_0=\sqrt{v_1^2+v_2^2}$ and
\footnotesize{\begin{eqnarray}
x_1=\frac{v_0 c_{\beta } \left(v_0 \left({\mathcal{M}}^{\pm}_{12} \left({\mathcal{M}}^{\pm}_{13} c_{\beta }+{\mathcal{M}}^{\pm}_{23} s_{\beta }\right)+{\mathcal{M}}^{\pm}_{13} m_{H^\pm_1}^2 s_{\beta }\right)+\sqrt{2} {\mathcal{M}}^{\pm}_{13} {\mathcal{M}}^{\pm}_{23} v_t\right)}{\sqrt{2} v_0 v_t \left(m_{H^\pm_1}^2 \left(M_{23} c_{\beta }+{\mathcal{M}}^{\pm}_{13} s_{\beta }\right)+{\mathcal{M}}^{\pm}_{12} \left({\mathcal{M}}^{\pm}_{13} c_{\beta }+{\mathcal{M}}^{\pm}_{23} s_{\beta }\right)\right)+v_0^2 m_{H^\pm_1}^2 \left(c_{\beta } m_{H^\pm_1}^2 s_{\beta }+{\mathcal{M}}^{\pm}_{12}\right)+2 {\mathcal{M}}^{\pm}_{13} {\mathcal{M}}^{\pm}_{23} v_t^2}\nonumber\\
\end{eqnarray}}
\footnotesize{\begin{eqnarray}
x_2=\frac{v_0 s_{\beta } \left({\mathcal{M}}^{\pm}_{23} \left(v_0 c_{\beta } m_{H^\pm_1}^2+\sqrt{2} {\mathcal{M}}^{\pm}_{13} v_t\right)+{\mathcal{M}}^{\pm}_{12} v_0 \left({\mathcal{M}}^{\pm}_{13} c_{\beta }+{\mathcal{M}}^{\pm}_{23} s_{\beta }\right)\right)}{\sqrt{2} v_0 v_t \left(m_{H^\pm_1}^2 \left({\mathcal{M}}^{\pm}_{23} c_{\beta }+{\mathcal{M}}^{\pm}_{13} s_{\beta }\right)+{\mathcal{M}}^{\pm}_{12} \left({\mathcal{M}}^{\pm}_{13} c_{\beta }+{\mathcal{M}}^{\pm}_{23} s_{\beta }\right)\right)+v_0^2 m_{H^\pm_1}^2 \left(c_{\beta } m_{H^\pm_1}^2 s_{\beta }+{\mathcal{M}}^{\pm}_{12}\right)+2 {\mathcal{M}}^{\pm}_{13} {\mathcal{M}}^{\pm}_{23} v_t^2}\nonumber\\
\end{eqnarray}}
\begin{eqnarray}
{\mathcal{N}}=\sqrt{1+x_1^2+x_2^2}
\end{eqnarray}
Furthermore
\begin{eqnarray}
\begin{array}{ccc}
\tan\theta_1^\pm=\frac{{\mathcal{C}}_{12}}{{\mathcal{C}}_{11}}=\frac{v_2}{v_1} \quad;\quad & \tan\theta_2^\pm=\frac{{\mathcal{C}}_{13}}{{\mathcal{C}}_{12}}s_{\theta_1^\pm} \quad;\quad & \tan\theta_3^\pm=\frac{{\mathcal{C}}_{23}}{{\mathcal{C}}_{33}}
\end{array}
\end{eqnarray}

\subsection{${\mathcal{CP}}_{odd}$ sector}
For the terms of odd sector, the rotation matrix ${\mathcal{O}}$ can be expressed in terms of the mixing angles $\beta_i$ as,
\begin{eqnarray}
{\mathcal{O}} =\left(\begin{array}{ccc}
{\mathcal{O}}_{11} & {\mathcal{O}}_{12} & {\mathcal{O}}_{13}\\
{\mathcal{O}}_{21} & {\mathcal{O}}_{22} & {\mathcal{O}}_{23} \\
{\mathcal{O}}_{31} & {\mathcal{O}}_{32} & {\mathcal{O}}_{33}
\end{array} \right)=\left( \begin{array}{ccc}
c_{\beta_1} c_{\beta_2} & s_{\beta_1} c_{\beta_2} & s_{\beta_2}\\
-(c_{\beta_1} s_{\beta_2} s_{\beta_3} + s_{\theta^\pm_1} c_{\beta_3})
& c_{\beta_1} c_{\beta_3} - s_{\beta_1} s_{\beta_2} s_{\beta_3}
& c_{\beta_2} s_{\beta_3} \\
- c_{\beta_1} s_{\beta_2} c_{\beta_3} + s_{\beta_1} s_{\beta_3} &
-(c_{\beta_1} s_{\beta_3} + s_{\beta_1} s_{\beta_2} c_{\beta_3})
& c_{\beta_2}  c_{\beta_3}
\end{array} \right)
\label{eq:mixingmatrix}
\end{eqnarray}

\begin{eqnarray}
&& {\mathcal{O}}_{11}=\frac{v_1 }{v},\hspace{3.1cm}{\mathcal{O}}_{12}=\frac{v_2}{v},\hspace{3.1cm}
{\mathcal{O}}_{13}=\sqrt{2}\frac{v_t}{v} \\
&& {\mathcal{O}}_{21}=\frac{y_1}{\sqrt{\mathcal{N}_1}},\hspace{2.8cm}{\mathcal{O}}_{22}=\frac{y_2}{\sqrt{\mathcal{N}_1}},\hspace{2.6cm}{\mathcal{O}}_{22}=\frac{1}{\sqrt{\mathcal{N}_1}}\\
&& {\mathcal{O}}_{31}={\mathcal{O}}_{21}[m^2_{A_1}\to m^2_{A_2}],
\hspace{1cm}
{\mathcal{O}}_{32}={\mathcal{O}}_{32}[m^2_{A_1}\to m^2_{A_2}],
\hspace{1cm}
{\mathcal{O}}_{33}={\mathcal{O}}_{23}[m^2_{A_1}\to m^2_{A_2}]
\end{eqnarray}
with $v_0=\sqrt{v_1^2+v_2^2}$ and,
\footnotesize{\begin{eqnarray}
\hspace{-1cm}y_1=\frac{v_0 c_{\beta } \left(v_0 \left({\mathcal{M}}^{odd}_{13} m_{A_1}^2 s_{\beta }+{\mathcal{M}}^{odd}_{12} \left({\mathcal{M}}^{odd}_{13} c_{\beta }+{\mathcal{M}}^{odd}_{23} s_{\beta }\right)\right)+2 {\mathcal{M}}^{odd}_{13} {\mathcal{M}}^{odd}_{23} v_t\right)}{2 v_0 v_t \left(m_{A_1}^2 \left({\mathcal{M}}^{odd}_{23} c_{\beta }+{\mathcal{M}}^{odd}_{13} s_{\beta }\right)+{\mathcal{M}}^{odd}_{12} \left({\mathcal{M}}^{odd}_{13} c_{\beta }+{\mathcal{M}}^{odd}_{23} s_{\beta }\right)\right)+v_0^2 m_{A_1}^2 \left(m_{A_1}^2 c_{\beta } s_{\beta }+{\mathcal{M}}^{odd}_{12}\right)+4 {\mathcal{M}}^{odd}_{13} {\mathcal{M}}^{odd}_{23} v_t^2}
\nonumber\\
\end{eqnarray}}

\footnotesize{\begin{eqnarray}
\hspace{-1cm}y_2=\frac{v_0 s_{\beta } \left({\mathcal{M}}^{odd}_{23} \left(v_0 m_{A_1}^2 c_{\beta }+2 {\mathcal{M}}^{odd}_{13} v_t\right)+{\mathcal{M}}^{odd}_{12} v_0 \left({\mathcal{M}}^{odd}_{13} c_{\beta }+{\mathcal{M}}^{odd}_{23} s_{\beta }\right)\right)}{2 v_0 v_t \left(m_{A_1}^2 \left({\mathcal{M}}^{odd}_{23} c_{\beta }+{\mathcal{M}}^{odd}_{13} s_{\beta }\right)+{\mathcal{M}}^{odd}_{12} \left({\mathcal{M}}^{odd}_{13} c_{\beta }+{\mathcal{M}}^{odd}_{23} s_{\beta }\right)\right)+v_0^2 m_{A_1}^2 \left(m_{A_1}^2 c_{\beta } s_{\beta }+{\mathcal{M}}^{odd}_{12}\right)+4 {\mathcal{M}}^{odd}_{13} {\mathcal{M}}^{odd}_{23} v_t^2}\nonumber\\
\end{eqnarray}}
\begin{eqnarray}
{\mathcal{N}_1}=\sqrt{1+y_1^2+y_2^2}
\end{eqnarray}
and 
\begin{eqnarray}
\begin{array}{ccc}
\tan\beta_1=\frac{{\mathcal{O}}_{12}}{{\mathcal{O}}_{11}}=\frac{v_2}{v_1} \quad,\quad & \tan\beta_2=\frac{{\mathcal{O}}_{13}}{{\mathcal{O}}_{12}}s_{\beta_1} \quad,\quad & \tan\beta_3=\frac{{\mathcal{O}}_{23}}{{\mathcal{O}}_{33}}
\end{array}
\end{eqnarray}
%

\section{Setting the model parameters}
\label{appendice:A}
There are currently two choices of input parameters implemented in 2HDMcT. 
%
To ease export formalities for second set, a simplified approximations should be adopted for such purpose. Indeed, the $\mathcal{C}_{ij}(i,j=1,2,3)$ and $\mathcal{O}_{ij}(i,j=1,2,3)$ are nearly the same as $v_t<<v_0$, which involves
\begin{eqnarray}
&&\theta^\pm_2=\beta_2=0
\label{eq:C0}\\
&&\sin\theta^\pm_1=\sin\beta_1=\frac{v_2}{ v_0}=\sin\beta
\label{eq:C1}\\
&&\cos\theta^\pm_1=\cos\beta_1=\frac{v_1}{ v_0}=\cos\beta
\label{eq:C2}
\end{eqnarray}
Using the Eqs. \ref{eq:C0}, \ref{eq:C1} and \ref{eq:C2}, the matrices $\mathcal{C}$ and $\mathcal{O}$
become,
\begin{eqnarray}
{\mathcal{C}} =\left( \begin{array}{ccc}
c_{\beta} & s_{\beta} & 0\\
-s_{\beta} c_{\theta^\pm_3} & c_{\beta} c_{\theta^\pm_3} & s_{\theta^\pm_3} \\
 s_{\beta} s_{\theta^\pm_3} & -c_{\beta} s_{\theta^\pm_3} & c_{\theta^\pm_3}
\end{array} \right)&\,\,\,\,\,,\,\,\,\,{\mathcal{O}} =\left( \begin{array}{ccc}
c_{\beta} & s_{\beta} & 0\\
-s_{\beta} c_{\beta_3} & c_{\beta} c_{\beta_3} &  s_{\beta_3} \\
s_{\beta} s_{\beta_3} & -c_{\beta} s_{\beta_3} & c_{\beta_3}
\end{array}\right)
\label{rota-matrix-odd-charged}
\end{eqnarray}
using again Eqs. \ref{eq:mHpmpm}, \ref{rota-matrix-charged}, \ref{rota-matrix-cp-odd}, \ref{rota-matrix-cp-even} and \ref{rota-matrix-odd-charged}, the non physical parameters can be expressed through the masses, mixing angles, $\mu_1$ and $m_{12}^2$. The input set becomes:
\begin{eqnarray}
\mathcal{P}_{II} = \left\{\mu_1,v_t,\alpha_{1,2,3},\theta^\pm_2,\beta_3,\tan\beta,m_{h_1,h_2,h_3},m_{A_1,A_2},m_{H_1^\pm,H_2^\pm},m_{H^{\pm\pm}}\right\}
\label{eq:set-para2}
\end{eqnarray}
One can then provide the following expressions for the non-physical parameters,
\small{\begin{flalign}
	\lambda_1=\frac{\text{sc}_{\beta }^2 \left(\right.-m_{A_2}^2 s_{\beta } s_{\beta _3}^2 \left(s_{\beta }+t_{\beta }\right)+2 c_{\alpha _1}^2 \left(c_{\alpha _2}^2 m_{h_1}^2+s_{\alpha _2}^2 \left(c_{\alpha _3}^2 m_{h_3}^2+m_{h_2}^2 s_{\alpha _3}^2\right)\right)}{2 v_0^2}\nonumber\\
	\frac{2 s_{\alpha _1}^2 \left(c_{\alpha _3}^2 m_{h_2}^2+m_{h_3}^2 s_{\alpha _3}^2\right)+\left(m_{h_2}^2-m_{h_3}^2\right) s_{2 \alpha _1} s_{\alpha _2} s_{2 \alpha _3}-2 m_{12}^2 t_{\beta }+2 \sqrt{2} \mu _1 v_t\left.\right)}{2 v_0^2},\hspace*{-0.6cm}&&
	\end{flalign}}
\vspace*{-0.6cm}
\small{\begin{flalign}
	\lambda_2=\frac{1}{2 v_0^2}\left(\right.\text{cs}_{\beta }^2 \left(\right.-m_{A_2}^2 c_{\beta } \left(c_{\beta }+1\right) s_{\beta _3}^2+2 c_{\alpha _1}^2 \left(c_{\alpha _3}^2 m_{h_2}^2+m_{h_3}^2 s_{\alpha _3}^2\right)+2 s_{\alpha _1}^2 \left(\right.c_{\alpha _2}^2 m_{h_1}^2\nonumber\\
	+s_{\alpha _2}^2
	\left(c_{\alpha _3}^2 m_{h_3}^2+m_{h_2}^2 s_{\alpha _3}^2\right)\left.\right)-2 m_{12}^2 \text{ct}_{\beta }+2 \sqrt{2} \mu _1 \text{ct}_{\beta }^2 v_t+\left(m_{h_3}^2-m_{h_2}^2\right) s_{2 \alpha _1} s_{\alpha _2} s_{2 \alpha _3}\left.\right)\left.\right)\hspace*{-1.1cm}&&
	\end{flalign}}
\vspace*{-0.6cm}
\small{\begin{flalign}
	\lambda_3=\frac{1}{2 v_0^2}\left(\right.m_{A_2}^2 s_{\beta _3}^2 \left(\text{sc}_{\beta }+1\right)+\text{cs}_{\beta } \text{sc}_{\beta } \left(\right.c_{\alpha _1}^2 \left(m_{h_3}^2-m_{h_2}^2\right) s_{\alpha _2} s_{2 \alpha _3}+s_{2 \alpha _1} \left(\right.c_{\alpha _2}^2 m_{h_1}^2\nonumber\\
	+c_{\alpha _3}^2 \left(m_{h_3}^2 s_{\alpha _2}^2-m_{h_2}^2\right)\left.\right)-2 m_{12}^2\left.\right)
	+4 c_{\theta^\pm_3}^2 m_{h_1}^2+2 \text{cs}_{2 \beta } \left(\right.\left(m_{h_2}^2-m_{h_3}^2\right) s_{\alpha _2} s_{2 \alpha _3} s_{\alpha _1}^2\hspace*{-0.5cm}\nonumber\\
	+s_{2 \alpha _1} s_{\alpha _3}^2 \left(m_{h_2}^2 s_{\alpha _2}^2-m_{h_3}^2\right)\left.\right)-2 \sqrt{2} \mu _1 \text{cs}_{\beta }^2 v_t-4 m_{h_2}^2 s_{\theta^\pm_3}^2 \text{sc}_{\beta }\left.\right),\hspace*{-0.9cm}\hspace*{3.2cm}&&
	\end{flalign}}
\vspace*{-0.6cm}
\small{\begin{flalign}
	\lambda_4=\frac{2 m_{A_1}^2 c_{\beta _3}^2-m_{A_2}^2 s_{\beta _3}^2 \left(\text{sc}_{\beta }-1\right)-4 c_{\theta^\pm_3}^2 m_{h_1}^2+2 m_{12}^2 \text{cs}_{\beta } \text{sc}_{\beta }-2 \sqrt{2} \mu _1 \text{cs}_{\beta }^2 v_t+4 m_{h_2}^2 s_{\theta^\pm_3}^2 \text{sc}_{\beta }}{2 v_0^2}&&
	\end{flalign}}
\vspace*{-0.6cm}
\small{\begin{flalign}
	\lambda_5=\frac{-2 m_{A_1}^2 c_{\beta _3}^2+m_{A_2}^2 s_{\beta _3}^2 \left(\text{sc}_{\beta }-1\right)+2 m_{12}^2 \text{cs}_{\beta } \text{sc}_{\beta }+2 \sqrt{2} \mu _1 \text{cs}_{\beta }^2 v_t}{2 v_0^2}&&
	\end{flalign}}
\vspace*{-0.6cm}
\begin{eqnarray}
\lambda_6&=&\frac{\text{sc}_{\beta } \left(\right.8 v_0 c_{\frac{\beta }{2}}^4 s_{\frac{\beta }{2}}^2 \left(m_{A_2}^2 c_{2 \beta _3}-m_{A_2}^2-4 c_{2 \theta^\pm_3} m_{h_2}^2+4 m_{h_2}^2\right)+c_{\alpha _1} s_{2 \alpha _2} v_t \left(\right.c_{2 \alpha _3} \left(m_{h_2}^2-m_{h_3}^2\right)}{4 v_0 v_t^2}\nonumber\hspace*{9cm}\\
&&\frac{2 m_{h_1}^2-m_{h_2}^2-m_{h_3}^2\left.\right)+2 c_{\alpha _2} \left(m_{h_3}^2-m_{h_2}^2\right) s_{\alpha _1} s_{2 \alpha _3} v_t\left.\right)}{4 v_0 v_t^2}
\end{eqnarray}
\vspace*{-0.6cm}
\begin{eqnarray}
\lambda_7&=&\frac{2 v_0 c_{\frac{\beta }{2}}^2 \left(m_{A_2}^2 c_{2 \beta _3}-m_{A_2}^2-4 c_{2 \theta^\pm_3} m_{h_2}^2+4 m_{h_2}^2\right)+\text{cs}_{\beta } v_t \left(\right.s_{\alpha _1} s_{2 \alpha _2} \left(\right.c_{2 \alpha _3} \left(m_{h_2}^2-m_{h_3}^2\right)}{4 v_0 v_t^2}\hspace*{9cm}\\
&&\frac{2 m_{h_1}^2-m_{h_2}^2-m_{h_3}^2\left.\right)+2 c_{\alpha _1} c_{\alpha _2} \left(m_{h_2}^2-m_{h_3}^2\right) s_{2 \alpha _3}\left.\right)}{4 v_0 v_t^2}
\end{eqnarray}
\vspace*{-0.6cm}
\small{\begin{flalign}
	\lambda_8=\frac{s_{\beta } \left(s_{\beta }+t_{\beta }\right) \left(m_{A_2}^2 s_{\beta _3}^2-2 m_{h_2}^2 s_{\theta^\pm_3}^2\right)}{v_t^2}&&
	\end{flalign}}
\vspace*{-0.6cm}
\small{\begin{flalign}
	\lambda_9=\frac{c_{\frac{\beta }{2}}^2 \left(m_{A_2}^2 \left(-c_{2 \beta _3}\right)+m_{A_2}^2+2 c_{2 \theta^\pm_3} m_{h_2}^2-2 m_{h_2}^2\right)}{v_t^2}&&
	\end{flalign}}
\vspace*{-0.6cm}
\small{\begin{flalign}
	\overline{\lambda} _{8}=\frac{2 v_0^2 c_{\frac{\beta }{2}}^6 s_{\frac{\beta }{2}}^2 \left(m_{A_2}^2(1-c_{2 \beta _3})+(8 c_{2 \theta^\pm_3}-8)m_{h_2}^2\right)+c_{\alpha _2}^2 v_t^2 \left(c_{\alpha _3}^2 m_{h_3}^2+m_{h_2}^2 s_{\alpha _3}^2\right)+v_t^2 \left(m_{h_1}^2 s_{\alpha _2}^2+2 m_{H^{\mp\pm}}^2\right)}{2 v_t^4}&&
	\end{flalign}}
\vspace*{-0.6cm}
\small{\begin{flalign}
	\bar{\lambda}_9=\frac{2 v_0^2 c_{\frac{\beta }{2}}^6 s_{\frac{\beta }{2}}^2 \left(m_{A_2}^2 c_{2 \beta _3}-m_{A_2}^2-4 c_{2 \theta^\pm_3} m_{h_2}^2+4 m_{h_2}^2\right)-m_{H^{\pm\pm}}^2 v_t^2}{v_t^4}&&
	\end{flalign}}
\vspace*{-0.6cm}
\small{\begin{flalign}
	\mu _2=\frac{m_{A_2}^2 s_{\beta }^2 s_{\beta _3}^2}{2 \sqrt{2} v_t}+\mu _1 \text{ct}_{\beta }^2&&
	\end{flalign}}
\vspace*{-0.6cm}
\small{\begin{flalign}
	\mu _3=\frac{m_{A_2}^2 \left(c_{\beta }+1\right) s_{\beta } s_{\beta _3}^2-2 \sqrt{2} \mu _1 \text{ct}_{\beta } v_t}{\sqrt{2} v_t}&&
\end{flalign}}
with $\text{t}_{x}=\tan x$, $\text{ct}_{x}=1/\tan x$, $\text{c}_{x}=\cos x$, $\text{s}_{x}=\sin x$, $\text{s}_{2x}=\sin 2x$, $\text{cs}_{x}=1/\cos x$, $\text{se}_{x}=1/\sin x$

\section{Unitarity Constraints Matrices}
\label{unita-appendix}
The unitarity constraints are derived in the basis of unrotated states, corresponding to the fields before electroweak symmetry breaking. The quartic scalar vertices have in this case a much simpler form than the complicated functions of $\lambda_i$,
$\alpha_1$, ${\mathcal{O}}_{ij}$ and ${\mathcal{C}}_{ij}$ obtained in the physical basis ($H^{\pm\pm}$, $H_1^\pm$, $H_2^\pm$, $G^\pm$, $H_1$, $H_2$, $H_3$, $A_1$, $A_2$ and $G^0$) of mass eigenstate fields.  The $S$-matrix for the physical fields
is related by a unitary transformation to the
$S$-matrix for the unrotated fields.

Close inspection shows that the full set of $2$-body scalar scattering processes leads to
a $40\times40$ $S$-matrix which can be decomposed into 7 block submatrices corresponding to mutually unmixed sets of channels with definite charge and CP states. One has the following submatrix dimensions,  
structured in terms of net electric charge in the initial/final states:   
${S}^{(1)}(18 \times 18)$, ${S}^{(2)}(10\times 10)$ and ${S}^{(3)}(3\times 3)$, corresponding to 
$0$-charge channels, ${S}^{(4)}(21 \times 21)$ corresponding to the $1$-charge channels, 
${S}^{(5)}(12 \times 12)$ corresponding to the $2$-charge channels, ${S}^{(6)}(3 \times 3)$ corresponding to the $3$-charge channels and finally ${S}^{(7)}(1 \times 1)$ corresponding to the $4$-charge channels.

The first submatrix ${\cal M}_1$  corresponds to scattering whose
initial and final states are one of the following :
($\phi_1^+\delta^-$, $\delta^+\phi_1^-$, $\phi_2^+\delta^-$, $\delta^+\phi_2^-$, $\phi_1^+\phi_2^-$, $\phi_2^+\phi_1^-$, $\rho_1\eta_0$, $\rho_2\eta_0$, $\rho_1\eta_2$, $\rho_0\eta_1$, $\rho_0\eta_2$, $\rho_2\eta_1$, $\eta_1\eta_0$, $\eta_2\eta_0$, $\eta_1\eta_2$, $\rho_1\rho_0$, $\rho_2\rho_0$, $\rho_1\rho_2$). With the help of Mathematica one finds, 
\begin{eqnarray}
\renewcommand{\arraystretch}{1.6}
{\cal M}_1=
\left(\begin{array}{c|c}
\,\displaystyle {\cal M}_1^{11}\,(7\times 7)  \,& \,\displaystyle {\cal M}_1^{12}\,(7\times 11)  \, \\
\hline
\,\displaystyle {\cal M}_1^{21}\,(11\times 7)  \,& \,\displaystyle {\cal M}_1^{22}\,(11\times 11)  \,
\end{array}
\right)
\end{eqnarray}
with,
\begin{eqnarray}
&&{\cal M}_1^{11}\,(7\times 7) = \hspace{0.32cm}
\left(
\begin{array}{ccccccc}
\displaystyle \tilde{\lambda}_{68} \,&\,\displaystyle  0 \,&\,\displaystyle  0 \,&\,\displaystyle  0 \,&\,\displaystyle  0 \,&\,\displaystyle  0  \,&\,\displaystyle -i\frac{\lambda_{8}}{2\sqrt{2}} \\	
\displaystyle 0 \,&\,\displaystyle  \tilde{\lambda}_{68} \,&\,\displaystyle  0 \,&\,\displaystyle  0 \,&\,\displaystyle  0 \,&\,\displaystyle  0  \,&\,\displaystyle +i\frac{\lambda_{8}}{2\sqrt{2}}  \\
\displaystyle 0 \,&\,\displaystyle  0 \,&\,\displaystyle  \tilde{\lambda}_{79} \,&\,\displaystyle  0 \,&\,\displaystyle  0 \,&\,\displaystyle  0  \,&\,\displaystyle 0  \\
\displaystyle 0 \,&\,\displaystyle  0 \,&\,\displaystyle  0 \,&\,\displaystyle  \tilde{\lambda}_{79} \,&\,\displaystyle  0 \,&\,\displaystyle  0  \,&\,\displaystyle 0  \\	
\displaystyle  0 \,&\,\displaystyle  0  \,&\,\displaystyle  0  \,&\,\displaystyle  0 \,&\,\displaystyle \lambda_{34}^+\,&\,\displaystyle 2\lambda_{5}\,&\,\displaystyle  0  \\	
\displaystyle  0 \,&\,\displaystyle  0  \,&\,\displaystyle  0  \,&\,\displaystyle  0 \,&\,\displaystyle 2\lambda_{5} \,&\,\displaystyle  \lambda_{34}^+  \,&\,\displaystyle  0  \\	
\displaystyle +i\frac{\lambda_{8}}{2\sqrt{2}} \,&\,\displaystyle -i\frac{\lambda_{8}}{2\sqrt{2}} \,&\,\displaystyle  0 \,&\,\displaystyle  0 \,&\,\displaystyle  0 \,&\,\displaystyle 0 \,&\,\displaystyle \lambda_{68}^+ \\	
\end{array}
\right)\nonumber\\
&&{\cal M}_1^{22}\,(11\times 11) = 
\left(
\begin{array}{ccccccccccc}
\displaystyle  \lambda_{79}^+  \,&\,\displaystyle 0  \,&\,\displaystyle 0  \,&\,\displaystyle 0  \,&\,\displaystyle 0  \,&\,\displaystyle  0 \,&\,\displaystyle 0  \,&\,\displaystyle 0  \,&\,\displaystyle 0  \,&\,\displaystyle 0  \,&\,\displaystyle 0  \\	

\displaystyle  0  \,&\,\displaystyle \lambda_{L}\,&\,\displaystyle  0  \,&\,\displaystyle  0  \,&\,\displaystyle \lambda_{5}\,&\,\displaystyle 0  \,&\,\displaystyle 0  \,&\,\displaystyle 0  \,&\,\displaystyle 0  \,&\,\displaystyle 0  \,&\,\displaystyle  0  \\	
\displaystyle 0  \,&\,\displaystyle 0  \,&\,\displaystyle  \lambda_{68}^+ \,&\,\displaystyle 0  \,&\,\displaystyle 0  \,&\,\displaystyle  0 \,&\,\displaystyle 0  \,&\,\displaystyle 0  \,&\,\displaystyle 0  \,&\,\displaystyle 0  \,&\,\displaystyle 0  \\	
\displaystyle  0  \,&\,\displaystyle 0  \,&\,\displaystyle 0  \,&\,\displaystyle  \lambda_{79}^+  \,&\,\displaystyle 0  \,&\,\displaystyle  0 \,&\,\displaystyle 0  \,&\,\displaystyle 0  \,&\,\displaystyle 0  \,&\,\displaystyle 0  \,&\,\displaystyle 0  \\	
\displaystyle  0  \,&\,\displaystyle \lambda_{5}\,&\,\displaystyle  0  \,&\,\displaystyle  0  \,&\,\displaystyle \lambda_{L}\,&\,\displaystyle 0  \,&\,\displaystyle 0  \,&\,\displaystyle 0  \,&\,\displaystyle 0  \,&\,\displaystyle 0  \,&\,\displaystyle  0  \\	
\displaystyle  0  \,&\,\displaystyle 0  \,&\,\displaystyle 0  \,&\,\displaystyle  0  \,&\,\displaystyle 0  \,&\,\displaystyle  \lambda_{68}^+ \,&\,\displaystyle 0  \,&\,\displaystyle 0  \,&\,\displaystyle 0  \,&\,\displaystyle 0  \,&\,\displaystyle 0  \\	
\displaystyle  0  \,&\,\displaystyle 0  \,&\,\displaystyle 0  \,&\,\displaystyle  0  \,&\,\displaystyle 0  \,&\,\displaystyle  0 \,&\,\displaystyle \bar{\lambda}_{79}^+  \,&\,\displaystyle 0  \,&\,\displaystyle 0  \,&\,\displaystyle 0  \,&\,\displaystyle 0  \\	
\displaystyle  0  \,&\,\displaystyle 0  \,&\,\displaystyle 0  \,&\,\displaystyle  0  \,&\,\displaystyle 0  \,&\,\displaystyle  0 \,&\,\displaystyle  0  \,&\,\displaystyle \lambda_{345}  \,&\,\displaystyle 0  \,&\,\displaystyle 0  \,&\,\displaystyle \lambda_{5}  \\	
\displaystyle  0  \,&\,\displaystyle 0  \,&\,\displaystyle 0  \,&\,\displaystyle  0  \,&\,\displaystyle 0  \,&\,\displaystyle   0  \,&\,\displaystyle 0  \,&\,\displaystyle 0  \,&\,\displaystyle \lambda_{68}^+   \,&\,\displaystyle 0  \,&\,\displaystyle 0  \\	
\displaystyle  0  \,&\,\displaystyle 0  \,&\,\displaystyle 0  \,&\,\displaystyle  0  \,&\,\displaystyle 0  \,&\,\displaystyle  0 \,&\,\displaystyle 0  \,&\,\displaystyle 0  \,&\,\displaystyle 0  \,&\,\displaystyle \bar{\lambda}_{79}^+  \,&\,\displaystyle 0  \\	
\displaystyle  0  \,&\,\displaystyle 0  \,&\,\displaystyle 0  \,&\,\displaystyle  0  \,&\,\displaystyle 0  \,&\,\displaystyle  0 \,&\,\displaystyle  0  \,&\,\displaystyle \lambda_{5}  \,&\,\displaystyle 0  \,&\,\displaystyle 0  \,&\,\displaystyle \lambda_{345}  \\	
\end{array}\right)\nonumber\\
&&{\cal M}_1^{12}\,(7\times 11) = \hspace{0.18cm}
\left(
\begin{array}{ccccccccccc}
\displaystyle  0  \,&\,\displaystyle  0  \,&\,\displaystyle +i\frac{\lambda_{8}}{2\sqrt{2}} \,&\,\displaystyle  0  \,&\,\displaystyle  0  \,&\,\displaystyle \frac{\lambda_{8}}{2\sqrt{2}} \,&\,\displaystyle  0  \,&\,\displaystyle  0  \,&\,\displaystyle \frac{\lambda_{8}}{2\sqrt{2}}  \,&\,\displaystyle  0 \,&\,\displaystyle  0 \\	
\displaystyle  0  \,&\,\displaystyle  0  \,&\,\displaystyle -i\frac{\lambda_{8}}{2\sqrt{2}} \,&\,\displaystyle  0  \,&\,\displaystyle  0  \,&\,\displaystyle \frac{\lambda_{8}}{2\sqrt{2}} \,&\,\displaystyle  0  \,&\,\displaystyle  0  \,&\,\displaystyle \frac{\lambda_{8}}{2\sqrt{2}}  \,&\,\displaystyle  0 \,&\,\displaystyle  0 \\	
\displaystyle  -i\frac{\lambda_{9}}{2\sqrt{2}}  \,&\,\displaystyle  0  \,&\,\displaystyle 0 \,&\,\displaystyle  +i\frac{\lambda_{9}}{2\sqrt{2}}  \,&\,\displaystyle  0  \,&\,\displaystyle 0 \,&\,\displaystyle  \frac{\lambda_{9}}{2\sqrt{2}}  \,&\,\displaystyle  0  \,&\,\displaystyle 0  \,&\,\displaystyle  \frac{\lambda_{9}}{2\sqrt{2}} \,&\,\displaystyle  0 \\	
\displaystyle  i\frac{\lambda_{9}}{2\sqrt{2}}  \,&\,\displaystyle  0  \,&\,\displaystyle 0 \,&\,\displaystyle  -i\frac{\lambda_{9}}{2\sqrt{2}}  \,&\,\displaystyle  0  \,&\,\displaystyle 0 \,&\,\displaystyle  \frac{\bar{\lambda}_{9}}{2\sqrt{2}}  \,&\,\displaystyle  0  \,&\,\displaystyle 0  \,&\,\displaystyle  \frac{\bar{\lambda}_{9}}{2\sqrt{2}} \,&\,\displaystyle  0 \\	
\displaystyle  0  \,&\,\displaystyle -i\frac{\lambda_{45}^-}{2} \,&\,\displaystyle  0  \,&\,\displaystyle  0  \,&\,\displaystyle  +i\frac{\lambda_{45}^-}{2} \,&\,\displaystyle  0  \,&\,\displaystyle  0  \,&\,\displaystyle \frac{\lambda_{45}^+}{2} \,&\,\displaystyle  0  \,&\,\displaystyle  0  \,&\,\displaystyle \frac{\lambda_{45}^+}{2} \\	
\displaystyle  0  \,&\,\displaystyle +i\frac{\lambda_{45}^-}{2} \,&\,\displaystyle  0  \,&\,\displaystyle  0  \,&\,\displaystyle  -i\frac{\lambda_{45}^-}{2} \,&\,\displaystyle  0  \,&\,\displaystyle  0  \,&\,\displaystyle \frac{\lambda_{45}^+}{2} \,&\,\displaystyle  0  \,&\,\displaystyle  0  \,&\,\displaystyle \frac{\lambda_{45}^+}{2} \\	
\displaystyle 0  \,&\,\displaystyle 0  \,&\,\displaystyle 0  \,&\,\displaystyle 0  \,&\,\displaystyle 0  \,&\,\displaystyle  0 \,&\,\displaystyle 0  \,&\,\displaystyle 0  \,&\,\displaystyle 0  \,&\,\displaystyle 0  \,&\,\displaystyle 0  \\	
\end{array}\right)\nonumber\\
&&{\cal M}_1^{12}\,(11\times 7) = \hspace{0.1cm} ({\cal M}_1^{12}\,(7\times 11))^\ast
\end{eqnarray}
where $\lambda_{ijk}=\lambda_i+\lambda_j+\lambda_k$, $\tilde{\lambda}_{ij}=\lambda_i + \frac{\lambda_j}{2}$, $\lambda_{ij}^{\pm}=\lambda_i \pm \lambda_j$ and $\lambda_L=\lambda_3+\lambda_4-\lambda_5$.

The second submatrix ${\cal M}_2$ corresponds to scattering with one of the following initial and final states:
$(\rho_1\eta_1$, $\rho_2\eta_2$, $\rho_0\eta_0)$. One finds that ${\cal M}_3$ given by :
\begin{eqnarray}
{\cal M}_2=\left(
\begin{array}{ccc}
\displaystyle \lambda_1& \lambda_5 & 0 \\
\displaystyle \lambda_5 & \displaystyle \lambda_2 & 0 \\
\displaystyle 0 & 0 & 2\bar{\lambda}_{89}^+\\
\end{array}
\right)
\end{eqnarray}

The third submatrix ${\cal M}_3$ corresponds to scattering with one of the following
initial and final states:
$(\phi_1^+\phi_1^-$, $\phi_2^+\phi_2^-$, $\delta^+\delta^-$, $\delta^{++}\delta^{--}$, $\frac{\eta_1\eta_1}{\sqrt{2}}$, $\frac{\eta_2\eta_2}{\sqrt{2}}$, $\frac{\eta_0\eta_0}{\sqrt{2}}$, $\frac{\rho_1\rho_1}{\sqrt{2}}$, $\frac{\rho_2\rho_2}{\sqrt{2}}$, $\frac{\rho_0\rho_0}{\sqrt{2}})$, where the $\sqrt{2}$ accounts for
identical particle statistics. One finds that ${\cal M}_2$ is given by:

\begin{eqnarray}
{\cal M}_3 = \left(
\begin{array}{cccccccccc}
2\lambda_{1} \,&\, \lambda_{34}^+ \,&\, \tilde{\lambda}_{68} \,&\, \lambda_{68}^+ \,&\, \frac{\lambda_1}{\sqrt{2}} \,&\, \frac{\lambda_3}{\sqrt{2}}  \,&\,\frac{\lambda_{6}}{\sqrt{2}} \,&\, \frac{\lambda_1}{\sqrt{2}}   \,&\, \frac{\lambda_3}{\sqrt{2}}  \,&\,\frac{\lambda_{6}}{\sqrt{2}} \\
\lambda_{34}^+ \,&\, 2\lambda_{2} \,&\, \tilde{\lambda}_{79} \,&\, \lambda_{79}^+ \,&\, \frac{\lambda_3}{\sqrt{2}}  \,&\, \frac{\lambda_2}{\sqrt{2}}  \,&\,\frac{\lambda_7}{\sqrt{2}} \,&\, \frac{\lambda_3}{\sqrt{2}}  \,&\, \frac{\lambda_2}{\sqrt{2}}  \,&\,\frac{\lambda_7}{\sqrt{2}} \\
\tilde{\lambda}_{68} \,&\, \tilde{\lambda}_{79} \,&\, 4\tilde{\bar{\lambda}}_{89} \,&\, 2\bar{\lambda}_{89} \,&\, \frac{\tilde{\lambda}_{68}}{\sqrt{2}} \,&\, \frac{\tilde{\lambda}_{79}}{\sqrt{2}}  \,&\,\sqrt{2}\bar{\lambda}_{89}\,&\, \frac{\tilde{\lambda}_{68}}{\sqrt{2}}  \,&\, \frac{\tilde{\lambda}_{79}}{\sqrt{2}}  \,&\,\sqrt{2}\bar{\lambda}_{89} \\
\lambda_{68}^+ \,&\, \lambda_{79}^+ \,&\, 2\bar{\lambda}_{89} \,&\, 4\bar{\lambda}_{89}^+ \,&\, \frac{\lambda_{6}}{\sqrt{2}} \,&\, \frac{\lambda_{7}}{\sqrt{2}}  \,&\,\sqrt{2}\bar{\lambda}_{8} \,&\, \frac{\lambda_{6}}{\sqrt{2}}  \,&\, \frac{\lambda_{7}}{\sqrt{2}}  \,&\,\sqrt{2}\bar{\lambda}_{8} \\
\frac{\lambda_1}{\sqrt{2}} \,&\, \frac{\lambda_3}{\sqrt{2}} \,&\, \frac{\tilde{\lambda}_{68}}{\sqrt{2}} \,&\, \frac{\lambda_{6}}{\sqrt{2}} \,&\, \frac{3\lambda_1}{2}  \,&\, \frac{\lambda_{345}}{2}  \,&\,\frac{\lambda_{68}^+}{2} \,&\, \frac{\lambda_1}{2}   \,&\, \frac{\lambda_{L}}{2}  \,&\,\frac{\lambda_{68}^+}{2} \\	
\frac{\lambda_3}{\sqrt{2}} \,&\, \frac{\lambda_2}{\sqrt{2}} \,&\, \frac{\tilde{\lambda}_{79}}{\sqrt{2}} \,&\, \frac{\lambda_{7}}{\sqrt{2}} \,&\, \frac{\lambda_{345}}{2} \,&\, \frac{3\lambda_2}{2}   \,&\,\frac{\lambda_{79}^+}{2} \,&\, \frac{\lambda_L}{2}  \,&\, \frac{\lambda_2}{2}   \,&\,\frac{\lambda_{79}^+}{2} \\

\frac{\lambda_{6}}{\sqrt{2}} \,&\, \frac{\lambda_7}{\sqrt{2}} \,&\, \sqrt{2}\bar{\lambda}_{89}^+ \,&\, \sqrt{2}\bar{\lambda}_{8} \,&\, \frac{\lambda_{68}^+}{2} \,&\, \frac{\lambda_{79}^+}{2}  \,&\,3\bar{\lambda}_{89}^+ \,&\, \frac{\lambda_{68}^+}{2}  \,&\, \frac{\lambda_{79}^+}{2}  \,&\,\bar{\lambda}_{89}^+ \\
\frac{\lambda_1}{\sqrt{2}}  \,&\, \frac{\lambda_3}{\sqrt{2}}  \,&\, \frac{\tilde{\lambda}_{68}}{\sqrt{2}} \,&\, \frac{\lambda_{6}}{\sqrt{2}} \,&\, \frac{\lambda_1}{2}  \,&\, \frac{\lambda_L}{2}  \,&\,\frac{\lambda_{68}^+}{2} \,&\, \frac{3\lambda_1}{2}   \,&\, \frac{\lambda_{345}}{2}  \,&\,\frac{\lambda_{68}^+}{2} \\
\frac{\lambda_3}{\sqrt{2}} \,&\, \frac{\lambda_2}{\sqrt{2}} \,&\, \frac{\tilde{\lambda}_{79}}{\sqrt{2}} \,&\, \frac{\lambda_{7}}{\sqrt{2}} \,&\, \frac{\lambda_{L}}{2} \,&\, \frac{\lambda_2}{2}   \,&\,\frac{\lambda_{79}^+}{2} \,&\, \frac{\lambda_{345}}{2}  \,&\, \frac{3\lambda_2}{2}   \,&\,\frac{\lambda_{79}^+}{2} \\
\frac{\lambda_6}{\sqrt{2}} \,&\, \frac{\lambda_7}{\sqrt{2}} \,&\, \sqrt{2}\bar{\lambda}_{89}^+ \,&\, \sqrt{2}\bar{\lambda}_{8} \,&\, \frac{\lambda_{68}^+}{2} \,&\, \frac{\lambda_{79}^+}{2}  \,&\,\bar{\lambda}_{89}^+ \,&\, \frac{\lambda_{68}^+}{2}  \,&\, \frac{\lambda_{79}^+}{2}  \,&\,3\bar{\lambda}_{89}^+	
\end{array}\right)
\end{eqnarray}
despite its apparently complicated structure, six eigenvalues for ${\cal M}_3$ come from a very long two polynomial equations of order 3. for solve these matrice we using the Jacobi method.

The fourth submatrix ${\cal M}_4$ corresponds to scattering with
initial and final states being one of the following $21$ sates:
($\rho_0\phi_1^+$, $\rho_1\phi_1^+$, $\rho_2\phi_1^+$, $\eta_0\phi_1^+$, $\eta_1\phi_1^+$, $\eta_2\phi_1^+$, $\rho_0\phi_2^+$, $\rho_1\phi_2^+$, $\rho_2\phi_2^+$, $\eta_0\phi_2^+$, $\eta_1\phi_2^+$, $\eta_2\phi_2^+$, $\rho_0\delta^+$, $\rho_1\delta^+$, $\rho_2\delta^+$, $\eta_0\delta^+$, $\eta_1\delta^+$, $\eta_2\delta^+$, $\delta^{++}\delta^-$, $\delta^{++}\phi_1^-$, $\delta^{++}\phi_2^-$). It reads that : 
\begin{eqnarray}
\renewcommand{\arraystretch}{1.6}
{\cal M}_4=
\left(
\begin{array}{c|c|c}
\,\displaystyle {\cal M}_4^{11}\,(7\times 7)  \,& \,\displaystyle {\cal M}_4^{12}\,(7\times 11)  \,& \,\displaystyle {\cal M}_4^{13}\,(7\times 11) \\
\hline
\,\displaystyle {\cal M}_4^{21}\,(7\times 7)  \,& \,\displaystyle {\cal M}_4^{22}\,(7\times 11)  \,& \,\displaystyle {\cal M}_4^{23}\,(7\times 11) \\
\hline
\,\displaystyle {\cal M}_4^{31}\,(7\times 7)  \,& \,\displaystyle {\cal M}_4^{32}\,(7\times 11)  \,& \,\displaystyle {\cal M}_4^{33}\,(7\times 11)
\end{array}
\right)
\label{eq:mat4}
\end{eqnarray}
with,
\begin{eqnarray}
\hspace{-0.3cm}{\cal M}_4^{11}\,(7\times 7) = 
\left(
\begin{array}{ccccccc}
 \lambda_6 \,&\,  0 \,&\,  0 \,&\,  0 \,&\,  0 \,&\,  0  \,&\, 0 \\	
 0 \,&\,  \lambda_1 \,&\,  0 \,&\,  0 \,&\,  0 \,&\,  0  \,&\, 0 \\
 0 \,&\,  0 \,&\,  \lambda_3 \,&\,  0 \,&\,  0 \,&\,  0  \,&\, 0 \\	
 0 \,&\,  0 \,&\,  0 \,&\,  \lambda_6 \,&\,  0 \,&\,  0  \,&\, 0 \\	
 0 \,&\,  0 \,&\,  0 \,&\,  0 \,&\,  \lambda_1 \,&\,  0  \,&\, 0 \\
 0 \,&\,  0 \,&\, 0 \,&\,  0 \,&\,  0 \,&\,   \lambda_3  \,&\, 0 \\
 0 \,&\,  0 \,&\, 0 \,&\,  0 \,&\,  0 \,&\,   0  \,&\, \lambda_7 	
\end{array}
\right),
{\cal M}_4^{22}\,(7\times 7) = 
\left(
\begin{array}{ccccccc}	
\lambda_3  \,&\,  0  \,&\, 0 \,&\,  0  \,&\,  0  \,&\, 0 \,&\,  0  \\
0  \,&\,  \lambda_2  \,&\, 0 \,&\,  0  \,&\,  0  \,&\, \frac{\lambda_{9}}{2\sqrt{2}} \,&\,  0  \\
0  \,&\,  0  \,&\, \lambda_{7} \,&\,  0  \,&\,  0  \,&\, 0 \,&\,  0 \\
0  \,&\,  0  \,&\, 0 \,&\,  \lambda_3  \,&\,  0  \,&\, 0 \,&\,  0 \\
0  \,&\,  0  \,&\, 0 \,&\,  0  \,&\,  \lambda_2  \,&\, i\frac{\lambda_{9}}{2\sqrt{2}} \,&\,  0  \\
0  \,&\,  \frac{\lambda_{9}}{2\sqrt{2}}  \,&\, 0 \,&\,  0  \,&\,
-i\frac{\lambda_{9}}{2\sqrt{2}}  \,&\, 2\bar{\lambda}_{89}^+ \,&\,  0\\
0  \,&\,  0  \,&\, 0 \,&\,  0  \,&\,  0  \,&\, 0 \,&\,  \tilde{\lambda}_{68}
\end{array}\right)\nonumber
\end{eqnarray}
\begin{eqnarray}
{\cal M}_4^{33}\,(7\times 7) = 
{\small\left(
\begin{array}{ccccccc}
\tilde{\lambda}_{79}  \,&\, 0  \,&\,  0 \,&\,  0  \,&\,  0 \,&\,  0 \,&\,  -\frac{\lambda_{9}}{2} \\ 
0  \,&\, 2\bar{\lambda}_{89}^+  \,&\,  0 \,&\,  0  \,&\,  -i\sqrt{2}\bar{\lambda}_{9} \,&\,  0 \,&\,  0\\
0  \,&\, 0  \,&\,  \tilde{\lambda}_{68} \,&\,  0  \,&\,  0 \,&\,  -i\frac{\lambda_{8}}{2} \,&\,  0\\
0  \,&\, 0  \,&\,  0 \,&\,  \tilde{\lambda}_{79}  \,&\,  0 \,&\,  0 \,&\,  -i\frac{\lambda_{9}}{2} \\
0  \,&\, i\sqrt{2}\bar{\lambda}_{9} \,&\,  0 \,&\,  0  \,&\,  2\bar{\lambda}_{89}^+  \,&\,  0 \,&\,  0\\
0  \,&\, 0  \,&\,  i\frac{\lambda_{8}}{2} \,&\,  0  \,&\,  0 \,&\,  \lambda_{68}^+ \,&\,  0\\
-\frac{\lambda_{9}}{2}  \,&\, 0  \,&\,  0   \,&\,  i\frac{\lambda_{9}}{2}  \,&\,  0 \,&\,  0  \,&\,  \lambda_{79}^+ 
\end{array}\right)}
\end{eqnarray}
and
\begin{eqnarray}
\hspace{-1.5cm}{\cal M}_4^{12}\,(7\times 7) = 
{\small\left(
\begin{array}{ccccccc}
0  \,&\,  0  \,&\, 0 \,&\,  0  \,&\,  0  \,&\, 0 \,&\,  \frac{\lambda_{8}}{2\sqrt{2}} \\
0  \,&\,  \frac{\lambda_{45}^+}{2}  \,&\, 0 \,&\,  0  \,&\,  i\frac{\lambda_{54}^-}{2}  \,&\, \frac{\lambda_{8}}{2\sqrt{2}} \,&\,  0  \\
\frac{\lambda_{45}^+}{2}  \,&\,  0  \,&\, 0 \,&\,  i\frac{\lambda_{45}^-}{2}  \,&\,  0  \,&\, 0 \,&\,  0\\
0  \,&\,  0  \,&\, 0 \,&\,  0  \,&\,  0  \,&\, 0 \,&\,  -i\frac{\lambda_{8}}{2\sqrt{2}} \\
0  \,&\,  i\frac{\lambda_{45}^-}{2}  \,&\, 0 \,&\,  0  \,&\,  \frac{\lambda_{45}^+}{2}  \,&\, i\frac{\lambda_{8}}{2\sqrt{2}} \,&\,  0 \\
i\frac{\lambda_{54}^-}{2}  \,&\,  0  \,&\, 0 \,&\,  \frac{\lambda_{45}^+}{2}  \,&\,  0  \,&\, 0 \,&\,  0\\
0  \,&\,  0  \,&\, 0 \,&\,  0  \,&\,  0  \,&\, 0 \,&\,  0  
\end{array}\right)},
{\cal M}_4^{13}\,(7\times 7) = 
{\small\left(
\begin{array}{ccccccc}
0  \,&\, 0  \,&\,  i\frac{\lambda_{8}}{2\sqrt{2}} \,&\,  0  \,&\,  0 \,&\,  0
\,&\,  0\\
0  \,&\, -i\frac{\lambda_{8}}{2\sqrt{2}}  \,&\,  0 \,&\,  0  \,&\,  -\frac{\lambda_{8}}{2} \,&\,  0 \,&\,  0\\
0  \,&\, 0  \,&\,  0 \,&\,  0  \,&\,  0 \,&\,  0 \,&\,  0\\
0  \,&\, 0  \,&\,  \frac{\lambda_{8}}{2\sqrt{2}} \,&\,  0  \,&\,  0 \,&\,  0 \,&\,  0\\
0  \,&\, \frac{\lambda_{8}}{2\sqrt{2}}  \,&\,  0 \,&\,  0  \,&\,  -i\frac{\lambda_{8}}{2} \,&\,  0 \,&\,  0\\
0  \,&\, 0  \,&\,  0 \,&\,  0  \,&\,  0 \,&\,  0 \,&\,  0\\
\frac{\lambda_{9}}{2\sqrt{2}}  \,&\, 0  \,&\,  0 \,&\,  i\frac{\lambda_{9}}{2\sqrt{2}}  \,&\,  0 \,&\,  0 \,&\,  0
\end{array}\right)}\nonumber
\end{eqnarray}
\begin{eqnarray}
{\cal M}_4^{23}\,(7\times 7) = 
{\small\left(
\begin{array}{ccccccccccccccccccccc}	
0  \,&\, 0  \,&\,  0 \,&\,  0  \,&\,  0 \,&\,  0 \,&\,  0\\
0  \,&\, -i\frac{\lambda_{9}}{2\sqrt{2}}  \,&\,  0 \,&\,  0  \,&\,  -\frac{\lambda_{9}}{2} \,&\,  0 \,&\,  0\\
-i\frac{\lambda_{9}}{2\sqrt{2}}  \,&\, 0  \,&\,  0 \,&\,  \frac{\lambda_{9}}{2\sqrt{2}}  \,&\,  0 \,&\,  0 \,&\,  0\\
0  \,&\, 0  \,&\,  0 \,&\,  0  \,&\,  0 \,&\,  0 \,&\,  0\\ 
0  \,&\, \frac{\bar{\lambda}_{9}}{2\sqrt{2}}  \,&\,  0 \,&\,  0  \,&\,  -i\frac{\bar{\lambda}_{9}}{2} \,&\,  0 \,&\,  0\\
0  \,&\, 0  \,&\,  0 \,&\,  0  \,&\,  -\sqrt{2}\bar{\lambda}_{9} \,&\,  0 \,&\,  0\\
0  \,&\, 0  \,&\,  0 \,&\,  0  \,&\,  0 \,&\,  -\frac{\lambda_{8}}{2} \,&\,  0
\end{array}\right)}
\end{eqnarray}
while the other off-diagonal lower elements in Eq. \ref{eq:mat4} are related to upper off-diagonal ones by switching $i\leftrightarrow j$ in ${\cal M}_4^{ij}\,(7\times 7) = ({\cal M}_4^{ji}\,(7\times 7))^\ast$ for $i>1\land i>j \land i\ne j$. 
\noindent
Solving the cubic polynomial equation given in Eq. \ref{eq:cubic-polynom}, we find:
\begin{eqnarray}
&& a_{17}=\frac{1}{3}\left( b+2\sqrt{p}\cos\left( \frac{\Lambda}{3} \right) \right) \label{eq:a17}\nonumber\\
&& a_{18}=\frac{1}{3}\left( b+2\sqrt{p}\cos\left( \frac{\Lambda+2\pi}{3} \right) \right) \label{eq:a18}\nonumber\\
&& a_{19}=\frac{1}{3}\left( b+2\sqrt{p}\cos\left( \frac{\Lambda-2\pi}{3} \right) \right) \label{eq:a19}
\end{eqnarray}
with
\begin{eqnarray}
&& b = 2\bar{\lambda}_8 + \bar{\lambda}_9+\lambda_1+\lambda_2 \nonumber\\
&& p = b^2 - 3\,r \nonumber\\
&& \Lambda = \cos^{-1}\left( \frac{q}{2\sqrt{p^3}}  \right) 
\end{eqnarray}
where $r$, $q$ stand for
\begin{eqnarray}
&& r=2\bar{\lambda}_8\lambda_1+\bar{\lambda}_9\lambda_1+2\bar{\lambda}_8\lambda_2+\bar{\lambda}_9\lambda_2+\lambda_1\lambda_2-\lambda_4^2-\lambda_8^2-\lambda_9^2\\
&& q= 2b^3 - 9\,b\,r - 27\,r_1
\end{eqnarray}
and $r_1=-2\bar{\lambda}_8\lambda_1\lambda_2-\bar{\lambda}_9\lambda_1\lambda_2+2\bar{\lambda}_8\lambda_4^2+\bar{\lambda}_9\lambda_4^2+\lambda_2\lambda_8^2-2\lambda_4\lambda_8\lambda_9+\lambda_1\lambda_9^2$

The fifth submatrix ${\cal M}_5$ corresponds to scattering with initial and final states being one of the following $12$ sates:
($\frac{\phi_1^+\phi_1^+}{\sqrt{2}}$, $\frac{\phi_2^+\phi_2^+}{\sqrt{2}}$, $\frac{\delta^+\delta^+}{\sqrt{2}}$, $\phi_1^+\phi_2^+$, $\phi_1^+\delta^+$, $\phi_2^+\delta^+$, $\delta^{++}\rho_0$, $\delta^{++}\rho_1$, $\delta^{++}\rho_2$, $\delta^{++}\eta_0$, $\delta^{++}\eta_1$, $\delta^{++}\eta_2$). It reads, 
\begin{eqnarray}
{\cal M}_5=\left(
\begin{array}{cccccccccccc}
\lambda_1 &  \lambda_5 & 0 &  0 &  0 &  0  & 0 &  0  &  0  & 0  &  0  &  0  \\
\lambda_5 \,&\, \lambda_2 \,&\, 0 \,&\, 0 \,&\, 0 \,&\, 0  \,&\,0 \,&\, 0  \,&\, 0  \,&\,0 \,&\, 0  \,&\, 0  \\
0 \,&\, 0 \,&\, 2\tilde{\bar{\lambda}}_{89} \,&\, 0 \,&\, 0 \,&\, 0  \,&\,0 \,&\, 0  \,&\, 0  \,&\,0 \,&\, 0  \,&\, 0  \\
0 \,&\, 0 \,&\, 0 \,&\, \lambda_{34}^+ \,&\, 0 \,&\, 0  \,&\,0 \,&\, 0  \,&\, 0  \,&\,0 \,&\, 0  \,&\, 0  \\
0 \,&\, 0 \,&\, 0 \,&\, 0 \,&\, \tilde{\lambda}_{68} \,&\, 0  \,&\,0 \,&\, -\frac{\lambda_{8}}{2}  \,&\, 0  \,&\,0 \,&\, -i\frac{\lambda_{8}}{2}  \,&\, 0  \\
0 \,&\, 0 \,&\, 0 \,&\, 0 \,&\, 0 \,&\,  \tilde{\lambda}_{79}  \,&\,0 \,&\, 0  \,&\, -\frac{\lambda_{9}}{2}  \,&\,0 \,&\, 0  \,&\, -i\frac{\lambda_{9}}{2}  \\
0 \,&\, 0 \,&\, 0 \,&\, 0 \,&\, 0 \,&\, 0  \,&\,2\bar{\lambda}_8 \,&\, 0  \,&\, 0  \,&\,0 \,&\, 0  \,&\, 0  \\
0 \,&\, 0 \,&\, 0 \,&\, 0 \,&\, -\frac{\lambda_{8}}{2}  \,&\, 0  \,&\,0 \,&\, \lambda_{6}  \,&\, 0  \,&\,0 \,&\, 0  \,&\, 0  \\
0 \,&\, 0 \,&\, 0 \,&\, 0 \,&\, 0 \,&\, -\frac{\lambda_{9}}{2}   \,&\,0 \,&\, 0   \,&\,\lambda_7 \,&\,0 \,&\, 0  \,&\, 0  \\
0 \,&\, 0 \,&\, 0 \,&\, 0 \,&\, 0 \,&\, 0  \,&\,0 \,&\, 0  \,&\, 0  \,&\, 2\bar{\lambda}_8 \,&\, 0  \,&\, 0  \\
0 \,&\, 0 \,&\, 0 \,&\, 0 \,&\, i\frac{\lambda_{8}}{2} \,&\, 0  \,&\,0 \,&\, 0  \,&\, 0  \,&\,0 \,&\, \lambda_6  \,&\, 0  \\
0 \,&\, 0 \,&\, 0 \,&\, 0 \,&\, 0  \,&\, i\frac{\lambda_{9}}{2}  \,&\,0 \,&\, 0  \,&\, 0  \,&\,0 \,&\, 0   \,&\, \lambda_7  \\
\end{array}
\right)
\end{eqnarray}
There are also triply charged states. The submatrix ${\cal M}_6$ corresponding to this case generates the scattering with
initial and final states being one of the following ($\delta^{++}\phi_1^+$, $\delta^{++}\phi_2^+$, $\delta^{++}\delta^+$), and is given by, 
\begin{eqnarray}
{\cal M}_6=\left(
\begin{array}{ccc}
\lambda_{68}^+ & 0 & 0 \\
0 & \lambda_{79}^+ & 0 \\
0 & 0 & 2\bar{\lambda}_{89} \\
\end{array}
\right)
\end{eqnarray}
and finally, it is easy to check that there is just one quadruply charged state $\frac{1}{\sqrt{2}}\delta^{++}\delta^{++}$, leading to
${\cal M}_7 = 4\,\bar{\lambda}_{89}^+$.

\section{Boundedness from below Constraints}
\label{bfb-appendix}
To proceed to the most general case, we adopt a different parameterization of the fields that will turn out to be
particularly convenient to entirely solve the problem. for that we combine both parameterizations used in \cite{Arhrib:2011uy} and \cite{Bonilla2015} and define:
\begin{eqnarray}
 r &\equiv& \sqrt{H_1^\dagger{H_1} + H_2^\dagger{H_2} + Tr\Delta^{\dagger}{\Delta}} \label{eq:para1}\\
 H_1^\dagger{H_1} &\equiv& r^2 \cos^2 \theta \sin^2 \phi  \label{eq:para2}\\
 H_2^\dagger{H_2} &\equiv& r^2 \sin^2 \theta \sin^2 \phi  \label{eq:para3}\\
 Tr\Delta^{\dagger}{\Delta} &\equiv& r^2 \cos^2 \phi  \label{eq:para4}\\
 Tr(\Delta^{\dagger}{\Delta})^2/(Tr\Delta^{\dagger}{\Delta})^2 &\equiv& \epsilon  \label{eq:para5}\\
  (H_1^\dagger{\Delta}{\Delta}^{\dagger}H_1)/(H_1^\dagger{H_1}Tr\Delta^{\dagger}{\Delta}) &\equiv&  \eta  \label{eq:para6}\\
  (H_2^\dagger{\Delta}{\Delta}^{\dagger}H_2)/(H_2^\dagger{H_2}Tr\Delta^{\dagger}{\Delta}) &\equiv&  \zeta \label{eq:para7}
\end{eqnarray}
\noindent
Obviously, when $H_1$, $H_2$ and $\Delta$ scan all the field space, the radius $r$ scans the domain $[0, \infty[$, the angle $\theta \in [0, 2 \pi]$ and the angle $\phi \in [0, \frac{\pi}{2}]$. Moreover, as $\frac{H_1^\dagger\cdot H_2}{|H_1||H_2|}$ is a product of unit spinor, it is a complex number $\alpha + i \beta$ such that $|\alpha + i \beta| \le\,1$. We can rewrite it in polar coordinates as $\alpha + i \beta = \xi e^{i\psi}$ with $\xi \in (0,1)$. We can also show that $\eta \in (0,1)$, $\zeta \in (0,1)$ and $\epsilon \in [\frac{1}{2},1]$.

\noindent
With this parameterization, one can cast $V^{(4)}(H_1,H_2,\Delta)$ in the following simple form,
\begin{eqnarray}
V^{(4)}(r,c^2_\theta,s^2_\phi,c_{2\psi},\xi,\epsilon,\eta,\zeta) &=& r^4\Big\{ \lambda_1c^4_\theta s^4_\phi + \lambda_2s^4_\theta s^4_\phi + \lambda_3 c^2_\theta s^2_\theta s^4_\phi + \lambda_4 c^2_\theta s^2_\theta s^4_\phi \xi^2 + \lambda_5c^2_\theta s^2_\theta s^4_\phi \xi^2 \cos2\psi \nonumber\\
&& + c^4_\phi (\bar{\lambda}_8 + \epsilon\bar{\lambda}_9) +  c^2_\theta c^2_\phi s^2_\phi (\lambda_6 + \eta\lambda_8) + s^2_\theta c^2_\phi s^2_\phi (\lambda_7 + \zeta\lambda_9)\Big\}
\label{eq:V4general2}
\end{eqnarray}
\noindent
To simplify we take :

\begin{eqnarray}
 x &\equiv& \cos^2\theta \\
 y &\equiv& \sin^2\phi  \\
 z &\equiv& \cos2\psi \in (-1,1)
\end{eqnarray}
which allows us to rewrite the potential in the final form :
\begin{eqnarray}
V^{(4)}/r^4 &=& \big\{\frac{\lambda_1}{2}\,x^2 + \frac{\lambda_2}{2}\,(1-x)^2 + \lambda_3\,x(1-x) + \lambda_4\,x(1-x)\xi^2 + \lambda_5\,x(1-x)\xi^2\,z \big\}\,y^2\nonumber\\
&+&  \big\{\bar{\lambda}_8 + \epsilon\bar{\lambda}_9\big\}\,(1-y)^2\nonumber\\
&+&  \big\{ (\lambda_6 + \eta\lambda_8)\,x + (\lambda_7 + \zeta\lambda_9)\,(1-x)\big\}\,y(1-y)
\label{eq:V4general3}
\end{eqnarray}
it is easy to find the constraint conditions by studying $V^{(4)}(x,y,z,\xi,\epsilon,\eta,\zeta) $ as a quadratic function using the fact that :
\begin{eqnarray}
f(y) =  a\,y^2 + b\,(1-y)^2 + c\,y\,(1 - y),\quad y \in (0,1)\quad \Leftrightarrow \quad a > 0,\, b > 0\,\,{\rm and}\,\,c +2\sqrt{ab} > 0\label{lemme}\nonumber\\
\end{eqnarray}
we can deduce the set of constraints as :
\begin{eqnarray}
 F_{I}(\xi,z) &\equiv& \frac{\lambda_1}{2}\,x^2 + \frac{\lambda_2}{2}\,(1-x)^2 + \lambda_3\,x(1-x) + \lambda_4\,x(1-x)\xi^2 + \lambda_5\,x(1-x)\xi^2\,z  > 0 \label{condA}\\
 F_{II}(\epsilon) &\equiv& \bar{\lambda}_8 + \epsilon\bar{\lambda}_9 > 0 \label{condB}  \\
 F_{III}(\eta,\zeta) &\equiv& (\lambda_6 + \eta\lambda_8)\,x + (\lambda_7 + \zeta\lambda_9)\,(1-x) > -2\sqrt{F_{I}(\xi,z)\,F_{II}(\epsilon)} \label{condC}
\end{eqnarray}
For $F_{I}(\xi,z) > 0$ one can use Eq \ref{lemme} again to get the ordinary 2${\rm HDM}$ BFB constraints taking into account if $\xi = {0;1}$ and $z={-1;1}$ :
\begin{eqnarray}
 \lambda_1\,,\,\lambda_2 &>& 0 \label{2hdm_1}\\
 \lambda_3 + \sqrt{\lambda_1\lambda_2} &>& 0 \label{2hdm_2}  \\
 \lambda_3 + \lambda_4 - |\lambda_5| + \sqrt{\lambda_1\lambda_2} &>& 0 \label{2hdm_3} 
\end{eqnarray}
For $F_{II}(\epsilon)$ which is monotonic function, the condition $0<F_{II}(\epsilon)$ is s equivalent to $0<F_{II}(\frac{1}{2})$ and $0<F_{II}(1)$. So that Eq. \ref{condB} becomes,
\begin{eqnarray}
\bar{\lambda}_8 + \bar{\lambda}_9 > 0 \,\,,\qquad \bar{\lambda}_8 + \frac{1}{2}\bar{\lambda}_9 > 0 \label{generic_condB}
\end{eqnarray}
For Eq. \ref{condC}, one can write it as:
\begin{eqnarray}
F_{III}(\eta,\zeta) + 2\sqrt{F_{I}(\xi,z)\,F_{II}(\epsilon)} > 0 \, \Leftrightarrow \, \left\{
\begin{aligned}
F_{III}(\eta,\zeta)\,&> 0&\text{and} && F_{I}(\xi,z) F_{II}(\epsilon) > 0 &&\text{(i)}\\
&&\text{or}\\
F_{III}(\eta,\zeta)\,&\leqslant 0&\text{and} && 4\,F_{I}(\xi,z) F_{II}(\epsilon) > F^2_{III}(\eta,\zeta) &&\text{(ii)}\\
\end{aligned}
\right.
\end{eqnarray}

\begin{figure}[!h]
\centering
\begin{minipage}{6.5cm}
\begin{center}
\includegraphics[height =5cm,width=6.5cm]{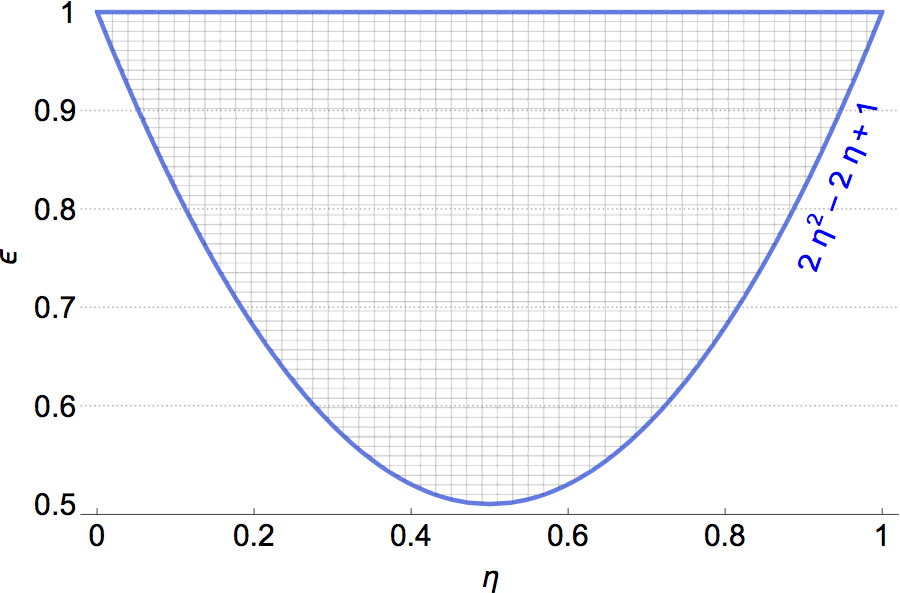}
\end{center}
\end{minipage}
\hspace{0.5cm}
\begin{minipage}{6.5cm}
\begin{center}
\includegraphics[height =5cm,width=6.5cm]{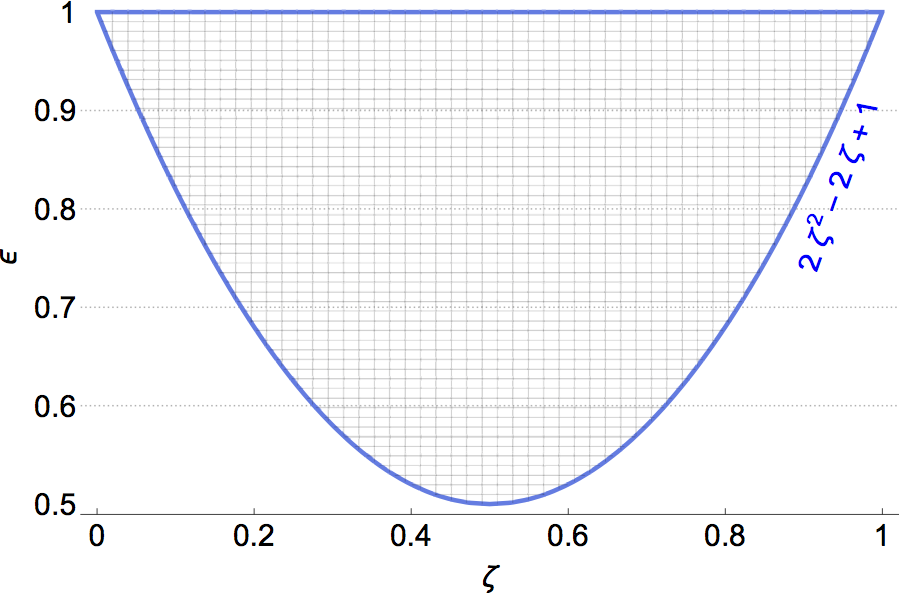}
\end{center}
\end{minipage}
\caption[titre court]{Dependence of parameters : $(\eta,\epsilon)$ (left) and $(\zeta,\epsilon)$ (right) showing the allowed shaded regions.}
\label{fig:etazetaeps1}
\end{figure}

\begin{itemize}
\item scenario (i) : starting with the fact that $x = \cos^2\theta > 0$ and $1-x =\sin^2\theta > 0$, thus $F_{III}(\eta,\zeta) > 0  \Rightarrow $ generic relations :
\begin{eqnarray} 
&&\lambda_6 + \eta\lambda_8 > 0  \,\,,\qquad\forall \eta \in [0, 1]\label{generic1_condC}\\
&&\lambda_7 + \zeta\lambda_9 > 0  \,\,,\qquad\forall \zeta \in [0, 1]\label{generic2_condC} 
\end{eqnarray}
We note here that $\eta$ and $\zeta$ are independent parameters  since there are no quartic coupling  linking together $H_1$, $H_2$ and $\Delta$. Consequently the two monotonic function in \ref{generic1_condC} and \ref{generic1_condC} leads as above to,
\begin{eqnarray}
\lambda_6  > 0 \,\,,\qquad \lambda_6 + \lambda_8 > 0 
\,\,,\qquad 
\lambda_7  > 0 \,\,,\qquad \lambda_7 + \lambda_9 > 0  
\label{generic3_condC}
\end{eqnarray}
\item scenario (ii) : this scenario imply both $(\lambda_6 + \eta\lambda_8)\,{\rm and}\,(\lambda_7 + \zeta\lambda_9) \le 0$ and leads to:
\begin{eqnarray}
&&\Big\{2\lambda_1(\bar{\lambda}_8 + \epsilon\bar{\lambda}_9) - (\lambda_6 + \eta\lambda_8)^2\Big\}\,x^2 + 
\Big\{2\lambda_2(\bar{\lambda}_8 + \epsilon\bar{\lambda}_9) - (\lambda_7 + \zeta\lambda_9)^2\Big\}\,(1-x)^2 \nonumber\\
&&+ \Big\{ 4(\lambda_3 + \lambda_4 \xi^2 + \lambda_5 \xi^2 z)(\bar{\lambda}_8 + \epsilon\bar{\lambda}_9) - 2(\lambda_6 + \eta\lambda_8)\,(\lambda_7 + \zeta\lambda_9) \Big\}x\,(1 - x) > 0\nonumber\\
\end{eqnarray}
 Applying the {\it lemme} given by Eq \ref{lemme}, we get the following generic new constraints,
\begin{eqnarray}
&& \lambda_6 + \eta\lambda_8 > -\sqrt{2\lambda_1(\bar{\lambda}_8 + \epsilon\bar{\lambda}_9)} \label{generic3_condC}\\
&& \lambda_7 + \zeta\lambda_9 > -\sqrt{2\lambda_2(\bar{\lambda}_8 + \epsilon\bar{\lambda}_9)} \label{generic4_condC}\\
&& 4(\lambda_3 + \lambda_4 \xi^2 + \lambda_5 \xi^2 z)(\bar{\lambda}_8 + \epsilon\bar{\lambda}_9) - 2(\lambda_6 + \eta\lambda_8)\,(\lambda_7 + \zeta\lambda_9) > \nonumber\\
&&\hspace{1cm}-2\sqrt{\bigg(2\lambda_1(\bar{\lambda}_8 + \epsilon\bar{\lambda}_9) - (\lambda_6 + \eta\lambda_8)^2\bigg)\bigg(2\lambda_2(\bar{\lambda}_8 +\epsilon\bar{\lambda}_9) - (\lambda_7 + \zeta\lambda_9)^2\bigg)}
\label{generic5_condC} 
\end{eqnarray}
\end{itemize}
To continue, some major adjustments are needed to complete the set of BFB constraints. Firstly, we note further that each of two parameters ($\eta$, $\epsilon$) and ($\zeta$, $\epsilon$) can not be anywhere in their small ranges bounded by the vertices $(0,\frac{1}{2})$ and $(1,1)$ each. Indeed, from Eqs. [\ref{eq:para5}-\ref{eq:para6}] and [\ref{eq:para5}-\ref{eq:para7}], it can be shown that the possible values of ($\eta$, $\epsilon$) and ($\zeta$, $\epsilon$) correspond respectively to:
\begin{eqnarray}
&& 2 \eta^2 - 2 \eta + 1 \le  \epsilon \le 1, \label{eq:etaepsilon} \\
&& 2 \zeta^2 - 2 \zeta + 1 \le  \epsilon \le 1 \label{eq:zetaepsilon} 
\end{eqnarray}
which define the shaded region depicted in Fig. \ref{fig:etazetaeps1}.

Returning to the equations \ref{generic3_condC} and \ref{generic4_condC}, they may be formulated as,
\begin{eqnarray}
0 < \lambda_6 + \eta\lambda_8 + \sqrt{2\lambda_1(\bar{\lambda}_8 + \epsilon\bar{\lambda}_9)} &\equiv& G_I(\eta,\epsilon)\label{new_generic3_condC}\\
0 < \lambda_7 + \zeta\lambda_9 + \sqrt{2\lambda_2(\bar{\lambda}_8 + \epsilon\bar{\lambda}_9)} &\equiv& G_{II}(\zeta,\epsilon)\label{new_generic4_condC}
\end{eqnarray}
and see how to get rid of $\eta$, $\zeta$ and $\epsilon$. The minimum of $G_I(\eta,\epsilon)$ and $G_{II}(\zeta,\epsilon)$ occurs at the border of the shaded regions in figure \ref{fig:etazetaeps1} since both functions are monotonic in both ($\eta$, $\epsilon$) and ($\zeta$, $\epsilon$) respectively. Otherwise, $0<G_I(\eta,\epsilon) \Leftrightarrow 0<\text{min}\,G_I(\eta,\epsilon)$ $\forall\,\eta,\epsilon$, and by the same way $0<G_{II}(\zeta,\epsilon) \Leftrightarrow 0<\text{min}\,G_{II}(\zeta,\epsilon)$ $\forall\,\zeta,\epsilon$. Such minimums that are part of the lines defined by $\epsilon=2\eta^2-2\eta+1$ and $\epsilon=2\zeta^2-2\zeta+1$. One redefines, 
\begin{eqnarray}
\widehat{G}_I(\eta)=G_I(\eta,2\eta^2-2\eta+1)\quad \text{and} \quad \widehat{G}_{II}(\zeta)=G_{II}(\zeta,2\zeta^2-2\zeta+1)
\end{eqnarray}
with the examination of $\widehat{G}_I(\eta)$ and $\widehat{G}_{II}(\zeta)$ convexities $\forall$ $\eta\in[0, 1]$ and $\eta\in[0, 1]$. Noticing here that
\begin{eqnarray}
\text{sgn}[\widehat{G}^{ ''}_I(\eta)]&=&\text{sgn}[4\lambda_1^2\bar{\lambda}_9(2\bar{\lambda}_8+\bar{\lambda}_9)] = \text{sgn}[\bar{\lambda}_9] \\
\text{sgn}[\widehat{G}^{ ''}_{II}(\zeta)]&=&\text{sgn}[4\lambda_2^2\bar{\lambda}_9(2\bar{\lambda}_8+\bar{\lambda}_9)] = \text{sgn}[\bar{\lambda}_9] 
\end{eqnarray}
where \ref{generic_condB} is taken into account.\\
Therefore, one can always find a values $\eta_0$ (resp. $\zeta_0$) for which $\widehat{G}^{'}_I(\eta_0)=0$ (resp. $\widehat{G}^{'}_{II}(\zeta_0)=0$), so that its may be a minimum only when $\widehat{G}^{''}_I(0)>0$ $\land$ $\widehat{G}^{''}_I(1)<0$ (resp. $\widehat{G}^{''}_{II}(0)>0$ $\land$ $\widehat{G}^{''}_{II}(1)<0$). This will be true if and only if $\bar{\lambda}_9\sqrt{2\lambda_1} \ge |\lambda_8|\sqrt{\bar{\lambda}_8+\bar{\lambda}_9}$ (resp. $\bar{\lambda}_9\sqrt{2\lambda_2} \ge |\lambda_9|\sqrt{\bar{\lambda}_8+\bar{\lambda}_9}$). In this case,
\begin{eqnarray}
\widehat{G}_I(\eta_0)=\lambda_6+\frac{1}{2}\lambda_8+\frac{1}{2}\sqrt{\bigg(4\lambda_1\bar{\lambda}_9-\lambda_8^2\bigg)\bigg(1+2\frac{\bar{\lambda}_8}{\bar{\lambda}_9}\bigg)}\label{eq:mineta0}\\
\widehat{G}_{II}(\zeta_0)=\lambda_7+\frac{1}{2}\lambda_9+\frac{1}{2}\sqrt{\bigg(4\lambda_2\bar{\lambda}_9-\lambda_9^2\bigg)\bigg(1+2\frac{\bar{\lambda}_8}{\bar{\lambda}_9}\bigg)}\label{eq:minzeta0}
\end{eqnarray}
We note also that the minimums may situate at $\eta=0 \lor \eta=1$ (resp. $\zeta=0 \lor \zeta=1$) from which we get the constraints that both $\widehat{G}_I(0)=\lambda_6+\sqrt{2\lambda_1 (\bar{\lambda}_8+\bar{\lambda}_9)}$ and $\widehat{G}_I(1)=\lambda_6+\lambda_8+\sqrt{2\lambda_1 (\bar{\lambda}_8+\bar{\lambda}_9)}$ should be positive quantities (resp. $\widehat{G}_{II}(0)=\lambda_7+\sqrt{2\lambda_2 (\bar{\lambda}_8+\bar{\lambda}_9)}$ and $\widehat{G}_{II}(1)=\lambda_7+\lambda_9+\sqrt{2\lambda_2 (\bar{\lambda}_8+\bar{\lambda}_9)}$).

As for the equation \ref{generic5_condC} its implies, 
\begin{eqnarray}
&&\hspace{-0.0cm} 0< g(\xi,z)(\bar{\lambda}_8 + \epsilon\bar{\lambda}_9) - 2(\lambda_6 + \eta\lambda_8)\,(\lambda_7 + \zeta\lambda_9) + \nonumber\\
&&\hspace{0cm} 2\sqrt{\bigg(2\lambda_1(\bar{\lambda}_8 + \epsilon\bar{\lambda}_9) - (\lambda_6 + \eta\lambda_8)^2\bigg)\bigg(2\lambda_2(\bar{\lambda}_8 +\epsilon\bar{\lambda}_9) - (\lambda_7 + \zeta\lambda_9)^2\bigg)}  \equiv G_{III}(\eta,\zeta,\epsilon)
\end{eqnarray}
where $g(\xi,z)=4(\lambda_3 + \lambda_4 \xi^2 + \lambda_5 \xi^2 z)$.
\begin{figure}[!h]
\centering
\begin{minipage}{8.5cm}
\begin{center}
\includegraphics[height =8.5cm,width=8.5cm]{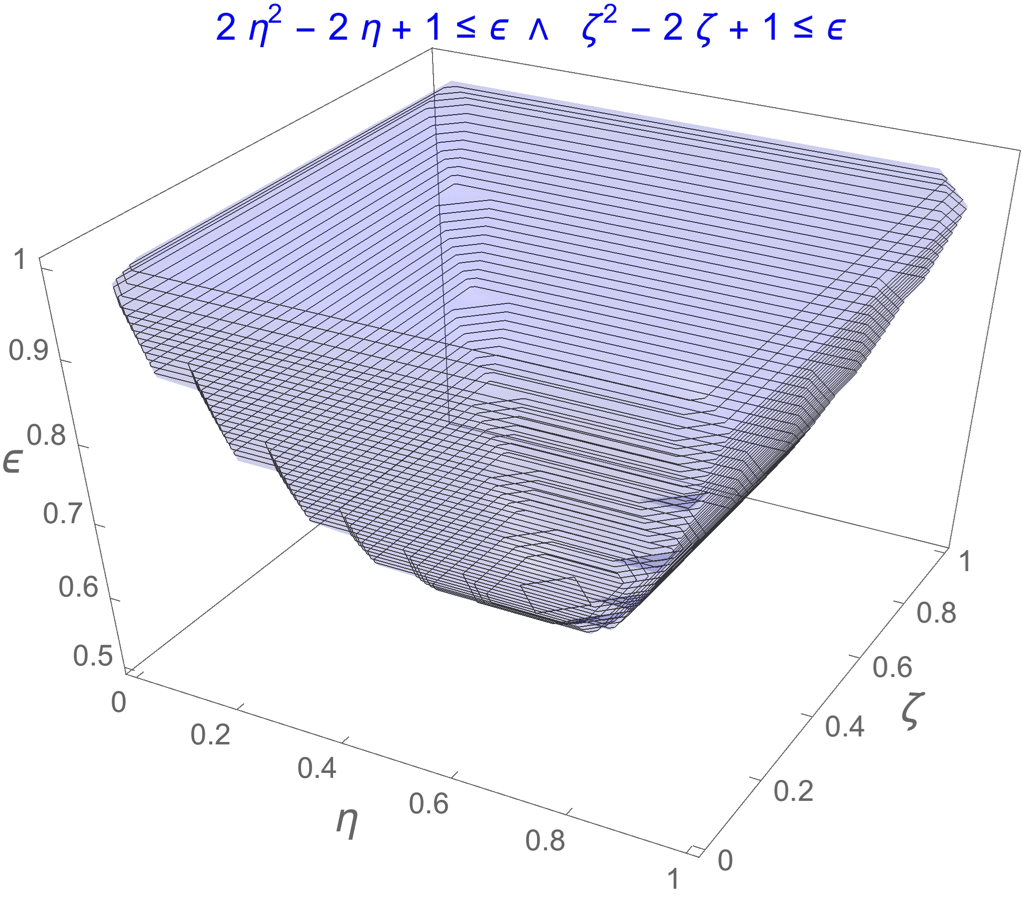}
\end{center}
\end{minipage}
\caption[titre court]{The simultaneous variation of $\epsilon$ as a function of $\eta$ and $\zeta$ showing the allowed dashed-line region.}
\label{fig:etazetaeps2}
\end{figure}

The minimum of $G_{III}(\eta,\zeta,\epsilon)$, as previously, occurs at the border of the dashed line in figure \ref{fig:etazetaeps2}, so that we can replace either $\epsilon \rightarrow 2 \eta^2 - 2 \eta + 1$ or $\epsilon \rightarrow 2 \zeta^2 - 2 \zeta + 1$ in which function of two independent variables could be simplified. If the first proposal was retained, the function $G_{III}(\eta,\zeta,\epsilon)$ becomes
\begin{eqnarray}
&&\widehat{G}_{III}(\eta,\zeta) \equiv g(\xi,z) \bigg(\bar{\lambda}_8 + \bar{\lambda}_9 (2 \eta^2 - 2 \eta + 1)\bigg) - 2(\lambda_6 + \eta\lambda_8)\,(\lambda_7 + \zeta\lambda_9)\\
&&+2\sqrt{\bigg(2\lambda_1\bigg(\bar{\lambda}_8 + \bar{\lambda}_9(2 \eta^2 - 2 \eta + 1)\bigg) - (\lambda_6 + \eta\lambda_8)^2\bigg)\bigg(2\lambda_2\bigg(\bar{\lambda}_8 +\bar{\lambda}_9(2 \eta^2 - 2 \eta + 1)\bigg) - (\lambda_7 + \zeta\lambda_9)^2\bigg)}\nonumber
\end{eqnarray}
In order to find out if such function is convex or not, differentiation is needed and one can easily check that:
\begin{eqnarray}
\Delta\widehat{G}_{III}(\eta,\zeta) \equiv 
\left(
\begin{array}{c} 
\displaystyle \frac{\partial\widehat{G}_{III}}{\partial\eta} \\
\displaystyle \frac{\partial\widehat{G}_{III}}{\partial\zeta}  
\end{array}\right) \equiv
\left(
\begin{array}{c} 
\displaystyle F_1(\lambda_i,\eta,\zeta) \\
\displaystyle F_2(\lambda_i,\eta,\zeta) 
\end{array}\right)
\end{eqnarray}
where the functions $F_1,F_2$ have a complex forms in $\lambda_i,\eta,\zeta$.

\noindent
Solving the above two stationary points give rise to one solution given by,
\begin{eqnarray}
\big(\eta_0,\zeta_0\big)_{1}, \big(\eta_0,\zeta_0\big)_{2}, \big(\eta_0,\zeta_0\big)_{3}\, \text{and}\, \big(\eta_0,\zeta_0\big)_{4} 
\label{eq:minimum}
\end{eqnarray}
Therefore, we examine the Hessian of $\widehat{G}_{III}$ as follows,
\begin{eqnarray}
H = \left(
\begin{array}{cc} 
\displaystyle \frac{\partial^2\widehat{G}_{III}}{\partial^2\eta} & \displaystyle \frac{\partial^2\widehat{G}_{III}}{\partial\eta\partial\zeta} \\
\displaystyle \frac{\partial^2\widehat{G}_{III}}{\partial\zeta\partial\eta} & \displaystyle \frac{\partial^2\widehat{G}_{III}}{\partial^2\zeta}
\end{array}\right)=
\left(
\begin{array}{cc} 
\displaystyle H_{11} & \displaystyle H_{12} \\
\displaystyle H_{21} & \displaystyle H_{22}
\end{array}\right)
\end{eqnarray}
where the matrix elements $H_{ij}=H_{ij}(\lambda_i,\eta,\zeta)$.

At the points given in \ref{eq:minimum}, the eigenvalues of H can be expressed in terms of $\lambda's$ by solving the following equation:
\begin{eqnarray}
|H-\Lambda \mathds{1}_{2}| = 
\begin{vmatrix}
H_{11} - \Lambda & H_{12} \\ 
H_{21} & H_{22} - \Lambda 
\end{vmatrix}
\end{eqnarray}
Hence, at this stage, and for $\widehat{G}_{III}$ to be convex, the eigenvalues $\Lambda_1$ and $\Lambda_2$ for each point should be positive quantities. Within sight of their long and complicated expressions,  we will not show them here, though they would be taken into account in our calculations. Furthermore, one must also make sure that for each point, both $0\le \eta_0, \zeta_0 \le 1$ for which we request that,
\begin{eqnarray}
\widehat{G}_{III}(\eta_0,\zeta_0)>0\nonumber
\end{eqnarray}
$\forall$ $\xi\in[0,1]$ and $z\in[-1,1]$.

\noindent
The remaining possibility is that the minimum of $\widehat{G}_{III}$ is at $(\eta,\zeta)=$ (0,0), (0,1), (1,0) or (1,1), from which we get the constraints that

\begin{eqnarray}
&& 0<(\bar{\lambda}_8+\bar{\lambda}_9)g(\xi,z)-2(\lambda_6+\lambda_8)(\lambda_7+\lambda_9)+2\sqrt{\bigg(2\lambda_1(\bar{\lambda}_8+\bar{\lambda}_9)-(\lambda_6+\lambda_8)^2\bigg)\bigg(2\lambda_2(\bar{\lambda}_8+\bar{\lambda}_9)-(\lambda_7+\lambda_9)^2\bigg)}\nonumber\\
&& 0<(\bar{\lambda}_8+\bar{\lambda}_9)g(\xi,z)-2(\lambda_6+\lambda_8)\lambda_7+2\sqrt{\bigg(2\lambda_1(\bar{\lambda}_8+\bar{\lambda}_9)-(\lambda_6+\lambda_8)^2\bigg)\bigg(2\lambda_2(\bar{\lambda}_8+\bar{\lambda}_9)-(\lambda_7+\lambda_9)^2\bigg)}\\
&& 0<(\bar{\lambda}_8+\bar{\lambda}_9)g(\xi,z)-2\lambda_6\lambda_7+2\sqrt{\bigg(2\lambda_1(\bar{\lambda}_8+\bar{\lambda}_9)-\lambda_6^2\bigg)\bigg(2\lambda_2(\bar{\lambda}_8+\bar{\lambda}_9)-\lambda_7^2\bigg)}\\
&& 0<(\bar{\lambda}_8+\bar{\lambda}_9)g(\xi,z)-2\lambda_6(\lambda_7+\lambda_9)+2\sqrt{\bigg(2\lambda_1(\bar{\lambda}_8+\bar{\lambda}_9)-\lambda_6^2\bigg)\bigg(2\lambda_2(\bar{\lambda}_8+\bar{\lambda}_9)-\lambda_7^2\bigg)}
\end{eqnarray}

\providecommand{\href}[2]{#2}\begingroup\raggedright\endgroup

\begin{thebibliography}{10}

\expandafter\ifx\csname natexlab\endcsname\relax\def\natexlab#1{#1}\fi
\expandafter\ifx\csname bibnamefont\endcsname\relax
\def\bibnamefont#1{#1}\fi
\expandafter\ifx\csname bibfnamefont\endcsname\relax
\def\bibfnamefont#1{#1}\fi
\expandafter\ifx\csname citenamefont\endcsname\relax
\def\citenamefont#1{#1}\fi
\expandafter\ifx\csname url\endcsname\relax
\def\url#1{\texttt{#1}}\fi
\expandafter\ifx\csname urlprefix\endcsname\relax\def\urlprefix{URL }\fi
\providecommand{\bibinfo}[2]{#2}
\providecommand{\eprint}[2][]{\url{#2}}



\bibitem{Aad:2012tfa}
G.~Aad {\it et al.} [ATLAS Collaboration],
Phys.\ Lett.\ B {\bf 716} (2012) 1.

\bibitem{Chatrchyan:2012ufa}
S.~Chatrchyan {\it et al.} [CMS Collaboration],
Phys.\ Lett.\ B {\bf 716} (2012) 30.

\bibitem{Patrignani:2016xqp}
C.~Patrignani {\it et al.} [Particle Data Group],
Chin.\ Phys.\ C {\bf 40} (2016) no.10, 100001.
  
\bibitem{Weinberg:1979sa}
S.~Weinberg,
Phys.\ Rev.\ Lett.\  {\bf 43} (1979) 1566.
    
\bibitem{Minkowski:1977sc}
P.~Minkowski,
Phys.\ Lett.\  B {\bf 67} (1977) 421.
  
\bibitem{GellMann:1980vs}
M.~Gell-Mann, P.~Ramond and R.~Slansky,
Conf.\ Proc.\ C {\bf 790927} (1979) 315
  
\bibitem{Yanagida:1979as} 
T.~Yanagida,
Conf.\ Proc.\ C {\bf 7902131}, 95 (1979).
  
\bibitem{Mohapatra:1979ia} 
R.~N.~Mohapatra and G.~Senjanovic,
Phys.\ Rev.\ Lett.\  {\bf 44}, 912 (1980).
  
\bibitem{Schechter:1980gr}
J.~Schechter and J.~W.~F.~Valle,
Phys.\ Rev.\ D {\bf 22} (1980) 2227.
  
\bibitem{Mohapatra:1980yp}
R.~N.~Mohapatra and G.~Senjanovic,
Phys.\ Rev.\ D {\bf 23} (1981) 165.
  
\bibitem{Lazarides:1980nt}
G.~Lazarides, Q.~Shafi and C.~Wetterich,
Nucl.\ Phys.\ B {\bf 181} (1981) 287.
\bibitem{Wetterich:1981bx}
C.~Wetterich,
Nucl.\ Phys.\ B {\bf 187} (1981) 343.
  
\bibitem{Schechter:1981cv}
J.~Schechter and J.~W.~F.~Valle,
Phys.\ Rev.\ D {\bf 25} (1982) 774.
  
\bibitem{Foot:1988aq}
R.~Foot, H.~Lew, X.~G.~He and G.~C.~Joshi,
Z.\ Phys.\ C {\bf 44} (1989) 441.
  

\bibitem{Arhrib:2011uy}
A.~Arhrib, R.~Benbrik, M.~Chabab, G.~Moultaka, M.~C.~Peyranere, L.~Rahili and J.~Ramadan,
Phys.\ Rev.\ D {\bf 84} (2011) 095005.
 
\bibitem{Arhrib:2011vc}
A.~Arhrib, R.~Benbrik, M.~Chabab, G.~Moultaka and L.~Rahili,
JHEP {\bf 1204} (2012) 136.
   
\bibitem{Chabab:2014ara}
M.~Chabab, M.~C.~Peyranere and L.~Rahili,
Phys.\ Rev.\ D {\bf 90} (2014) no.3,  035026.
   
\bibitem{Chabab:2015nel}
M.~Chabab, M.~C.~Peyranere and L.~Rahili,
Phys.\ Rev.\ D {\bf 93} (2016) no.11, 115021.


\bibitem{Arhrib:2014nya}
A.~Arhrib, R.~Benbrik, G.~Moultaka and L.~Rahili,
arXiv:1411.5645 [hep-ph].
 
    
\bibitem{Camargo:2018uzw}
D.~A.~Camargo, A.~G.~Dias, T.~B.~de Melo and F.~S.~Queiroz,
arXiv:1811.05488 [hep-ph].
  
  
\bibitem{Chen-Nomura-2014}
C.~H.~Chen and T.~Nomura,
Phys.\ Rev.\ D {\bf 90} (2014) no.7,  075008
 

 
\bibitem{Arhrib:2000is}
A.~Arhrib,'Unitarity constraints on scalar parameters of the standard and two Higgs doublets model,'
hep-ph/0012353.
  
\bibitem{Akeroyd:2000wc} 
A.~G.~Akeroyd, A.~Arhrib and E.~M.~Naimi,
Phys.\ Lett.\ B {\bf 490} (2000) 119.

 \bibitem{veltman75}
 D.~Ross and M.~Veltman, 
 Nucl.\ Phys.\ B {\bf 95} (1975) 135;
 M.~Veltman, 
 Nucl.\ Phys.\ B {\bf 123} (1977) 89. 

\bibitem{pdg2016}
C.~Patrignani {\it et al.} [Particle Data Group],
Chin.\ Phys.\ C {\bf 40} (2016) no.10,  100001.


\bibitem{negU}
A.~Goudelis, B.~Herrmann and O.~Stal,
JHEP {\bf 1309} (2013) 106

\bibitem{sharma2012}
L.~Lavoura and L.~F.~Li,
Phys.\ Rev.\ D {\bf 49} (1994) 1409.


\bibitem{HiggsBounds}
P.~Bechtle, O.~Brein, S.~Heinemeyer, G.~Weiglein and K.~E.~Williams,
Comput.\ Phys.\ Commun.\  {\bf 182} (2011) 2605.
;
P.~Bechtle, O.~Brein, S.~Heinemeyer, G.~Weiglein and K.~E.~Williams,
Comput.\ Phys.\ Commun.\  {\bf 181} (2010) 138.

\bibitem{masse_higgs}
G.~Aad {\it et al.} [ATLAS and CMS Collaborations],
Phys.\ Rev.\ Lett.\  {\bf 114} (2015) 191803.


\bibitem{HiggsSignals}
P.~Bechtle, S.~Heinemeyer, O.~Stal, T.~Stefaniak and G.~Weiglein,
Eur.\ Phys.\ J.\ C {\bf 74} (2014) no.2,  2711.





\bibitem{ATLAS1}
G.~Aad {\it et al.} [ATLAS Collaboration],
JHEP {\bf 1601} (2016) 032


\bibitem{ATLAS2}
G.~Aad {\it et al.} [ATLAS Collaboration],
Eur.\ Phys.\ J.\ C {\bf 76} (2016) no.1,  45


\bibitem{ATLAS3}
V.~Khachatryan {\it et al.} [CMS Collaboration],
JHEP {\bf 1510} (2015) 144


\bibitem{ATLAS4}
''Search for a high-mass Higgs boson decaying to a pair of W bosons in pp collisions at
$\sqrt(s)=13$ TeV with the ATLAS detector,'' Tech. Rep. ATLAS-CONF-2016-021, CERN,
Geneva, Apr, 2016. \eprint{http://cds.cern.ch/record/2147445.}


\bibitem{ATLAS5}
''Search for ZZ resonances in the qq final state in pp collisions at $\sqrt{s}$=13 TeV with the
ATLAS detector'', Tech. Rep. ATLAS-CONF-2016-016, CERN, Geneva, Mar, 2016.  \eprint{https://cds.cern.ch/record/2141005.} 


\bibitem{ATLAS6}
"Search for high-mass resonances decaying into a Z boson pair in the $ll\nu\nu$ final state in pp
collisions at $\sqrt{s}$=13 TeV with the ATLAS detector,'' Tech. Rep. ATLAS-CONF-2016-012,
CERN, Geneva, Mar, 2016. \eprint{http://cds.cern.ch/record/2140833}


\bibitem{ATLAS7} 
''Search for new phenomena in the Z($\rightarrow$ll)+$E^{miss}_T$ final state at $\sqrt{s}$=13 TeV with the ATLAS detector,'' Tech. Rep. ATLAS-CONF-2016-056,
CERN, Geneva, Aug, 2016. \eprint{https://cds.cern.ch/record/2206138}

\bibitem{ATLAS88}
''Search for high mass Higgs to WW with fully
leptonic decays using 2015 data,'' Tech. Rep. CMS-PAS-HIG-16-023, CERN, Geneva, 2016.
\eprint{{https://cds.cern.ch/record/2205151}}
\bibitem{Sushi1}
R.~V.~Harlander, S.~Liebler and H.~Mantler,
Comput.\ Phys.\ Commun.\  {\bf 184} (2013) 1605


\bibitem{Sushi2}
R.~V.~Harlander, S.~Liebler and H.~Mantler,
Comput.\ Phys.\ Commun.\  {\bf 212} (2017) 239


\bibitem{ATLAS10}
''Searches for heavy ZZ and ZW resonances in the
llqq and vvqq final states in pp collisions at $\sqrt(s)=13$ TeV with the ATLAS detector,''
Tech. Rep. ATLAS-CONF-2016-082, CERN, Geneva, Aug, 2016.
\eprint{{http://cds.cern.ch/record/2206275.}}


\bibitem{ATLAS11}
''Study of the Higgs boson properties and search
for high-mass scalar resonances in the $H\rightarrow ZZ^{*} \rightarrow 4l$ decay channel at $\sqrt(s)=13$ TeV with
the ATLAS detector,'' Tech. Rep. ATLAS-CONF-2016-079, CERN, Geneva, Aug, 2016. 
\eprint{{http://cds.cern.ch/record/2206253.}}

\bibitem{ATLAS12}
''Search for diboson resonance production in the
$lvqq$ final state using pp collisions at $\sqrt(s)=13$ with the ATLAS detector at the LHC,''
Tech. Rep. ATLAS-CONF-2016-062, CERN, Geneva, Aug, 2016.
\eprint{{http://cds.cern.ch/record/2206199.}}
\bibitem{ATLAS13}
''Search for a high-mass Higgs boson decaying to a
pair of W bosons in pp collisions at $\sqrt(s)=13$ TeV with the ATLAS detector,'' Tech. Rep.
ATLAS-CONF-2016-074, CERN, Geneva, Aug, 2016.
\eprint{{http://cds.cern.ch/record/2206243.}}


\bibitem{ATLAS14}
"Search for additional heavy neutral Higgs and
gauge bosons in the ditau final state produced in
36.1 $fb^{-1}$ of pp collisions at $\sqrt{s}= 13\,TeV$ with the
ATLAS detector,'' Tech. Rep. ATLAS-CONF-2017-050, CERN, July, 2017. 
\eprint{https://cds.cern.ch/record/2273866}

\bibitem{ATLAS15}
''Search for a neutral MSSM Higgs boson decaying
into $\tau\tau$ with 12.9 $fb^{-1}$
of data at $\sqrt{s}=13$ TeV,'' Tech. Rep. CMS-PAS-HIG-16-037, CERN,
Geneva, 2016. 
\eprint{http://cds.cern.ch/record/2231507.}


\bibitem{ATLAS16}
''Search for scalar diphoton resonances with
15.4 $fb^{-1}$ of data collected at $\sqrt{s}=13$ TeV in 2015 and 2016 with the ATLAS detector,''
Tech. Rep. ATLAS-CONF-2016-059, CERN, Geneva, Aug, 2016.
\eprint{http://cds.cern.ch/record/2206154}



\bibitem{Bonilla2015}
C.~Bonilla, R.~M.~Fonseca and J.~W.~F.~Valle,
Phys.\ Rev.\ D {\bf 92} (2015) no.7,  075028
\end{thebibliography}
\end{document}